# Deep Learning for Mortgage Risk

Justin A. Sirignano, Apaar Sadhwani, Kay Giesecke*

September 15, 2015; this version: March 8, 2018†

**Abstract**

We develop a deep learning model of multi-period mortgage risk and use it to analyze an unprecedented dataset of origination and monthly performance records for over 120 million mortgages originated across the US between 1995 and 2014. Our estimators of term structures of conditional probabilities of prepayment, foreclosure and various states of delinquency incorporate the dynamics of a large number of loan-specific as well as macroeconomic variables down to the zip-code level. The estimators uncover the highly nonlinear nature of the relationship between the variables and borrower behavior, especially prepayment. They also highlight the effects of local economic conditions on borrower behavior. State unemployment has the greatest explanatory power among all variables, offering strong evidence of the tight connection between housing finance markets and the macroeconomy. The sensitivity of a borrower to changes in unemployment strongly depends upon current unemployment. It also significantly varies across the entire borrower population, which highlights the interaction of unemployment and many other variables. These findings have important implications for mortgage-backed security investors, rating agencies, and housing finance policymakers.

*Sirignano (jasirign@illinois.edu) is from the University of Illinois at Urbana-Champaign. Sadhwani (apaar@google.com) is from Google Brain, and Giesecke (giesecke@stanford.edu) is from Stanford University. The majority of this work was completed while Sirignano and Sadhwani were doctoral students at Stanford.

†The authors gratefully acknowledge support from the National Science Foundation through Methodology, Measurement, and Statistics Grant No. 1325031 as well as from the Amazon Web Services in Education Grant award. We are very grateful to Michael Ohlrogge, Andreas Eckner, Jason Su, and Ian Goodfellow for comments. Chris Palmer, Richard Stanton, and Amit Seru, our discussants at the Macro Financial Modeling Winter 2016 Meeting, provided insightful comments on this work, for which we are very grateful. We are also grateful for comments from the participants of the Macro Financial Modeling Winter 2016 Meeting, the 7th General AMaMeF and Swissquote Conference in Lausanne, the Western Conference on Mathematical Finance, the Machine Learning in Finance Conference at Columbia University, the Consortium of Data Analytics in Risk Symposium, and seminar participants at Columbia University, UC Berkeley, Northwestern University, New York University, UT Austin, Imperial College London, Fannie Mae, Freddie Mac, Federal Housing Finance Administration, Federal Reserve Board, International Monetary Fund, Federal Reserve Bank of San Francisco, Morgan Stanley, J.P. Morgan, Georgia State University, Payoff, Bank of England, and Winton Capital Management. We are also grateful to Powerlytics, Inc. for providing access to income data. Luyang Chen provided excellent research assistance, for which we are very grateful.

# 1 Introduction

The empirical mortgage literature identifies a number of variables that help predict mortgage credit and prepayment risk, including borrower credit score and income, loan-to-value ratio, loan age, interest rates, and housing prices.[1] To ensure econometric tractability, researchers often impose restrictions on the empirical models they use to study the role of various risk factors. Importantly, the relationship between factors and mortgage performance is typically constrained to be of a pre-specified form, with the standard choice being linear. The mortgage performance data, however, do not support such restrictions. They indicate the presence of nonlinear effects. For example, Figure 1 highlights the complex relationship that exists between the empirical prepayment rate and the prepayment incentive, given by the initial mortgage rate minus the market rate.[2] Consider the sensitivity of prepayment rates to changes in incentive, which is a measure of the economic importance of the incentive variable. The sensitivity varies significantly, both in magnitude as well as sign, depending on the incentive. The widely-used linear empirical models can be mis-specified because they pretend the sensitivity is a constant. The sensitivity estimates generated by these models can therefore misrepresent the influence of risk factors. This can make it difficult to draw valid economic conclusions from these models regarding the influence on borrower behavior of key variables such as interest rates, unemployment, and housing prices, which play a major role in housing finance markets and the wider economy.

This paper proposes a nonlinear approach to address this important issue. We develop a deep learning model of mortgage credit and prepayment risk in which the relationship between risk factors and loan performance is not predicated on a pre-specified form as in prior empirical models. In our approach, this relationship is entirely dictated by the data themselves, minimizing model mis-specification and bias of variable sensitivity estimates. Any type of behavior is permitted, including nonlinear behavior such as interactions between multiple variables. An unprecedented dataset of over 120 million mortgages enables us to accurately estimate this behavior. The data include prime and subprime loans in more than 30,000 zip-codes across the nation, a wide range of mortgage products, and detailed origination and monthly performance records for each loan. We study over 3.5 billion loan-month observations that span 1995 to 2014, and examine the role of a broad set of novel and conventional risk factors, including loan and borrower-specific variables as well as time-varying macroeconomic and demographic variables down to the zip-code level (see Tables 1, 5, and 7 for a complete list of variables).

[1] See, for example, Campbell & Dietrich (1983), Cunningham & Capone (1990), Curley & Guttentag (1974), Deng, Quigley & Van Order (2000), Gau (1978), Green & Shoven (1986), Schwartz & Torous (1993), Titman & Torous (1989), Vandell (1978), Vandell (1992), von Furstenberg (1969), Webb (1982), and others.

[2] This is based on loan performance data described in Section 2.

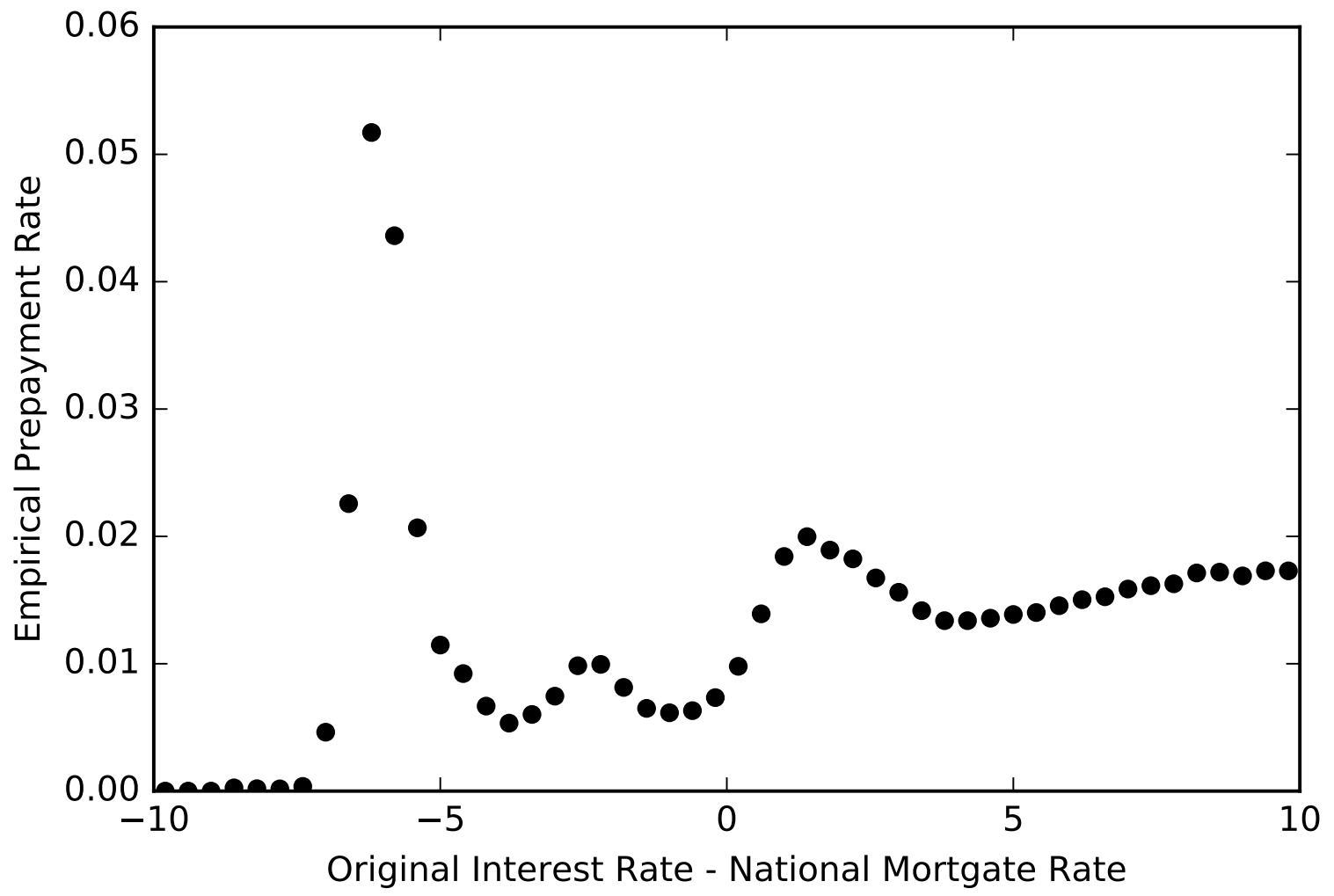

Figure 1: Empirical monthly prepayment rate vs. prepayment incentive.

Our empirical analysis reveals that many variables have a highly nonlinear influence on borrower behavior which prior work does not address. Prepayment events are especially affected. Variable interactions, including those between more than two factors, are found to represent a significant component of the nonlinear effects. An interaction between variables occurs when the sensitivity of loan performance to a variable also depends on one or more other variables. As an important example, consider the impact of state unemployment rates, which we find to have the greatest explanatory power for borrower behavior among all variables studied. Figure 2 shows fitted conditional prepayment probabilities versus state unemployment rates, for a borrower with a FICO credit score of 630 (the median for subprime borrowers in our data set) as well as a borrower with a FICO score of 730 (the median for prime borrowers). The relationship between prepayment and unemployment is highly nonlinear, and strongly depends on the borrower's credit score, indicating an interaction between unemployment and credit score. All else equal, high-FICO borrowers are estimated to prepay at significantly higher rates than low-FICO borrowers in all unemployment scenarios. The "prepayment gap" between high- and low-FICO borrowers tends to widen as unemployment grows. With low unemployment between 5 and 7 percent, low- and high-FICO borrowers are equally sensitive to changes in unemployment. However, with moderate unemployment between 7 and 11 percent, low-FICO borrowers are significantly more sensitive to changes in unemployment than high-FICO borrowers. While high-FICO borrower prepayment is essentially flat in this unemployment range, low-FICO borrower prepayment decreases significantly with unemployment rising from 7 to 11 percent. The prepayment sensitivities of

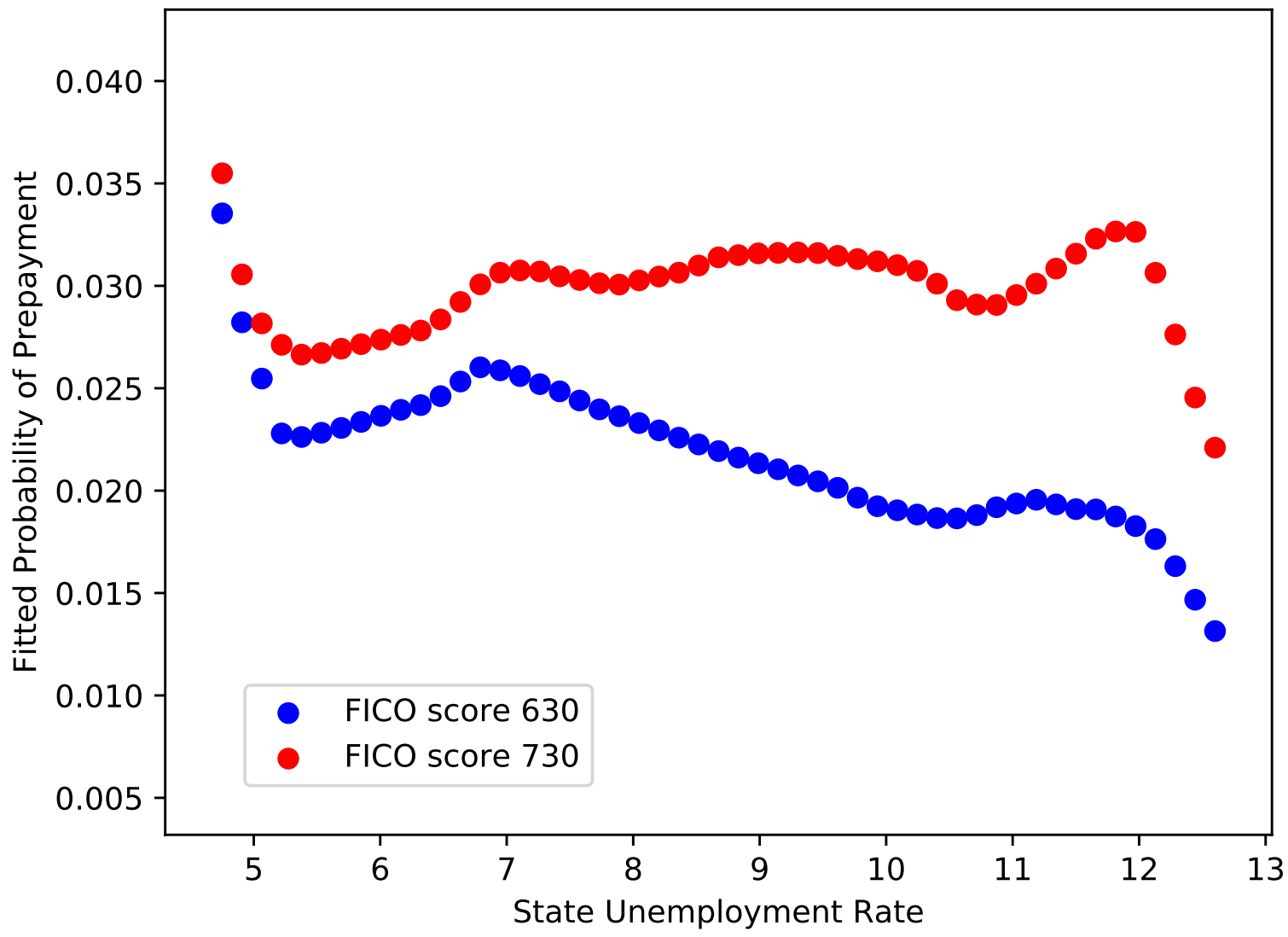

Figure 2: Fitted monthly prepayment probability vs. state unemployment.

high- and low-FICO borrowers converge as unemployment rises above 11 percent.

Unemployment is found to interact in nontrivial ways with a range of other variables including loan-to-value ratios, mortgage rates, and house price appreciation. The effects are not limited to prepayment risk but also pertain to the credit risk of a mortgage borrower. For example, Figure 3 shows fitted conditional delinquency probabilities versus zip-code level house price appreciation since loan origination, for two state unemployment scenarios. As expected, the probability of a borrower to fall behind payment drops when house prices appreciate and borrower equity increases in value. However, the behavior is not linear. The sensitivity of the delinquency rate to changes in house prices strongly depends on the appreciation already realized. The higher that appreciation, the smaller is the reduction in the delinquency rate in response to additional price increases. Moreover, the behavior of the delinquency rate as a function of house price appreciation strongly depends on state unemployment (an interaction effect). Unsurprisingly, the delinquency rate increases with state unemployment. Interestingly, however, when state unemployment is high, the delinquency rate drops much more in response to an appreciation in house prices than when unemployment is low. The gap between delinquency rates in different unemployment scenarios narrows as house prices increase, and vanishes completely when prices have doubled. This means that labor market conditions hardly matter for borrower behavior after house prices have sufficiently appreciated. This, in turn, suggests that home equity can insulate borrowers from labor market shocks.

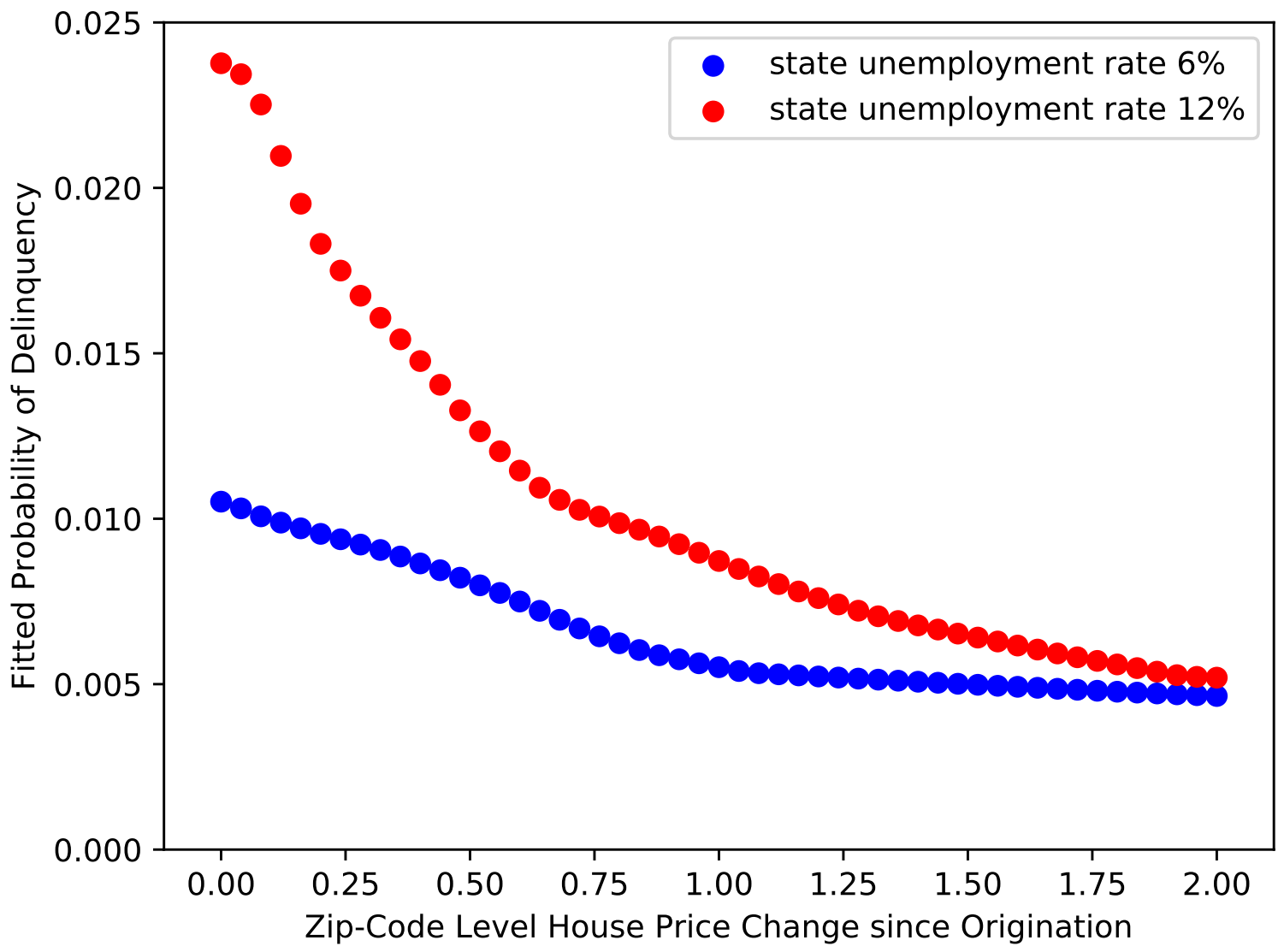

Figure 3: Fitted monthly delinquency probability vs. house price appreciation.

Our findings yield important new insights into the interplay of borrower behavior and the macroeconomy. They differ from the findings of Campbell & Dietrich (1983), Cunningham & Capone (1990), Deng (1997), Elul, Souleles, Chomsisengphet, Glennon & Hunt (2010), Foote, Gerardi, Goette & Willen (2010) and others. These prior studies highlight loan-level variables such as loan-to-value ratio, loan age, and credit score as major predictors of borrower behavior. They assign a relatively limited role to unemployment. Foote et al. (2010), for example, find no evidence for the influence of unemployment on prepayment. Our results, which are based on a much longer sample period, imply that unemployment dominates the aforementioned and all other variables considered in terms of explanatory power for borrower behavior. This suggests a much tighter connection between housing finance markets and the macroeconomy than previously thought.

The important role of unemployment implies that the loan-to-loan correlation due to the exposure of borrowers to economic cycles can be substantial. This source of loan-to-loan correlation is distinct from the foreclosure contagion channel studied by Anenberg & Kung (2014), Campbell, Gigli & Pathak (2011), Towe & Lawley (2013), and others. We control for the contagion channel by including as a variable the lagged foreclosure rate at the zip-code level. The significance of loan-to-loan correlation due to the common exposure of borrowers to economic cycles emphasizes the need for mortgage-backed security investors as well as mortgage lenders to diversify mortgage risk geographically, beyond the conventional borrower characteristics highlighted in the literature.

The nonlinear nature of the relationship between unemployment and borrower behavior entails that the sensitivity of a borrower to changes in unemployment strongly depends on the prevailing unemployment rate as well as the borrower's characteristics. For example, a 10% drop in unemployment from 7% to 6.3% affects a given borrower differently than one from 10% to 9%. Moreover, due to multiple variable interactions, the effect varies very significantly across the entire borrower population, not just across borrowers with different loan-to-value ratios and credit scores as suggested by Elul et al. (2010) and Foote et al. (2010) in the case of borrower default. This is important for mortgage-backed security investors, who need to account for these effects when hedging their positions against macroeconomic volatility. Rating agencies need to address the nonlinear behavior when assessing the exposure of mortgage securities to adverse macroeconomic conditions. Many prior articles studying the role of unemployment, for example Campbell & Dietrich (1983), Cunningham & Capone (1990), Deng (1997) and others, ignore the pronounced nonlinear behavior by assuming that the sensitivity of borrowers to changes in unemployment is constant and independent of all other variables including borrower characteristics.

Our empirical results are based on a deep learning model for mortgage states over multiple periods. The model harnesses the unprecedented size of our sample set and the large number of risk factors we examine (272 in total). It is designed to capture the nonlinear relationships between variables and mortgage state probabilities present in the data. The model distinguishes between multiple states, including current, 30 days past due, 60 days past due, 90+ days past due, foreclosure, REO (real-estate-owned), and prepaid. It offers likelihood estimators for the term structure of the full conditional transition probability matrix for these states. The estimators incorporate the significant time-series variation of the explanatory variables over the 20-year sample period as well as their future movements. They are shown to yield superior out-of-sample predictions of multi-period mortgage risk at the individual loan level as well as the mortgage pool level. We also demonstrate that they enable the selection of performing mortgage investment portfolios. The tests indicate the ability of the deep learning model to capture the stand-alone risk of a loan as well as the substantial correlation that exists between the loans in a portfolio. The model's predictive accuracy suggests its usefulness for several important applications, including the valuation of mortgage-backed securities as in Curley & Guttentag (1977), Schwartz & Torous (1989) and Stanton & Wallace (2011).[3]

[3] The valuations generated by our model would harness the detailed data available for each of the underlying loans. This contrasts with alternative "top-down" valuation approaches such as Schwartz & Torous (1989), McConnell & Singh (1994) and Boudoukh, Whitelaw, Richardson & Stanton (1997), who directly model the aggregate behavior of a pool without reference to the pool constituents.

Our deep learning model for mortgage state probabilities is a nonlinear extension of the familiar logistic regression model, which is widely used in the empirical mortgage literature[4] and beyond. It can be thought of as a logistic regression of recursively specified basis functions that nonlinearly transform the explanatory variables and are learned from the data. The model can also be represented by an interconnected set of input, output, and "hidden" nodes, which is often called a neural network.[5] The input nodes represent the explanatory variables while the output nodes represent the conditional probabilities of the different mortgage states (current, 30 days late, prepaid, foreclosed, etc.). The hidden nodes connect the input and output nodes, and represent the nonlinear transformations of input variables. Given enough hidden nodes, a neural network can approximate arbitrarily well the true mapping between explanatory variables and conditional mortgage state probabilities.[6] This of course includes approximating nonlinear relations and interactions such as the product and division of variables.

In particular, we examine *deep* neural networks, which have multiple layers of hidden nodes. Deep architectures enable sparser representations of complex relationships than shallow networks with few hidden layers.[7] Our fitting experiments with networks of different depth indicate a strong preference for deeper networks, highlighting the existence of highly nonlinear relationships and variable interactions in the mortgage data. The optimal network architecture, determined via cross-validation methods, has five layers of hidden nodes, each having between 140 and 200 nodes. We develop computationally efficient maximum likelihood fitting algorithms that take advantage of recent advances in GPU parallel and cloud computing. Overfitting is tightly controlled by regularization, dropout, and ensemble modeling, and as a result is found to be insignificant.

The remainder of the introduction discusses the related literature. Section 2 examines our dataset and performs some exploratory analyses that will inform the specification of our deep learning model in Section 3. Section 4 discusses likelihood estimation for the deep learning model. Section 5 reports our empirical results. Section 6 examines the out-of-sample behavior of the deep learning model. Section 7 offers concluding remarks. There are several technical appendices.

---

[4]See Campbell & Dietrich (1983), Cunningham & Capone (1990), and more recently, Agarwal, Amromin, Ben-David, Chomsisengphet & Evanoff (2011), Agarwal, Chang & Yavaz (2012), Jiang, Nelson & Vytlacil (2014), and Rajan, Seru & Vig (2015).

[5]For a broad introduction to deep learning, see White (1992) and Goodfellow, Bengio & Courville (2016).

[6]More precisely, a neural network can approximate arbitrarily well continuous functions on compact sets, see Hornik, Stinchcombe & White (1989) and Hornik (1991).

[7]See Mallat (2016), Telgarsky (2016), Eldan & Shamir (2016), Bengio & LeCun (2007), and Montufar, Pascanu, Cho & Bengio (2014).

## 1.1 Related Literature

There is a substantial empirical literature on mortgage delinquency and prepayment risk. In early work, von Furstenberg (1969) establishes the influence on home mortgage default rates of variables such as income, loan age, and loan-to-value ratio. Gau (1978), Vandell (1978), Webb (1982), Campbell & Dietrich (1983) and others examine additional variables. Commercial mortgage default is studied by Titman & Torous (1989) and Vandell (1992), among others. Curley & Guttentag (1974) is an early study of prepayment rates. Green & Shoven (1986) and Richard & Roll (1989) examine the influence of interest rates on prepayments. Cunningham & Capone (1990), Schwartz & Torous (1993), and Deng (1997) analyze the influence on default and prepayment of several loan-level and macro-economic variables, recognizing the "competing" nature of default and prepayment events. Deng et al. (2000) analyze the extent to which option theory can explain default and prepayment behavior. Schwartz & Torous (1989) pioneered the use of empirical pool-level prepayment models for the pricing of agency mortgage-backed securities. More recently, Stanton & Wallace (2011) use empirical models of default and prepayment to price private-label MBS. Chernov, Dunn & Longstaff (2016) estimate market-implied risk-neutral prepayment rates and relate them to various explanatory variables.

This study of mortgage credit and prepayment risk represents a significant departure from earlier work, in several respects. Our empirical analysis is based on an unprecedented dataset of 120 million prime and subprime mortgages observed over the period 1995–2014. The aforementioned prior work has examined much smaller samples (tens to hundreds of thousands of loans), focusing on particular geographic regions, time periods, economic regimes, loan products, borrower profiles, and a limited set of loan-level and macro-economic risk factors. It is unclear to what extent the empirical findings of these earlier studies can be generalized. Our dataset includes about 70 percent of all US mortgages originated between 1995 and 2014, and is the most comprehensive mortgage data set studied to date. It covers all product types, including fixed-rate, adjustable-rate, hybrid, balloon, and other types of loans, and tracks their performance during several economic cycles. With samples spanning two decades and spread across over 30,000 zip codes, we are in a position to study the joint influence on mortgage risk of a broad set of novel and conventional risk factors that describe a variety of borrower and product characteristics as well as economic and demographic conditions down to the zip-code level. Our results highlight the importance of the local economic conditions that borrowers face, after controlling for the influence of the factors that prior studies have identified as significant predictors of mortgage risk. In particular, we identify state unemployment rates as the factor with the greatest explanatory power for borrower behavior. Some prior studies, focusing on short sample periods and few controls, found unemployment to be dominated by borrower variables (e.g., Deng (1997), Elul et al. (2010)) while others found little or no evidence for the influence of unemployment rates. For example, Foote et al. (2010) found unemployment to be insignificant for prepayment. Our findings, being based on the richest mortgage performance and covariate data sets studied to date, settle the important role played by unemployment. They provide strong evidence of the tight connection between housing finance markets and the macroeconomy.

Our econometric model addresses the nonlinear relationships between borrower behavior and risk factors that prior work has largely ignored. Most previous papers use logistic regression (Campbell & Dietrich (1983), Cunningham & Hendershott (1986), Elul et al. (2010), Agarwal et al. (2011), and others) and Cox proportional hazard models (Green & Shoven (1986), Deng et al. (2000), Stanton & Wallace (2011), and others).[8] These models are also standard in the empirical corporate credit literature, see Campbell, Hilscher & Szilagyi (2008), Duffie, Saita & Wang (2007), Shumway (2001), and several others. The Cox models sometimes include a nonparametric baseline hazard function that can capture a nonlinear influence of loan age on mortgage risk. Both Cox and logistic regression models sometimes include quadratic or other nonlinear transformations of certain variables. For example, Agarwal et al. (2012) use the squared loan age as risk factor in addition to loan age itself, Elul et al. (2010) discretize continuous variables such as the loan-to-value ratio, and Foote et al. (2010) include certain pairwise interaction terms. Unlike these extensions of linear models, our deep learning model is inherently nonlinear. It captures *all* nonlinear effects, including variable interactions of any order, that exist in the data. Our model eliminates bias due to linear model mis-specification, and yields accurate estimates of the economic significance of the explanatory variables. Moreover, it eliminates the need for the econometrician to identify and specify the nonlinearities ahead of time. The practice of using certain nonlinear transformations in a linear model requires the identification of the variables to be transformed and the specification of the transformations to be used. To this end, the researcher must systematically explore a potentially very large number of possible relationships between variables and outcomes as well as interactions that might exist between variables. This approach is impractical with more than just a few variables. Our deep learning model automatically identifies all nonlinearities that are present in the data, even in high-dimensional settings with many risk factors.

The fitted deep learning model provides a detailed picture of the nonlinear effects governing borrower behavior. We see exactly how important variables such as unemployment rates, housing prices, and credit scores jointly influence borrower behavior. This includes understanding the sensitivity of borrowers to changes in variables, and how this sensitivity varies with all other factors. For example, we find that the sensitivity of borrowers to move-ments in unemployment rates is not constant as the linear relationship assumed by Campbell & Dietrich (1983), Cunningham & Capone (1990), Deng (1997) and others would suggest, but varies significantly with unemployment itself as well as a host of other variables. This highlights the nonlinear effects pertaining to unemployment. In particular, unemployment strongly interacts with many other risk factors, not just with loan-to-value ratios and credit scores in the context of default as suggested by Elul et al. (2010) and Foote et al. (2010), respectively. Our findings uncover the nature of the relationship between housing finance markets and the macroeconomy, and how it depends upon borrower and other characteristics.

[8]Other approaches include Poisson regression (Schwartz & Torous (1993)), kernel regression (Maxam & LaCour-Little (2001)) and radial basis function networks (Episcopos, Pericli & Hu (1998)).

We analyze the behavior of borrowers at an unprecedented level of granularity. We distinguish between multiple states, including current, 30 days behind payment, 60 days behind payment, 90+ days behind payment, foreclosure, REO, and prepaid, and estimate the full conditional transition probability matrix for these states. The aforementioned papers only consider transitions from current to prepayment or default, usually meaning a severe delinquency (such as 60 days or more late). However, this simplified treatment ignores important transitions between other delinquency states such as 30 days late, 60 days late, 90+ days late, and foreclosure. Our data indicates that transitions between these states are in fact frequent; see Tables 8–10. For example, a meaningful number of loans enter foreclosure but eventually return to current. Similarly, many loans are consistently behind payment but do not ever enter foreclosure. This behavior often matters. For example, during periods of delinquency Fannie Mae and Freddie Mac suffer disruptions to the cashflow from a loan that need to be considered when analyzing their capital needs (see Frame, Gerardi & Willen (2015)). Our model enables an econometric treatment of this behavior and our empirical results offer insights into the factors influencing it.

Several authors have used shallow neural networks in other areas of financial economics. Bansal & Viswanathan (1993) approximate the pricing kernel using a neural network. Hutchinson, Lo & Poggio (1994) pioneered the use of neural networks for nonparametric option pricing. Brown, Goetzmann & Kumar (1998) use neural networks to predict stock markets. Swanson & White (1997) propose the use of neural networks for macroeconomic forecasting; Donaldson & Kamstra (1996) use them for dividend projections, and Elliott & Timmermann (2008) discuss other applications in economic forecasting. Lee, White & Granger (1993) construct tests for neglected nonlinearities in time series models using neural networks. White (1989), Granger (1995), and Kuan & White (1994) study nonlinear or neural network modeling of financial time series. Khandani, Kim & Lo (2010) and Butaru, Chen, Clark, Das, Lo & Siddique (2016) examine other machine learning models of financial default. Recent applications of deep learning in financial economics include Sirignano (2016), who models limit order books and Dixon, Klabjan & Bang (2016), who model market movements. Heaton, Polson & Witte (2016) use deep learning for portfolio selection.

# 2 The Data

Our dataset includes data for over 120 million mortgages as well as local and national economic factors. The mortgage dataset includes highly-detailed characteristics for each loan and month-by-month loan performance. We complement this dataset with extensive local economic data such as housing prices, incomes, and unemployment rates.

## 2.1 Loan Performance and Feature Data

The mortgage data was licensed from CoreLogic, who collects the data from mortgage originators and servicers. It is the most comprehensive mortgage data set studied to date. It covers roughly 70% of all mortgages originated in the US and contains mortgages from over 30,000 zip codes across the US. The mortgages' origination dates range from January 1995 to June 2014. The dataset includes 25 million subprime and 93 million prime mortgages.[9] The loan data is divided into (1) loan features at origination and (2) performance data, which we describe below.

Each mortgage has a number of detailed features at origination, such as borrower FICO score, original loan-to-value (LTV) ratio, original debt-to-income (DTI) ratio, original balance, original interest rate, product type, type of property, prepayment penalties, zip code, state, and many more. Many of the variables are categorical with many categories (some with up to 20 categories). Table 1 provides a complete list of the features we consider. Only relatively small subsets of these variables have been considered in prior work. Tables 2 through 4 provide summary statistics for FICO score, original LTV ratio, original interest rate, and original balance. The median FICO score of subprime borrowers is 630, while that of prime borrowers is 730. The median interest rate of subprime loans is 7.8 percent, while that for prime loans is 5.8 percent.

Month-by-month performance for each mortgage is reported between 1995 and 2014. This includes how many days behind payment the mortgage currently is, the current interest rate, current balance, whether the mortgage is real estate owned (REO), is in foreclosure, or has paid off. It also includes variables representing borrower behavior over the recent past, such as the familiar burnout factor and novel factors such as the number of delinquencies (30 days late, 60 days late) over the past 12 months, which have not been considered before. Table 5 provides the full list of performance features.

[9] We adopt CoreLogic's designation of loans as subprime vs. prime. These designations are based on their categorizations by the originators and servicers who provide the mortgage data to CoreLogic. Loan characteristics such as FICO score, documentation status, and product features such as negative amortization are often used in practice to distinguish between subprime and prime loans. Our approach reflects the way that these loans are viewed by the key economic actors in the mortgage market.

The dataset covers various mortgage products including, for example, fixed rate mortgages, adjustable rate mortgages (ARMs), hybrid mortgages, and balloon mortgages.[10] Table 6 lists the fraction of mortgages in each product category. The vast majority of the prime mortgages are fixed-rate (86%), followed by ARMs (9%) and hybrids (4%). 48% of the subprime mortgages are fixed-rate, 29% are ARMs, and 9% are hybrids. Prior work has typically focused on a particular product such as fixed-rate loans.

Every monthly observation from each of the loans constitutes a data sample. After cleaning the data as described in Appendix A, there are roughly 3.5 billion monthly observations remaining. 90% of the samples are for prime mortgages and the remaining are for subprime. The samples cover the period January, 1995 to May, 2014. Each sample (i.e., monthly observation) has 272 explanatory variables as well as the outcome for that month (i.e., if the loan is current, 30 days delinquent, 60 days delinquent, etc.). Of these explanatory variables, 234 are loan-level feature and performance variables, 25 are indicators for missing features (see Appendix A), and 13 are economic variables which are introduced next.

## 2.2 Local and National Economic Factors

We complement the loan-level data described above by data for local and national economic factors which may influence loan performance. Table 7 lists the factors we consider. We use a mortgage's zip code to match a mortgage with local factors such as the monthly housing price in that zip code. Housing prices are obtained from Zillow and the Federal Housing Administration (FHA). Zillow housing prices are at the five-digit zip code level and cover roughly 10,000 zip codes. In order to cover less populated areas not covered by the Zillow dataset, we also include FHA housing prices which cover all three-digit zip codes. The monthly national mortgage rate is obtained from Freddie Mac, is also included as a factor.[11] Unemployment rates at the county level for each year and state unemployment rates for each month are obtained from the Bureau of Labor Statistics.[12] Our data also includes the yearly median income in each zip code, which was acquired from the data provider Powerlytics. Moreover, we include a dummy variable for the vintage year. Finally, the granular geographic data is used to construct the lagged default and prepayment rates in each zip code across the US, using the historical data for all mortgages. The inclusion of these rates allows us to capture a potential contagion effect where defaults of mortgages increase the likelihood of default for nearby surviving mortgages. Such a feedback mechanism has been supported by several recent empirical papers; see Agarwal, Ambrose & Yildirim (2015), Anenberg & Kung (2014), Campbell et al. (2011), Harding, Rosenblatt & Yao (2009), Lin, Rosenblatt & Yao (2009), Towe & Lawley (2013), and others.

[10] A fixed rate mortgage has constant interest and principal payments over the lifetime of the mortgage. An ARM has interest payments which fluctuate with some other index interest rate (such as the Treasury rate) plus some fixed margin. A hybrid mortgage has a period with a fixed rate followed by a period with an adjustable rate. Hybrid mortgages can also have other features such as interest rate caps. A balloon mortgage only partially amortizes; a portion of the loan principal is due at maturity.

[11] The monthly national mortgage rate used in this paper is an average of 30 year fixed rates for first-lien prime conventional conforming home purchase mortgages with a loan-to-value of 80 percent.

[12] We match counties and zip codes in order to associate each mortgage with a particular county.

## 2.3 Mortgage States and Transitions

Mortgages are allowed to transition between 7 states: current, 30 days delinquent, 60 days delinquent, 90+ days delinquent, foreclosed, REO, and paid off. $X$ days delinquent simply means the mortgage borrower is $X$ days behind on their payments to the lender. We use the standard established by the Mortgage Bankers Association of America for determining the state of delinquency. A mortgage is determined to be 1 month delinquent if no payment has been made by the last day of the month and the payment was due on the first day of the month. REO stands for real estate owned property. When a foreclosed mortgage does not sell at auction, the lender or servicer will assume ownership of the property. Paid off can occur from a mortgage prepaying, maturing (this is very rare since the mortgages in the dataset are almost entirely originated in the 2000s), a shortsale, or a foreclosed mortgage being sold at auction to a third party (this is again rare in comparison to prepayments, which form the bulk of the paid off events in the dataset).[13]

The state transition matrix for the monthly transitions between states are given in Tables 8, 9, and 10 for subprime, prime, and all mortgages, respectively. The state transition matrix records the empirical frequency of the different types of transitions between states. For the calculation of these transition matrices and the remainder of the paper, REO and paid off are treated as absorbing states.[14] That is, we stop tracking the mortgage after the first time it enters REO or paid off. The transition matrices highlight that mortgages frequently transition back and forth between current and various delinquency states. Disruptions in cashflow to the lender or servicer are common due to the mortgage being behind payment. Similarly, even loans that are extremely delinquent may return to current; the transition from foreclosed back to current is actually a relatively frequent occurrence.[15] A foreclosure could get cured via paying the outstanding balance, there could be a pre-auction sale that covers all or some of the amount outstanding, or there could be a sale at the foreclosure auction that covers all or some of the amount outstanding. Any of these will register as a foreclosure to paid off transition. Mortgages can also transition directly from current, 30 days delinquent, 60 days delinquent, or 90+ days delinquent to REO via a "deed in lieu of foreclosure".[16]

[13]A foreclosed loan sold at auction may or may not be sold for a loss. The CoreLogic dataset makes no distinction between the two events.

[14]In some states in the USA there are laws that allow the mortgage borrower to reclaim their mortgage even after it has entered REO. However, such events are exceedingly rare.

[15]Many servicers follow a "dual path servicing approach" where they foreclose on the borrower as a threat in order to force the borrower to become current on payments.

## 2.4 Nonlinear Effects

The relationships between state transition rates and explanatory variables (i.e., loan-level features and economic factors) are often highly nonlinear. For instance, Figure 1 shows the empirical monthly prepayment rate versus the "incentive to prepay", initial interest rate minus national mortgage rate.[17] A higher interest rate on the loan (as compared to the national mortgage rate) should encourage the borrower to seek better terms by refinancing the loan, implying that an upward trend should be observed in the graph. The observed data, however, point to more complicated underlying mechanisms, such as the presence of prepayment penalties or the lack of refinancing options due to other factors such as low FICO scores. For example, a low initial interest rate may have been facilitated by prepayment penalties or points upfront, which will be disincentives to prepaying. Figure 4 shows the empirical monthly prepayment rate versus the FICO score. The propensity to prepay is less for borrowers with lower FICO scores but it plateaus once the score crosses a threshold of about 500 points. Figure 5 shows the empirical monthly prepayment rate versus the time since origination. Several spikes in the rate occur at $1, 2,$ and 3 years. These might be due to the expiration of prepayment penalties or adjustable rate and hybrid mortgages having rate resets. Many of the subprime mortgages started with low teaser rates and would later jump to higher rates; borrowers would refinance to avoid these rate jumps. Figure 6 plots the empirical monthly prepayment rate versus the loan-to-value (LTV) ratio. One should expect this curve to have a downward slope since a loan with high LTV will have lesser opportunities to refinance due to a large loan amount relative to the value of the asset. Each of these charts displays significant nonlinear relationships between the variable and the empirical prepayment rate. This reinforces the need for a loan performance model that is capable of addressing such relationships.

[16] A "deed in lieu of foreclosure" is when the loan is in default and the borrower gives ownership of the property directly to the lender, thereby forgoing foreclosure.

[17] A more accurate proxy for the incentive to prepay would be the current interest rate minus the mortgage rate. However, a large portion of the mortgages in the dataset are missing the current interest rate so the initial interest rate was used instead to achieve greater coverage.

# 3 Deep Learning Model

We propose a dynamic nonlinear model for the performance of a pool of mortgage loans over time. We adopt a discrete-time formulation for periods $0, 1, \dots, T$ (e.g., months).[18] We enumerate the possible mortgage states (current, 30 days delinquent, etc.), and let $\mathcal{U} \subset \mathbb{N}$ denote the set of these states. The variable $U_t^n \in \mathcal{U}$ prescribes the state of the $n$-th mortgage at time $t$ after origination. A mortgage will transition between the various states over its lifetime. For instance, a trajectory of the state process might be:

$$U_0^n = 1 \text{ (current)}, \quad U_1^n = 2 \text{ (30 days late)}, \quad U_2^n = 1 \text{ (current)}, \quad U_3^n = 5 \text{ (paid off)}.$$

We allow the dynamics of the state process to be influenced by a vector of explanatory variables $X_t^n \in \mathbb{R}^{d_X}$ which includes the mortgage state $U_t^n$. In our empirical implementation, $X_t^n$ represents the original and contemporary loan-level features in Tables 1 and 5, and the contemporary local and national economic factors in Table 7.[19] We specify a probability transition function $h_\theta : \mathcal{U} \times \mathbb{R}^{d_X} \to [0, 1]$ satisfying

$$\mathbb{P}[U_t^n = u \,|\, \mathcal{F}_{t-1}] = h_\theta(u, X_{t-1}^n), \qquad u \in \mathcal{U}, \tag{1}$$

where $\theta$ is a parameter to be estimated. Equation (1) gives the marginal conditional probability for the transition of the $n$-th mortgage from its state $U_{t-1}^n$ at time $t-1$ to state $u$ at time $t$ given the explanatory variables $X_{t-1}^n$. The family of conditional probabilities give a conditional transition probability matrix, which is the conditional counterpart of the empirical transition matrix reported in Table 10. Note that the conditional probabilities will be correlated across loans if $X_{t-1}^n$ includes variables that are common to several loans. This formulation allows us to capture loan-to-loan correlation due to geographic proximity and common economic factors.

We propose to model the transition function $h_\theta$ by a *neural network*. Let $g$ denote the standard softmax function:

$$g(z) = \left( \frac{e^{z_1}}{\sum_{k=1}^K e^{z_k}}, \dots, \frac{e^{z_K}}{\sum_{k=1}^K e^{z_k}} \right), \quad z = (z_1, \dots, z_K) \in \mathbb{R}^K, \tag{2}$$

where $K = |\mathcal{U}|$.[20] The vector output of the function $g$ is a probability distribution on $\mathcal{U}$.

[18] We fix a probability space $(\Omega, \mathcal{F}, \mathbb{P})$ and an information filtration $(\mathcal{F}_t)_{t=0,1,\dots,T}$.

[19] As usual, categorial variables are encoded in terms of indicator functions (dummy variables).

[20] Certain transitions are not allowed in the dataset (e.g., current to 60 days delinquent). Although such a transition is theoretically allowed in the formulation (2), the transition probabilities of transitions which do not occur in the dataset will be driven to zero during training.

The specification $h_\theta(u,x) = (g(Wx+b))_u$, where $W \in \mathbb{R}^K \times \mathbb{R}^{d_X}$, $b \in \mathbb{R}^K$, and $V_u$ is the $u$-th element of the vector $V$, gives a logistic regression model.[21] Here, the link function $g$ takes a *linear* function of the covariates $x$ as its input. The output $h_\theta(u,x)$ varies only in the constant direction given by $W$. A standard approach to achieve a more complex model with greater flexibility is to replace $x$ in the specification with a nonlinear function of $x$. Let $\phi : \mathbb{R}^{d_X} \to \mathbb{R}^{d_\phi}$ and set $h_\theta(u,x) = (g(W\phi(x)+b))_u$, where $W \in \mathbb{R}^K \times \mathbb{R}^{d_\phi}$ and $b \in \mathbb{R}^{d_\phi}$. This is a logistic regression of the basis functions $\phi = (\phi_1, \ldots, \phi_{d_\phi})$. For instance, polynomials, step functions, or splines could be chosen as the basis functions. It is important to recognize that, even if the basis functions are nonlinear in the input space, the logistic regression model remains a link function of a model which is *linear* in the parameters $\theta$. The logistic model may perform poorly if the chosen basis functions are not appropriate for the problem. Instead of fixing a set of basis functions $\phi$ ahead of time, a neural network learns these feature functions directly from the data. The function $\phi$ is replaced by a parameterized function $\phi_\theta$ where $\theta$ is estimated from data. A neural network is composed of a sequence of nonlinear operations (or "layers"). Each operation takes the output from the previous layer and applies (1) a linear function and then (2) an element-wise nonlinearity. As a whole, a neural network is a flexible function, highly nonlinear in the parameters, which can learn the best feature functions $\phi_\theta$ for the problem.

Define the nonlinear transformation $\phi_\theta(x)$ as $h_{\theta,L-1}(x)$. A multi-layer neural network repeatedly passes linear combinations of learned basis functions through simple nonlinear link functions to produce a highly nonlinear function. Formally, the output $h_{\theta,l} : \mathbb{R}^{d_X} \to \mathbb{R}^{d_l}$ of the $l$-th layer of the neural network is:

$$h_{\theta,l}(x) = g_l(W_l^\top h_{\theta,l-1}(x) + b_l), \tag{3}$$

where $W_l \in \mathbb{R}^{d_l} \times \mathbb{R}^{d_{l-1}}$, $b_l \in \mathbb{R}^{d_l}$, and $h_{\theta,0}(x) = x$. For $l = 1, \ldots, L-1$, the nonlinear transformation $g_l(z) = (\sigma(z_1), \ldots, \sigma(z_{d_l}))$ for $z = (z_1, \ldots, z_{d_l}) \in \mathbb{R}^{d_l}$ and $g_L(z)$ is given by the softmax function $g(z)$ defined in (2). Note that $d_L = K = |\mathcal{U}|$. The function $\sigma : \mathbb{R} \to \mathbb{R}$ is a simple nonlinear link function; typical choices are sigmoidal functions, tanh, and rectified linear units (i.e., $\max(x,0)$). The final output of the neural network is given by:

$$h_\theta(u,x) = (h_{\theta,L}(x))_u = (g(W_L^\top h_{\theta,L-1}(x) + b_L))_u. \tag{4}$$

The parameter specifying the neural network is

$$\theta = (W_1, \ldots, W_L, b_1, \ldots, b_L), \tag{5}$$

[21]Campbell & Dietrich (1983), Cunningham & Hendershott (1986), Elul et al. (2010), Agarwal et al. (2011), and many others use logistic regression models to analyze mortgage performance.

where $L$ is the number of layers in the neural network. At each layer $l$, the output $h_{\theta,l}(x)$ is a simple nonlinear link function $g_l$ of a linear combination of the nonlinear basis functions $h_{\theta,l-1}(x)$, where the nonlinear basis function $h_{\theta,l-1}(x)$ must be learned from data via the parameter $\theta$. The output $h_{\theta,l}(x)$ from the $l$-th layer of the neural network becomes the basis function for the $(l+1)$-th layer.

The layers between the input at layer $l = 0$ and the output at layer $l = L$ are referred to as the hidden layers. Thus, the neural network $h_\theta$ has $L - 1$ hidden layers. A neural network with zero hidden layers ($L = 1$) is a logistic regression model. More hidden layers allow for the neural network to fit more complex patterns. Each subsequent layer extracts increasingly nonlinear features from the data. Early layers pick up simpler features while later layers will build upon these simple features to produce more complex features. The $l$-th layer has $d_l$ outputs where each output is an affine transformation of the output of layer $l - 1$ followed by an application of the nonlinear function $\sigma$. This composition of functions is called a hidden unit, or simply, a unit, since it is the fundamental building block of neural networks. The number of units in the $l$-th layer is $d_l$ and the complexity of any layer (and the complexity of the features it can extract) increases with the number of units in that layer. Thus, increased complexity can be achieved by increasing either the number of units or the number of layers. Given enough units, a neural network can approximate arbitrarily well continuous functions on compact sets (Hornik 1991). This of course includes approximating arbitrarily well interactions such as the product and division of features. The advantage of more layers (as opposed to simply adding more units to existing layers) is that the later layers learn features of greater complexity by utilizing features of the lower layers as their inputs. Moreover, *deep neural networks*, i.e., networks with three or more hidden layers, typically need exponentially fewer units than shallow networks or logistic regressions with basis functions; see Bengio & LeCun (2007) and Montufar et al. (2014).[22]

(1) is a dynamic model and therefore gives transition probabilities between the states over multiple periods (2 month, 6 month, 1 year, etc.). The transition probability matrix for 1-month ahead transitions is specified by the transition function in (1). The transition probability matrix for $t$-months ahead is simply the expectation of the product of the transition probability matrices at months $0, 1, \ldots, t-1$. Note that the transition probability matrices at months $t = 1, 2, \ldots, t-1$ are random due to their dependence on the random covariates $X_t^n$. To compute these expectations, a time-series model for $X_t^n$ needs to be formulated and Monte Carlo samples from $X_t^n$ need to be generated. An alternate approach, which is advantageous for reducing the computational burden and can be accurate for shorter time horizons, is that the economic covariates in $X_t^n$ are frozen at $t = 0$. That is, only the state of the mortgage and deterministic elements of $X_t^n$ (e.g., the balance of a fixed rate mortgage and time to maturity) are allowed to evolve over time. Then, the transition probability matrix for a horizon $t > 1$ is the product of the *deterministic* transition probability matrices at months $0, 1, \ldots, t-1$. The two approaches are implemented in Section 5.[23]

[22]The number of layers and the number of neurons in each layer, along with other hyperparameters of the model, are chosen by the standard approach of cross-validation. Section C provides the details.

Our formulation captures loan-to-loan correlation due to geographic proximity and common economic factors. Pool-level quantities, such as the distribution of the prepayment rate for a given pool, can also be computed via standard Monte Carlo simulation. The cashflow from a pool is the sum of the cashflows from the individual loans. Thus, one simply needs to simulate all of the individual loans based on the fitted model and then aggregate the individual cashflows.[24] If the economic covariates are frozen at time $t = 0$, the pool-level distribution can be approximated in closed-form via a Poisson approximation or the central limit theorem. Such approximations are accurate (for the distribution where covariates are frozen) even for pools with only a few hundred loans.

We have considered alternative model architectures. For instance, one could individually model transitions from each particular initial state with a neural network; such an approach would require fitting $K$ different neural networks. Another approach would be to have separate models for each product (fixed-rate vs. ARM, etc.) or borrower class (prime vs. subprime). Clearly, our neural network architecture is more parsimonious, which is a desirable characteristic. However, in addition to parsimony, there is a statistical motive for our architecture. Neural networks learn via their hidden layers recognizing, and abstracting, nonlinear features from the data (i.e., nonlinear functions of the initial input). Different transitions may strongly depend upon the same nonlinear features. Similarly, different types of products are likely to depend on some of the same nonlinear features. For instance, it is likely that there are many similar factors driving the transitions current $\to$ paid off and 30 days delinquent $\to$ paid off. In our neural network architecture, all transitions are modeled by the same neural network, which has the advantage that more data can be used to better estimate the nonlinear factors which drive multiple types of transitions.

# 4 Likelihood Estimation

This section discusses the estimation of the parameter (5) specifying the deep learning model by the method of maximum likelihood. We are given observations of $X_t = (X_t^1, \ldots, X_t^N)$ at each time $t = 0, \ldots, T$ where $N$ is the number of mortgages which are observed. We let $X_t^n = (U_t^n, L_t^N, V_t^n)$, where $U_t^n \in \mathcal{U}$ is the state of the $n$-th mortgage, and $L_t^n$ includes the lagged default and prepayment rates in the zip code of the $n$-th mortgage.[25] The vector $V_t^n \in \mathbb{R}^{d_Y}$ includes the remaining contemporary local and national economic factors in Table 7, as well as the original and contemporary loan-level features in Tables 1 and 5. We make the standard assumption[26] that the variables $V_t^n$ are exogenous in the sense that the law of $V = (V_0, \ldots, V_T)$, where $V_t = (V_t^1, \ldots, V_t^N)$ does not depend on the parameter $\theta$ specifying the law of the observed mortgage states $U = (U_0, \ldots, U_T)$, where $U_t = (U_t^1, \ldots, U_t^N)$. Therefore, the likelihood problem for $V$ can be treated separately from that for $U$.

[23]An alternative approach to the dynamic model (1) is to fit a model for each different time horizon, as in Campbell et al. (2008). Many static models could be fitted for each of the time horizons (1 month, 2 months, 6 months, 1 year, 1.5 years, 2 years, etc.). Fitting so many models is computationally expensive and the two approaches mentioned above do not incur this cost.

[24]Large portfolios can be rapidly simulated using methods from Sirignano & Giesecke (2015). Fast optimal selection of loan portfolios can be performed using methods from Sirignano, Tsoukalas & Giesecke (2016).

Although the model framework (1) is a dynamic model where the function $h_\theta$ may assume a very complicated form, the likelihood of the observed states $U$ takes an analytical form. The likelihood of $U$ depends only on the observed value of $V$ and is independent of $V$'s exact form or parameterization since $V$ is exogenous. Letting $L = (L_0, \ldots, L_T)$ and writing informally, the log-likelihood function for $\theta$ given $V$ is

$$\mathcal{L}_{T,N}(\theta) = \log \mathbb{P}_\theta\big[U, L|V\big] = \log \mathbb{P}_\theta\big[L|U, V\big]\mathbb{P}_\theta\big[U|V\big] = \log \mathbb{P}_\theta\big[U|V\big],$$

where we use the fact that $\mathbb{P}_\theta\big[L|U, V\big] = 1$ since $L_t = (L_t^1, \ldots, L_t^N)$ is a deterministic function of $U_0, \ldots, U_t$. Under the standard assumption that the variables $U_t^1, \ldots, U_t^N$ are conditionally independent given $X_{t-1}$, we have[27]

$$\begin{aligned}\mathcal{L}_{T,N}(\theta) = \sum_{t=1}^{T} \log \mathbb{P}_\theta\big[U_t \big| U_{t-1}, V_{t-1}\big] &= \sum_{t=1}^{T}\sum_{n=1}^{N} \log \mathbb{P}_\theta\big[U_t^n \big| U_{t-1}^n, V_{t-1}^n\big] \\ &= \sum_{n=1}^{N}\sum_{t=1}^{T} \log h_\theta(U_t^n, X_{t-1}^n).\end{aligned}$$

A maximum likelihood estimator (MLE) $\hat{\theta} = \theta_{T,N}$ for the parameter $\theta$ satisfies

$$\theta_{T,N} \in \arg\max_{\theta \in \Theta} \mathcal{L}_{T,N}(\theta). \tag{6}$$

The asymptotic properties of the MLEs have been studied before. Under certain conditions, the estimators are consistent and asymptotically normal; see White (1989*a*) and White (1989*b*). Sussmann (1992) and Albertini & Sontag (1993) study identifiability.

Neural networks tend to be low-bias, high-variance models. We use several methods to address overfitting, including regularization, dropout, and ensemble modeling. A standard $\ell^2$ regularization term is included in the objective function in addition to the log-likelihood $\mathcal{L}_{T,N}(\theta)$. The $\ell^2$ term represents the sum of the squares of the parameters. Secondly, we use dropout in each of the layers. Dropout is a widely-used technique in deep learning that has proven to be very successful; see Srivastava, Hinton, Krizhevsky & Sutskever (2014). During fitting, hidden units are randomly removed from the network. This prevents complex "fictitious" relationships forming between different neurons since neuron $i$ cannot depend upon neuron $j$ being present. Finally, we also build an ensemble of neural networks. This simply means that we fit a set of randomly initialized neural networks on datasets obtained by bootstrapping from the original datasets. Typically, each neural network reaches a different local minimum due to each being trained with a different random initial starts and random sequence of bootstrapped samples. Variance (i.e., overfitting) of an individual neural network's prediction can be reduced by taking the prediction as the average of the neural networks' predictions. The averaged prediction, or ensemble prediction, has lower variance since the idiosyncratic variance for each neural network is averaged out.

[25]In general, $L_t^n$ could include any variables describing the aggregate lagged performance of the mortgages.
[26]See Duffie et al. (2007), Campbell et al. (2008), and many others.
[27]This expression assumes that every mortgage is originated at time $t = 0$. The modification for the case where mortgages have different origination dates is straightforward.

Appendix B discusses the implementation of the MLE. We develop fitting algorithms that can deal very efficiently with the large number of samples and explanatory variables we observe. The algorithms harness recent advances in GPU parallel computing and run on a cluster of Amazon Web Services nodes.

Appendix C details our cross-validation approach to the selection of the hyperparameters, which include the number of layers and number of neurons per layer, the type of the activation function $\sigma$, the size of the regularization penalty, and several other parameters governing the fitting algorithm (see Appendix B). The optimal network architecture has five hidden layers, with 200 units in the first hidden layer and 140 units in each subsequent one. The rectified linear unit activation function $\sigma(x) = \max(0, x)$ was found to yield better performance and faster convergence than the sigmoid $\sigma(x) = 1/(1 + e^{-x})$.

The training set includes all the data before May 1, 2012. The validation set, which is used for the selection of hyperparameters (see Appendix C), is May 1, 2012 until October 31, 2012. Once the hyperparameters are chosen, the model is re-fitted on the combined training and validation sets. The final trained model is then tested out-of-sample on the test set, which is from November 1, 2012 until May 31, 2014. All explanatory variables are normalized by their means and standard deviations (which are calculated using data only from the training set).

# 5 Empirical Results

The fitted deep learning model is used to understand the relationship between explanatory factors and borrower behavior. Our analysis shows that many highly nonlinear relationships exist. Furthermore, borrower behavior is found to have nontrivial dependencies on the nonlinear interaction between multiple factors.

## 5.1 Explanatory Power of Variables

We begin by studying the explanatory power of the different factors for the behavior of borrowers. To this end we consider the behavior of the out-of-sample negative average log-likelihood with respect to changes to the composition of the set of factors. The negative average log-likelihood $\frac{1}{N}\mathcal{L}_{T,N}(\hat{\theta})$ is a standard measure of fit, which is sometimes called the cross-entropy error or simply the loss. We measure by how much the loss increases when a variable is removed as an explanatory variable. Variables that have large explanatory power, and whose information is not also largely contained in the other remaining covariates, will produce large increases in the loss if they are removed.

Table 11 reports the results.[28] The state unemployment rate is the variable with by far the highest explanatory power among all variables, emphasizing the importance of local economic conditions for borrower behavior. Standard loan-level variables such as credit score and loan-to-value ratio, which were highlighted by Campbell & Dietrich (1983), Cunningham & Capone (1990), Curley & Guttentag (1974), and others as major predictors of borrower behavior, have less explanatory power. This finding suggests a much tighter connection between housing finance markets and the macroeconomy than previously thought. Earlier studies have assigned a relatively limited role to unemployment, see Campbell & Dietrich (1983), Cunningham & Capone (1990), Deng (1997), Elul et al. (2010), Foote et al. (2010) and others. Relative to our analysis, these earlier studies are based on much shorter sample periods, selected loan products (such as 30 year fixed-rate loans) and borrower profiles (such as prime borrowers), and much smaller samples. Thus, these earlier findings are not necessarily inconsistent with ours. However, our results provide a more definitive picture regarding the role of unemployment than prior results because our results are based on an unprecedented sample that spans several economic cycles over two decades.

The dominant role of unemployment suggests that the loan-to-loan correlation due to the exposure of borrowers to economic cycles can be substantial. This source of loan-to-loan correlation is distinct from the foreclosure contagion channel studied by Anenberg & Kung

[28] For clarity, we exclude from Table 11 the variable representing the current state as well as the dummies representing missing values (see Appendix A).

(2014), Campbell et al. (2011), Towe & Lawley (2013), and others.[29] Contagion entails a foreclosure having negative spillover effects on neighboring properties that increase the likelihood of additional foreclosures. We control for the contagion channel by including the lagged default rate for prime and subprime borrowers at the zip-code level as explanatory variables (see Section 2.2). The results in Table 11 suggest that these variables have some explanatory power, which is consistent with the existence of a contagion channel. The fact that unemployment plays a critical role even after controlling for the contagion channel provides evidence for the prevalence of additional loan-to-loan correlation due to the exposure of borrowers to the local economy. This, in turn, emphasizes the need for mortgage-backed security investors to diversify loan risk geographically, beyond the conventional borrower characteristics highlighted in the literature.

## 5.2 Economic Significance of Variables

We now turn to analyzing the economic significance of the different explanatory factors for borrower behavior. The economic significance of a particular variable is measured by the magnitude of the derivative of a fitted transition probability with respect to the variable (averaged over the data). The derivative is over a representative sample drawn from the dataset rather than a single point. Specifically, we calculate the sensitivity (with respect to $j$-th variable) of the fitted probability for a transition from state $u$ to $v$ as:

$$\mathbb{E}\left[\left|\frac{\partial}{\partial x_j} h_{\hat{\theta}}(V,X)\right|\middle| V=v, U=u\right]. \tag{7}$$

A sensitivity of value $z$ for a given variable means that the probability for a transition from state $u$ to $v$ will approximately change (in magnitude) by $z\Delta$ if that variable is changed by a small amount $\Delta$. As explained in Appendix D, the sensitivity (7) can be estimated directly from the dataset and the fitted model.

In our model formulation, the sensitivity is governed by multiple model parameters that represent the nonlinear connections between a variable and a transition probability. It is instructive to contrast that with the linear formulations widely used in earlier studies of borrower behavior (see the references in Section 1.1). In the case of a linear model, a single coefficient governs the sensitivity. When nonlinear relations are present, as in our data (see Section 5.3 below), these coefficients/sensitivities can severely misrepresent the true economic significance of variables.[30]

[29] See Azizpour, Giesecke & Schwenkler (2017) for a discussion of the sources of loan-to-loan correlation.

[30] To see this, consider the linear regression $f(x;\alpha,\beta)=\alpha+\beta x$ fitted to data produced from the function $y=x^2$ on $x\in[-1,1]$. The least-squares estimators are $\hat{\alpha}=\frac{1}{3}$ and $\hat{\beta}=0$. This suggests that there is *no relationship* between $y$ and the covariate $x$. However, clearly there is a very strong relationship, which would

Table 12 reports sensitivities for a transition from current to paid off (i.e., prepayment). The sensitivities indicate the strong economic significance for prepayment of original and current outstanding loan balance.[31] Other economically significant variables include original interest rate, interest rate differentials, house price appreciation, loan age, FICO score, lagged prepayment rates, and state unemployment.[32] The importance of loan balance variables is interesting in light of earlier studies of prepayment such as Green & Shoven (1986), Cunningham & Capone (1990) and Richard & Roll (1989). These studies focus on the influence of interest rate differentials, premium burnout, loan age, LTV ratio and several other variables, all of which are included in our analysis. Our results indicate that loan balance variables in fact overshadow all those previously considered variables in terms of economic significance. Our results also firmly establish the economic significance of unemployment, a variable which earlier studies such as Cunningham & Capone (1990) and Foote et al. (2010) found to have no significant influence on prepayment.[33]

Table 13 reports sensitivities for a transition from current to 30 days delinquent.[34] What stands out is the significant role of variables that describe borrower behavior over the recent past. These variables include the number of delinquencies (30, 60, and 90+ days late) over the past year as well as the number of times current over the past year. The number of times the borrower was 30 days delinquent in the last year dominates all other variables in terms of economic significance for a transition from current to 30 days delinquent. The strong influence of these variables indicates that borrower behavior is strongly path dependent. Prior work on mortgage default risk such as von Furstenberg (1969), Gau (1978), Vandell (1978), Webb (1982), Campbell & Dietrich (1983), Elul et al. (2010), Foote et al. (2010) has not analyzed the influence of path-dependent borrower behavior variables, and instead focused on the role of standard loan and borrower characteristics such as FICO score, interest rates, and LTV ratios. Our results suggest that these standard variables are less influential predictors of mortgage delinquency than previously thought.

---

have been identified if a nonlinear model (such as a neural network) was used.

[31]We compare the sensitivities implied by the linear logistic regression model (i.e., a 0-layer network) with those in Table 12 (the values are available upon request). The logistic regression model significantly *understates* the sensitivity and hence economic importance of original and current outstanding loan balance, and *overstates* the sensitivity for the interest rate and interest rate differentials. Our analysis in Section 5.3 below shows that prepayment has a nonlinear relationship with all these variables. A linear model such as logistic regression can produce inaccurate sensitivities for such nonlinear relationships.

[32]Unemployment is not the most economically significant variable. However, this is not inconsistent with our earlier finding that unemployment is the dominant variable in terms of explanatory power. In this section we are considering a particular transition and the sensitivity of that transition to small changes in a variable such as unemployment. In Section 5.1 above, we consider the ability of variables to explain the observed data jointly for all transitions, not just a particular one.

[33]Note that these earlier studies are based on much smaller samples and shorter sample periods.

[34]The sensitivities for other transitions are available upon request.

## 5.3 Nonlinear Effects

This section studies nonlinear relationships between borrower behavior and variables. In particular, we examine the "one-dimensional" relationships between borrower behavior and the most influential real-valued variables that were identified above. The interactions between multiple variables and borrower behavior will be analyzed in subsequent sections.

### 5.3.1 Prepayment Behavior

Figures 2 and 7 show the relationship between prepayment and some of the most influential variables. In each plot, the fitted prepayment probability's dependence on a particular variable is examined, keeping all other variables constant. The other covariates are fixed at their average values in the dataset, thus representing the "average borrower/loan." Most of the relationships in Figures 2 and 7 are highly nonlinear. They reveal new and important patterns in borrower behavior. Having discussed the significant effects associated with unemployment already in Section 1, below we focus on the relationship between prepayment and some of the other influential variables identified in Table 12, including loan balance variables, interest rates and interest rate differentials, and house price appreciation.

The fitted prepayment probability is a decreasing function of the current outstanding balance, which is the most influential variable for prepayment. Borrowers with relatively low current balances are quite likely to prepay, with prepayment probabilities topping 60% for the smallest balances. This could suggest that borrowers prefer closing out their mortgages towards the end of the lifetime of the loan, when they have the means to do so, rather than continue making monthly payments until maturity. The prepayment probability decreases very quickly to about 10% with the current balance increasing to about $8,000. The likelihood of prepayment decreases at a much slower rate for balances increasing beyond that amount, and is relatively flat for balances larger than $15,000. This means that borrower behavior is relatively insensitive to changes in the current outstanding balance for sufficiently high balances. Borrower behavior changes very significantly once the balance reaches a level of about $8,000.

The behavior of the fitted prepayment probability as a function of original loan balance is markedly different. The prepayment probability is non-monotonic with regards to the original loan balance, which is the second most influential variable for prepayment. The probability increases roughly linearly until the original loan balance reaches a level of about $350,000. The rate of change then decreases, with the probability peaking at around 5% for original loan balances around $600,000. It then levels off and finally decreases for very large original loan balances. Large loan balances may be harder to refinance, and borrowers with large balances may not care much about the benefits of a refinancing in light of the effort it takes to close the transaction (assuming it is optimal to refinance).

The current interest rate minus the national mortgage rate indicates the incentive of the mortgage holder to prepay, and is another highly influential variable for prepayment. The prepayment probability is highly nonlinear as a function of this quantity. For negative values, the probability is almost 0 as the mortgage holder has no incentive to prepay. Near to 0%, the probability suddenly jumps to 3%. In the range $2.5 - 5\%$, the probability linearly increases. Then, for values greater than 5%, the probability increases at a much faster rate. Above the threshold of 5%, the advantage of prepaying may outweigh prepayment penalties that certain mortgages have.

The fitted prepayment probability is an approximately piece-wise linear function of house price appreciation in a borrower's zip-code since origination. After home prices double a borrower is 50% more likely to prepay. This is consistent with a behavior where borrowers sell to realize significant price gains and move into more expensive homes. The sensitivity of borrowers to additional price appreciation is somewhat lower. The likelihood of prepayment levels off for price appreciation beyond 250%.

### 5.3.2 Delinquency Behavior

Figures 3 and 8 show the relationship between delinquency and some of the most influential variables. In each plot, the fitted probability of a transition from current to 30 days delinquent is plotted versus a particular variable, keeping all other variables fixed at their average values in the dataset. The plots reveal that many of the variables have a highly nonlinear influence on delinquency. They point to several new and interesting patterns in delinquency behavior. Having discussed the role of house price appreciation already in Section 1 (see Figure 3 and the attendant discussion), below we focus on the relationship between delinquency and some of the other influential variables identified in Table 13, including variables that describe recent borrower behavior.

The fitted delinquency probability is an increasing function of the number times a borrower was 30 days delinquent during the past year, which is the most influential variable for delinquency. (The behavior of delinquency with respect to the number of times a borrower was 60 days delinquent is similar.) Without delinquencies during the past year, the likelihood of a delinquency is under 0.5%. With a single delinquency, the likelihood increases to about 4%, which represents a very significant percentage increase in borrower credit risk. With two delinquencies during the past year, the likelihood increases to about 7%, and with three the likelihood stands at 12%. This behavior indicates the path-dependent nature of mortgage credit risk. It is consistent with borrowers "getting used" to being behind payment after falling behind payment for the first time. Delinquency loses its stigma after the borrower has fallen behind payment for the first time. The path-dependent behavior is also consis-tent with the existence of borrowers who have a hard time making their monthly mortgage payments and who fall behind payment multiple times a year.

The fitted delinquency probability is a decreasing function of the original loan balance. For a \$100,000 loan, the likelihood of a delinquency is around 3.5% while for a \$200,000 loan, the likelihood is 1.5%. For loans of \$300,000 and larger, the likelihood of delinquency is flat at around 0.1%, consistent with the fact that borrowers for larger loans are typically better off financially.

## 5.4 Interactions between Variables

Borrower behavior is a high-dimensional function of the explanatory variables. We wish to understand how borrower behavior *simultaneously* depends upon *multiple* variables, i.e., how different variables interact to influence a certain state transition. To this end we estimate cross partial derivatives of the fitted transition probabilities, which measure how the effect of a shift in one variable depends on the size of the shift in another variable. Specifically, we measure the economic significance of the interaction between covariates $i$ and $j$ for a transition from state $u$ to $v$ by the derivative

$$\mathbb{E}\Big[\Big|\sum_{i,j=1}^{2}\frac{\partial^2}{\partial x_i \partial x_j}h_{\hat{\theta}}(V,X)\Big|\Big|V=v,U=u\Big]. \tag{8}$$

This derivative can be generalized to measure higher-order interactions. Appendix D provides a finite-difference estimator for (8) and a third-order extension.

### 5.4.1 Prepayment

Tables 14 and 15 present the most influential pairs of variables and triplets of variables for prepayment. We see that original interest rate, state unemployment, loan balance variables, and FICO score, which we have identified above as the most influential variables for prepayment on a stand-alone basis, also strongly interact.

To develop a detailed understanding of the interactions, in Figure 9 we present contour plots describing the relationship between prepayment and some of the most influential pairs of variables, and in Figure 10 we present contour plots describing the relationship between prepayment and triplets of variables. The contour plots represent the joint effects on prepayment of multiple variables. In our nonlinear framework, they replace the analysis of the coefficients of the dummy variables encoding variable interactions in a linear framework such as Elul et al. (2010) and Foote et al. (2010). The contour plots uncover, for the first time in the empirical mortgage literature, the complex interplay of many variables.

Consider the interaction between current outstanding balance and original loan balance, which are the two economically most significant variables for prepayment. For any given value of the original loan balance, the likelihood of prepayment is a decreasing function of the current outstanding balance. However, the behavior strongly depends upon the original loan balance. The higher the original loan balance, the larger the likelihood of prepayment, for any given value of the current outstanding balance. Borrowers who took out relatively large loans and whose current outstanding balance is relatively small are the most likely to prepay. Those borrowers tend to be more creditworthy, and therefore can more easily obtain refinancing for relatively small current outstanding loan balances (they have already paid down a substantial portion of the loan).

Figure 9 also shows the relationship between current outstanding balance and the prepayment incentive, as measured by the current interest rate minus national mortgage rate. For any given value of the incentive, prepayment becomes more likely as the current outstanding balance decreases. For any given value of the current outstanding balance, as the prepayment incentive increases, prepayment becomes more likely. For very large outstanding balances, even if the prepayment incentive is large, borrowers do not prepay. This observation is consistent with the fact that larger loans are typically more difficult to refinance. For negative prepayment incentives, even if the outstanding balance is very small, borrowers also do not prepay. While borrowers with low current balances might want to close out their mortgage as argued in Section 5.3.1 above, even if they do have the required liquidity they might recognize the disincentive to prepay. Borrowers with small outstanding balances and high prepayment incentives are the most likely to prepay.

The relationship between original term of the loan and the number of times that a borrower was current during the last year is also shown in Figure 9. The prepayment probability is an increasing function of the number of times current, for any loan term. Borrowers with 30 year loans which were current during each of the past 12 months are the most likely to prepay. Borrowers with 30 year loans which were not current at all during the 12 months are the least likely to prepay.

Three-dimensional interactions can be analyzed via the stacked contour plots in Figures 10. For example, consider Figure 11, which focuses on one of the panels in Figure 10. Figure 11 displays the three-dimensional interaction between the original interest rate, original interest rate - national mortgage rate, and state unemployment rate. In favorable economic scenarios when the unemployment rate is low (the bottom contour plot in the stack), prepayment probabilities strongly depend upon the original interest and the original interest - national mortgage rate. Under adverse economic conditions, when the unemployment rate is high (the top contour plot in the stack), the prepayment probability is very low and is in fact insensitive to the original interest rate and original interest - national mortgage rate.

That is, very few borrowers will prepay no matter how great the financial incentive there is for prepaying.

### 5.4.2 Delinquency

Tables 16 and 17 present the most influential pairs of variables and triplets of variables for 30-day delinquency. We find that original interest rate, interest rate differentials, original loan term, FICO score, loan balance variables, and past delinquency behavior, which we have identified above as influential variables for delinquency on a stand-alone basis, also strongly interact.

Figure 12 presents contour plots describing the relationship between delinquency and different pairs of covariates. The plots offer insights into the joint effects on delinquency behavior of multiple influential variables. Consider the relationship between past delinquency behavior, current outstanding balance, and transitions to delinquency. The probability of a transition from current to 30 days delinquent increases if the borrower has been delinquent in the past. For example, if the borrower has been delinquent more than 8 times in the past 12 months, then a transition from current to 30 days delinquent occurs at an almost 20% monthly rate. The behavior strongly depends upon the current outstanding balance. Specifically, the probability of becoming delinquent increases as the current outstanding balance increases. A larger outstanding balance places significant financial stress on the borrower. Borrowers with large outstanding balances and who have frequently been delinquent in the past are very likely to return to delinquency.

Figure 12 also includes a contour plot describing the relationship between FICO score, current outstanding balance, and transitions to delinquency. As the FICO score decreases, a transition to delinquency becomes more likely. As the current outstanding balance increases, a transition to delinquency also becomes more likely. This is to be expected since borrowers will find it more difficult to service larger loans. Even borrowers with high FICO scores have a nontrivial probability of becoming delinquent for very large loans. Borrowers with low FICO scores and large outstanding balances are very likely to become delinquent. In any given month, a transition from current to 30 days delinquent occurs at a 10% rate for this group of borrowers. In contrast, borrowers with high FICO scores and small outstanding balances almost never become delinquent.

Another of the contour plots in Figure 12 describes interaction of original interest rate and FICO score. There are several regimes; within each regime, the delinquency behavior is very different. For very high FICO scores ($> 800$), borrowers rarely become delinquent no matter how high the original interest rate is. That is, in the high FICO score regime, the probability of delinquency is insensitive to the interest rate. For low FICO scores ($< 650$), the probability of delinquency strongly depends upon the interest rate. In this regime, the likelihood of delinquency increases as the interest rate increases. Notably, there is a nonlinear relationship within the low FICO score regime. The sensitivity to the interest rate is smaller in the low interest rate, low FICO score regime than in the high interest rate, low FICO score regime.

Figure 13 displays the interactions between triplets of variables. Figure 14 focuses on one of the panels of Figure 13. It displays the relationship between FICO score, original interest rate, number of times current in last 12 months, and transitions to delinquency. When the borrower is rarely current, the borrower's FICO score is low, and the interest rate is high, then the borrower is highly likely to fall behind payment again (the likelihood exceeds 25% on a monthly basis). Furthermore, there is a very nonlinear relationship between FICO score and interest rate, and this relationship strongly depends on the number of times current. In the high FICO score regime ($> 800$), the probability of delinquency is insensitive to the interest rate. For low FICO scores ($< 650$), the probability of delinquency strongly depends upon the interest rate. If the loan is more frequently current, the probability of delinquency decreases, across all FICO scores and original interest rates. Furthermore, the delinquency probability quickly asymptotes to a small value as the FICO score increases or the original interest rate decreases.

# 6 Out-of-Sample Analysis

This section examines the out-of-sample behavior of the deep learning model. The results illustrate the model's goodness-of-fit and predictive accuracy. They also provide insights into the nature and significance of the nonlinear effects associated with borrower behavior. In particular, the results suggest that a prepayment event involves the strongest nonlinear effects among all events. This is shown to have import implications for mortgage investment management and the behavior of pool-level mortgage risk.

## 6.1 Goodness-of-Fit

We begin by considering the goodness-of-fit of the deep learning model. Since model parameters were fitted to maximize the log-likelihood, we use the negative average log-likelihood (loss) as a measure of the overall goodness-of-fit of a model across all possible state transitions. Table 18 reports the in- and out-of-sample loss for neural networks with 0, 1, 3, 5, and 7 hidden layers, as well as an ensemble model composed of eight 5-layer networks.

The results indicate the behavior of the goodness-of-fit as a function of the depth of the network. As expected, the more complex the network the better the in-sample fit. However, deeper networks do not always yield better out-of-sample fit due to higher model capacity;

there are several other factors at play, such as over-fitting and difficulty in estimating the model, that affect the out-of-sample performance. We also observe that the use of dropout significantly improves out-of-sample fit for networks with greater depths. This shows that dropout provides effective regularization and addresses over-fitting for networks with larger numbers of layers. The characteristic U-shaped behavior of the out-of-sample loss in Table 18 suggests that the optimal number of hidden layers (in the sense of out-of-sample goodness-of-fit) is five when using dropout and three when dropout is not applied.[35] The goodness-of-fit optimality of deep networks provides strong evidence for the existence of complex nonlinearities in the mortgage performance data.

We also observe that the ensemble model, which is composed of eight 5-layer networks, outperforms all other formulations in terms of fit. Figure 15 shows the out-of-sample loss versus the number of models in the ensemble. Including just eight networks in the ensemble significantly improves out-of-sample fit, indicating that ensemble modeling effectively reduces over-fitting. Although larger ensembles lead to marginal improvements in fit, the computational cost (which increases linearly with the number of neural networks used) may not justify using larger ensembles in practice. Henceforth, we only consider ensembles of eight independently fitted networks.

Finally, Table 18 reports test statistics for likelihood ratio tests of 0- vs. 1-layer models, 1- vs. 3-layer models, 3- vs. 5-layer models, and 5- vs. 7-layer models. All tests are highly significant (with $p$-values of less than 0.01). Moreover, the behavior of the test statistic suggests that the most significant improvement of model fit is obtained by permitting nonlinear relationships between borrower behavior and explanatory variables. Going from a linear model (0-layer network) to the simplest nonlinear model (1-layer network) generates the most significant improvement in the test statistic, which provides additional evidence for the existence of nonlinear relationships in the data.

## 6.2 Loan-Level Predictive Accuracy

We next consider Receiver Operating Characteristic (ROC) curve and Area Under Curve (AUC), which are standard measures of predictive accuracy for a binary classifier. The binary classifier generates an estimate of the probability that the input sample belongs to the positive class (e.g., 30 days past due). The ROC curve plots the true positive rate versus the false positive rate as the discriminative threshold is varied between 0 and 1. The AUC is the area under the ROC curve and a higher value shows an improved ability of the classifier to discern between the two classes (the maximum value of the AUC is 1, corresponding to perfect forecasts). Alternatively, the AUC can be interpreted as the probability that the model generates a larger value for a sample randomly chosen from the positive class than for a sample randomly chosen from the negative class.

[35] See Goodfellow et al. (2016) for an excellent discussion of the optimal depth and number of hidden units.

We consider the out-of-sample AUC for the ROC curve for the transition of a mortgage between two states (paid off, current, 30 days delinquent, 60 days delinquent, 90+ days delinquent, foreclosure, and REO). Specifically, the AUC for transition $u \to v$ is the AUC for the two-way classification of whether the mortgage is in state $v$ or not in state $v$ at a 1-month horizon conditional on the mortgage currently being in state $u$. Figures 16, 17, 18, 19, and 20 report the AUCs for networks with 0, 1, 3, and 5 hidden layers, and an ensemble of eight 5-layer networks. They give a complete picture of model performance in different states and for different types of transitions. They also yield interesting insights into the behavior of the nonlinear effects associated with transitions.

We see that in general, the greater the depth of the network, the larger are the out-of-sample AUCs. The improvements in performance are most striking for transitions to paid off (i.e., prepayments). This suggest that prepayments involve highly nonlinear effects, more so than other transition events. Table 19 reports the AUCs for prepayment, and Figure 21 provides several of the corresponding ROC curves. The prepayment AUCs of the 1-layer neural network represent improvements between 3 percent (for transitions from 30 days delinquent to paid off) and 30 percent (for transitions from 90+ days delinquent to paid off) over the AUCs of the linear 0-layer network. This suggests that the nonlinear effects associated with transitions from severe delinquency to prepaid are the strongest among all transitions to prepaid.

## 6.3 Mortgage Investment Portfolios

To analyze the implications of nonlinear effects for investment management, we consider the out-of-sample investment performance of mortgage portfolios constructed using networks of different depths. Consider an investor who seeks to design a loan portfolio with uninterrupted cashflow. An example is a financial institution which originates loans and retains some loans on their balance sheet. Another example is an asset manager who constructs a loan investment portfolio. Delinquency often produces a loss of cashflow while prepayments lead to early cashflows that might have to be reinvested at lower interest rates. Uninterrupted cashflows require loans which are both unlikely to be delinquent and unlikely to prepay. This is equivalent to designing a portfolio of loans which are highly likely to remain current. Given an available pool of loans to select a portfolio from, loans can be ranked by a model-implied likelihood that they remain current. For a portfolio of size $N$, one then chooses the $N$ loans with the highest probabilities of remaining current.

The above approach can be used to evaluate the importance for investment management of capturing nonlinear effects in mortgage state transitions. We form two portfolios of $N$ loans from an available pool of 100,000 mortgages (randomly chosen from the test dataset). The first portfolio is chosen using the nonlinear 5-layer network and the second portfolio is chosen using the linear 0-layer network. A portfolio is chosen by ranking the available loans according to their probability of being current as predicted by the model and then choosing the $N$ loans mostly likely to be current.[36] Figure 22 shows how the portfolios perform out-of-sample over 1 month and 1 year time horizons.[37] The portfolio generated by the nonlinear 5-layer network significantly outperforms the portfolio generated by the linear 0-layer model in terms of prepayment rates. This finding is consistent with our earlier finding in Section 6.2 that prepayment events involve strong nonlinear effects, which the 0-layer model fails to capture. Table 20 (Table 21) reports the percent of the two portfolios with size $N = 20,000$ in each state (REO, paid off, etc.) at a 1 month (1 year) time horizon. At a 1 year horizon, the 5-layer network portfolio has a significantly lower prepayment rate than the 0-layer network portfolio. This feature of the 5-layer network portfolio directly translates into improved return for an investor. Conservatively assume that prepayment results in a loss of 5% of notional, foreclosure and REO produce losses of 40% of notional, and $m$ months delinquent leads a loss of $\frac{m}{360} \times 100\%$ of notional. Then, the 5-layer network portfolio has a 46% smaller loss than the 0-layer network portfolio at a 1 year time horizon. These results indicate the significant importance for investment management of capturing nonlinear effects in borrower behavior, especially prepayment.

## 6.4 Pool-Level Accuracy

We finally evaluate the accuracy of out-of-sample predictions of pool-level risk, which is especially relevant to mortgage-backed security investors. The expected delinquency rates, prepayment rate, and return of a mortgage pool is easily calculated by simply summing the expectations of the individual mortgages. We examine the pool-level predictions of the 5-layer network for 2,000 pools created from 2 million mortgages in our test dataset. Each pool contains 1,000 mortgages. Pools are created by rank ordering the loans according to a given loan characteristic (e.g., the interest rate) and then sequentially placing the loans in pools of size 1,000.[38] This produces pools with varying levels of risk. Four cases are examined. We create pools by rank ordering according to FICO score, interest rate, LTV ratio, and the predicted probability that the loan is current. Figures 23, 24, 25, and 26 show the pool-level prediction of prepayment for a 1 year time horizon. The 1 year transition probabilities and the pool-level prediction are produced using the method described in Section 3 where the time-varying covariates are frozen. The diagonal line represents a perfect prediction. The prediction generated by the 5-layer model is fairly accurate; for comparison we also show the prediction produced by the linear 0-layer model.[39]

[36]The same approach can be used to rank loans according to other criteria. For instance, if one wanted to account for both the interest rate and the risk of the loan, the expected return for each loan could be calculated for each model. Then, the loans could be ranked according to their expected returns.

[37]The 1 year transition probabilities are produced using the method described in Section 3 where the time-varying covariates (e.g., unemployment rates) are frozen.

[38]The loans with the top thousand highest interest rates are placed in the first pool, loans with the 1001-2000th highest interest rates are placed in the second pool, etc.

Figures 27 and 28 show the pool-level distribution of prepayment for the 5-layer neural network for several different pools. Each pool has 10,000 mortgages and the time horizon is 1 year. The national mortgage rate was simulated forward and all other time-varying covariates were frozen. For illustrative purposes, we use a simple AR(4) model that was fitted to historical data for the national mortgage rate obtained from Freddie Mac.[40] This formulation accounts for correlated prepayment behavior due to the common exposure of borrowers to future movements of the mortgage rate. Note that the actual observed prepayments in Figures 27 and 28 falls in the center of the distribution generated by the 5-layer network while it falls in the tail of the distribution of the linear 0-layer model, which is included here for comparison. Thus, in these cases, the 5-layer network-produced prepayment distribution accurately captured the out-of-sample outcome at pool-level.

To obtain a more comprehensive understanding of pool-level accuracy, we consider 50 test portfolios with 10,000 mortgages each obtained by slicing a pool of 500,000 mortgages. Table 22 reports average statistics (over the 50 test portfolios) for the prepayment distributions produced by the 5-layer neural network, and gives a sense of the out-of-sample prediction error. The statistics for 0-layer model are included for comparison. The 5-layer network-produced distribution tends to have less variance with the mean of the distribution closer to the observed number of prepayments, thereby predicting pool-level prepayments more accurately. These pool-level results suggest that nonlinear effects are not only associated with individual borrower behavior, but also with correlated borrower behavior, especially correlated prepayment events.

[39]The 0-layer model's pool-level predictions in Figures 23, 24, 25, and 26 appear to be systematically biased upwards (i.e., above the diagonal line). This systematic bias is due to the predictions for all pools depending upon the same covariates. For instance, if one prediction is biased upwards due to the realized value of the national mortgage rate, it is likely that all pool predictions will be biased upwards since they all depend upon the same realization of the national mortgage rate.

[40]The fitted parameters are $[0.6687, 1.3514, -0.5131, 0.2410, -0.0838]$. The lag was chosen using the partial autocorrelation plot. Of course, more complex models could be chosen alternatively.

# 7 Conclusion

This paper analyzes the behavior of mortgage borrowers using an unprecedented dataset of origination and monthly performance records for over 120 million mortgages originated across the US between 1995 and 2014. The analysis is based on a nonlinear deep learning model of multi-period borrower state transitions that incorporates the influence on borrower behavior of a large number of loan- and borrower-specific as well as economic and demographic variables at national, state, county and zip-code levels.

Our empirical findings yield a range of important new insights into the behavior of mortgage borrowers. The relationship between borrower behavior and risk factors is found to be highly nonlinear, which questions many linear models studied in prior work. Interaction effects, where the impact of a variable depends on the values of other variables, are ubiquitous. We find evidence suggesting that prepayments are most affected; they involve the strongest nonlinear effects among all events. The major drivers of prepayment, which include original and current outstanding loan balances even before standard factors such interest rates and interest rate spreads, strongly interact. Jumbo loan borrowers whose current outstanding balance is relatively small are the most likely to prepay.

Our results also highlight the importance of local economic conditions for borrower behavior. State unemployment is found to have the greatest explanatory power among all factors. This indicates a tighter connection of housing finance markets and the macroeconomy than previously thought. We find that the sensitivity of a borrower to changes in unemployment is not constant as often assumed in prior work, but strongly depends upon current unemployment. The sensitivity also significantly varies across the entire borrower population, highlighting an interaction effect between unemployment and many standard borrower characteristics that were previously studied only individually.

Our empirical results have significant implications for mortgage-backed security (MBS) investors. The dominant role of unemployment highlights the exposure of borrowers to economic cycles, which can be a substantial source of loan-to-loan correlation. The prevalence of this correlation emphasizes the need for MBS investors to diversify mortgage risk geographically, beyond the conventional borrower characteristics highlighted in the literature. The nonlinear nature of the influence on borrower behavior of many variables including unemployment has implications for the rating and hedging of MBS. Investors need to account for the nonlinear effects when constructing hedging positions against macroeconomic volatility. Rating agencies need to address the nonlinear behavior when assessing the exposure of MBS investors with respect to adverse macroeconomic conditions.

# A Data Cleaning

A fraction of the mortgages have missing data, including missing data for some of the key features such as FICO, LTV ratio, original interest rate, and original balance. The missing data is a result of reporting errors by the originator or servicer, or incomplete information provided by the borrower at the time of origination. Key features that are missing are more likely to be the result of reporting errors. For instance, original balance and original interest rate are required details in any mortgage contract. If they are missing, it must be due to a reporting error and not the borrower failing to provide this information. Similarly, FICO score and LTV ratio are almost universally available for mortgage borrowers. Other features, however, may simply not have been provided by the borrower at the time of origination. An example is debt-to-income ratio, which is often not available for subprime borrowers. We take the following approach towards missing data. We insist that any sample must have at least FICO, LTV ratio, original interest rate, and original balance. Samples missing one of these variables are removed. Missing data for other features, which are more likely to be due to incomplete information provided by the borrower, are typically encoded as an additional indicator variable (1 if it's missing, 0 if it's not missing). This standard approach eliminates the need to remove the corresponding samples, and allows us to measure the implications of missing features. For a discussion of this approach, see Gelman & Hill (2007, Chapter 25).

There are also certain events reported in the CoreLogic dataset which are errors. For instance, monthly mortgage transitions from current to 60 days delinquent or from 30 days delinquent to 90+ days delinquent are not possible. Errors of this type are very infrequent in the dataset and we remove those samples where such errors occur. Mortgages can also have their "servicing released" or their state may be reported as "unknown", "status no longer provided", or "invalid data". "Servicing released" means the servicer which previously reported the data to CoreLogic for that particular mortgage no longer services that mortgage and therefore no longer reports data for it. The mortgage state being "unknown", "status no longer provided", or "invalid data" could be due to a range of clerical/software errors. Whenever a mortgage is in any of these states or has its "servicing released", we exclude any subsequent monthly data from our sample.

# B Implementation of MLE

There are significant computational hurdles to training models due to the large size of our dataset as well as due to the size of the deep neural networks. The dataset includes the individual characteristics of each loan as well as monthly updates on loan performance. We include 272 features for each mortgage. Since our models are dynamic, there is a sample for each month of data. In total, we train over roughly 3.5 billion samples, which is almost 2 terabytes of data, and only a fraction of the dataset can be loaded into (RAM) memory at any one time. Further, many of the deep neural networks contain tens of thousands of free parameters. Estimating these parameters in order to fit the model requires computing the gradients using backpropagation, which is a time and memory intensive procedure. Fitting just one such model on our data using typical computing resources (e.g., using MATLAB or R on a desktop with conventional CPU) would require weeks of training time, which makes fitting and iterating through models impractically slow. In contrast, we train up to 10 models simultaneously in a span of few days. This is made possible by using several tools that harness both optimized hardware as well as computational tricks, which we describe in the remainder of the section.

While training, every data sample undergoes the same series of transformations through the layers of the neural network, which makes the procedure very amenable to parallelization. Accelerated training can be achieved by employing Graphics Processing Units (GPU) which enable performing several thousand simple operations, such as matrix multiplication, simultaneously. We harness the power of GPUs, which provide more than a 10x speedup over CPU, to address the problem of a large dataset. Moreover, to iterate faster it is important to be able to train multiple models simultaneously. Therefore, we set up a cluster computing environment where each model is trained independently on individual nodes (powered by GPUs) and all nodes have access to a central data server. This avoids the need for replicating data on individual nodes and enables efficient training. We achieve this practically by using Amazon Web Services (AWS), which is a cloud computing platform that allows flexible scaling of compute resources. In our implementation, we use up to 10 single-GPU nodes, where each GPU contains 1,536 CUDA cores and 4 GB of memory. The bandwidth of the central data server allows up to 15 nodes to fetch data simultaneously to train their models.

There are several other software optimizations that help make the training faster. We use a specialized deep learning library Torch, which has been developed by Facebook and Google and uses the Lua programming language. Such specialized libraries optimize the commonly used operations for neural networks and have fast routines written in $C$ that speed up training by an order of magnitude. Further, we use single precision floating point operations (instead of double precision) throughout our code. This has no practical effect on the parameter estimates and it halves the memory requirements and leads to substantial speed up in the computations.

Gradient descent for fitting models is impractical due to the size of the data. We use the standard machine learning method of minibatch gradient descent with momentum; see Ngiam, Coates, Lahiri, Prochnow, Le & Ng (2011) for a discussion of minibatch gradient descent for deep neural networks. In minibatch gradient descent, gradient steps are sequentially taken using subsets of the dataset. A block of the data is loaded into memory and the gradient of the objective function is calculated on this block of data. A step is then taken in the direction of the minibatch gradient with the step size determined by the "learning rate". Another block of data is then loaded into memory and the process repeats. The size of this block of data, referred to as the batch size, and the learning rate are optimized in order for fast (but stable) convergence of the model parameters.

In order that the minibatch gradients are unbiased, blocks of data must be drawn at random from the entire dataset. If gradients are biased, training may not converge and accuracy may be lost. A typical issue with mortgage and other types of loan data is that it is not stored randomly, but instead split into categories such as geographic region, time period, and loan type. Due to the size of the data, randomly scrambling the data can be computationally challenging. The original CoreLogic dataset needs to be reorganized for model fitting. The original dataset provided by CoreLogic is divided into static data (origination features) and dynamic data (monthly loan performance). The static data itself is divided into separate geographic regions (e.g., Pacific, Northeast, Southeast). The dynamic data is divided into geographic regions and then into months. In order to create a training sample, one has to match the static data for a loan with all of the monthly updates for that loan in the dynamic data files. In addition, one has to randomly order the training samples such that there is no bias towards a particular origination time or geographic location. Matching static data with the dynamic data via a search through these different subsets is impractical due to the size of the dataset. In order join the static and dynamic data, we create a hash table whose keys are the loan IDs and whose values are the destination folders $1, \ldots, L$ (randomly chosen). This hash table is used to randomly distribute the loans to the folders. Secondly, we use another hash table to match static data with dynamic data with these destination folders in order to avoid a search.

# C Hyperparameter Selection

Neural networks have a number of hyperparameters which need to be chosen. The standard approach to choosing these hyperparameters is to cross-validate them via a validation set. We train neural networks with different hyperparameters on the training set and compare the log-likelihood on the validation set. In particular, we cross-validate the number of layers and number of neurons per layer. The more layers and more neurons, the more complex the neural network is and the better able it is to fit complex relationships. However, with more complexity, there is also a higher chance of overfitting. We also cross-validate the size of the $\ell^2$ penalty, the learning rate schedule, batch size (see Appendix B), and the type of nonlinearity via a sparse grid search. For each grid point, we train a neural network and record its validation error. We then choose the hyperparameters at the grid point with the lowest validation error.

Several learning rate schedules were tested. The learning rate schedule is critical for training the neural network. If the learning rate is too high, there can be large oscillations that may drive the estimates away from the optimal value. If the learning rate is too low, the neural network will learn very slowly. The chosen learning rate schedule is:

$$\text{Learning rate} \quad = \quad \frac{LR_0}{1 + t/800}, \tag{9}$$

where the initial learning rate $LR_0 = 0.1$ and $t$ is the epoch number. The half-life (i.e., the number of epochs until the learning rate is reduced by half) is 800. Each epoch contains approximately 1.5 million training samples. The batchsize is $4,000$, meaning approximately 375 gradient steps are taken per epoch.

Cross-validation leads to the choice of 5 hidden layers, with 200 units in the first hidden layer and 140 units in each subsequent one. The rectified linear unit (ReLU) nonlinearity (i.e., $\sigma(x) = \max(0, x)$) was found to yield better performance and faster convergence than the sigmoid nonlinearity ($\sigma(x) = 1/(1 + e^{-x})$).

# D Finite-Difference Approximation of Sensitivities

This appendix provides finite-difference estimators for the sensitivities considered in in Section 5.4. The sensitivity (7) can be estimated directly from the dataset using the formula

$$\text{Sensitivity}(u, v, j) = \frac{1}{|M_u|} \sum_{(n,t) \in M_u} \left| \frac{\partial h_{\hat{\theta}}(v, x)}{\partial x_j} \right|_{x = X_t^n} \tag{10}$$

where $M_u = \{(n, t) : U_t^n = u, 1 \leq t \leq T, 1 \leq n \leq N\}$, $x_j$ is the $j$-th element of $x$, and $\hat{\theta}$ is the MLE. Here $X_t^n$ and $U_t^n$ are the vector of explanatory variables and the state of mortgage $n$ at time $t$, respectively. Note that the quantity in (10) aggregates over only the relevant mortgages and times, namely those in $M_u$. This allows computing the probability that this mortgage attains state $v$ at time $t + 1$, which in turn facilitates computing the sensitivity for transition from state $u$ to $v$. The formula (10) is the sensitivity across the entire dataset rather than the sensitivity at a single representative point.

We now develop a finite-difference approximation for the sensitivity (8). Importantly, the finite-difference formulas can be used to analyze the sensitivity of nonlinear functions which are piecewise linear (such as neural networks with ReLU units). Let

$$\begin{aligned}
\text{Interaction}(u,v,i,j) &= \frac{1}{|M_u|}\sum_{(n,t)\in M_u}\Big|\big(h_{\hat\theta}(v,\cdot,x_i+\Delta_i,\cdot,x_j+\Delta_j,\cdot)-h_{\hat\theta}(v,\cdot,x_i,\cdot,x_j,\cdot)\big)\\
&- \big(h_{\hat\theta}(v,\cdot,x_i+\Delta_i,\cdot,x_j,\cdot)-h_{\hat\theta}(v,\cdot,x_i,\cdot,x_j,\cdot)\big)\\
&- \big(h_{\hat\theta}(v,\cdot,x_i,\cdot,x_j+\Delta_j,\cdot)-h_{\hat\theta}(v,\cdot,x_i,\cdot,x_j,\cdot)\big)\Big|_{x=X_t^n}
\end{aligned} \tag{11}$$

The formula (11) measures how much of the change in the fitted transition probability $h_{\hat\theta}$ cannot be explained by independent shifts $x_i \to x_i+\Delta_i$ and $x_j \to x_j+\Delta_j$. It is a *second-order sensitivity.* If $h_{\hat\theta}$ is smooth and $\Delta_i=\Delta_j=\Delta \ll 1$, $\frac{2}{\Delta^2}\text{Interaction}(u,v,i,j)$ is a finite-difference estimator for

$$\mathbb{E}\Big[\big|\sum_{i,j=1}^{2}\frac{\partial^2 h_{\hat\theta}}{\partial x_i\partial x_j}(V,X)\big|\Big|V=v,U=u\Big] \tag{12}$$

If there is no interaction (for example, loan performance depends linearly on the covariates), $\text{Interaction}(u,v,i,j)$ of course equals 0.

Similarly, the importance of the interaction between three covariates $i,j,k$ for a transition from state $u$ to $v$ is measured by:

$$\begin{aligned}
\text{Interaction}(u,v,i,j,k) &= \frac{1}{|M_u|}\sum_{(n,t)\in M_u}\Big|\big(h_{\hat\theta}(v,\cdot,x_i+\Delta_i,x_j+\Delta_j,x_k+\Delta_k)-h_{\hat\theta}(v,\cdot,x_i,x_j,x_k)\big)\\
&- \big(h_{\hat\theta}(v,\cdot,x_i+\Delta_i,x_j+\Delta_j,x_k)-h_{\hat\theta}(v,\cdot,x_i,x_j,x_k)\\
&- \big(h_{\hat\theta}(v,\cdot,x_i+\Delta_i,x_j,x_k+\Delta_k)-h_{\hat\theta}(v,\cdot,x_i,x_j,x_k)\\
&- \big(h_{\hat\theta}(v,\cdot,x_i,x_j+\Delta_j,x_k+\Delta_k)-h_{\hat\theta}(v,\cdot,x_i,x_j,x_k)\big)\Big|_{x=X_t^n}
\end{aligned} \tag{13}$$

The formula (13) measures the *third-order interactions.* It detects interactions between three covariates which are not explained by the sum of the pairwise interactions. If $h_{\hat\theta}$ is smooth and $\Delta_i=\Delta_j=\Delta \ll 1$, $\frac{6}{\Delta^3}\text{Interaction}(u,v,i,j,k)$ is a finite-difference estimator for

$$\mathbb{E}\Big[\big|\sum_{\alpha+\beta+\zeta=3}\frac{\partial^3 h_{\hat\theta}}{\partial x_i^{\alpha}\partial x_j^{\beta}\partial x_k^{\zeta}}(V,X)\big|\Big|V=v,U=u\Big]. \tag{14}$$

If there are only pairwise interactions (for example, loan performance depends quadratically on the covariates), $\text{Interaction}(u,v,i,j,k)$ of course equals 0. If $\text{Interaction}(u,v,i,j,k)\neq 0$, it indicates that there is significant nonlinearity beyond even a quadratic model. More impor-tantly, it indicates which triplets of covariates are interacting most strongly with each other. The formula (13) can be generalized to measure even higher order interactions (fourth, fifth, etc.); however, this paper only empirically investigates second- and third-order interactions.

| Feature | Values |
|---|---|
| FICO score | Continuous |
| Original debt-to-income ratio | Continuous |
| Original loan-to-value ratio | Continuous |
| Original interest rate | Continuous |
| Original balance | Continuous |
| Original term of loan | Continuous |
| Original sale price | Continuous |
| Buydown flag | True, False |
| Negative amortization flag | True, False |
| Occupancy Type | Owner occupied, second home, non-owner occupied or investment property, other |
| Prepayment penalty flag | True, False |
| Product type | See Table 6 |
| Loan purpose | Purchase, Refinance Cash-out, Refinance No Cash Out, Second mortgage, Refinance Cash Out Unkown, Construction Loan, Debt Consolidation Loan, Home Improvement Loan, Education Loan, Medical Loan, Vehicle Purchase, Reverse Mortgage, Other |
| Documentation | Full documentation, Low or minimal documentation, No asset or income verification, Other |
| Lien type | 1st Position, 2nd Position, 3rd Position, 4th Position, Other |
| Channel | Retail Branch, Wholesale Bulk, Mortgage Broker, Realtor Originated, Relocation Corporate, Relocation Mortgage Broker, Builder, Direct Mail, Other Direct, Internet, Other Retail, Mortgage Banker, Correspondent, Other |
| Loan type | Conventional Loan, VA Loan, FHA Loan, Other Government Loan, Affordable Housing Loan, Pledged Asset Loan, Other |
| Number of units | 1,2,3,4,5 |
| Appraised value < sale price? | True, False |
| Initial Investor Code | Portfolio Held, Securitized Other, GNMA/Ginnie Mae, GSE |
| Interest Only Flag | True, False |
| Margin for ARM mortgages | Continuous |
| Periodic rate cap | Continuous |
| Periodic rate floor | Continuous |
| Periodic pay cap | Continuous |
| Periodic pay floor | Continuous |
| Lifetime rate cap | Continuous |
| Lifetime rate floor | Continuous |
| Rate reset frequency | 1,2,3, ... (months) |
| Pay reset frequency | 1,2,3, ... (months) |
| First rate reset period | 1,2,3, ... (months since origination) |
| Convertible flag | True, False |
| Pool insurance flag | True, False |
| Alt-A flag | True, False |
| Prime flag | True, False |
| Subprime flag | True, False |
| Geographic state | CA, FL, NY, MA, etc. |
| Vintage (origination year) | 1995, 1996, ..., 2014 |

Table 1: Loan-level features at origination (from CoreLogic).

| Feature | Mean | Median | Min | Max | 25% Quantile | 75% Quantile |
|---|---|---|---|---|---|---|
| FICO | 634 | 630 | 300 | 900 | 580 | 680 |
| Original LTV | 74 | 80 | 0 | 200 | 68 | 90 |
| Original interest rate | 7.8 | 7.8 | 0 | 30 | 6.3 | 9.6 |
| Original balance | 160,197 | 124,000 | 7 | 318,750 | 68,850 | 210,000 |

Table 2: Summary statistics for some mortgage features in subprime data.

| Feature | Mean | Median | Min | Max | 25% Quantile | 75% Quantile |
|---|---|---|---|---|---|---|
| FICO | 720 | 730 | 300 | 900 | 679 | 772 |
| Original LTV | 74 | 79 | 0 | 200 | 63 | 90 |
| Original interest rate | 5.8 | 5.8 | 0 | 20.6 | 4.9 | 6.6 |
| Original balance | 190,614 | 151,353 | 1 | 6,450,000 | 98,679 | 238,000 |

Table 3: Summary statistics for some mortgage features in prime data.

| Feature | Mean | Median | Min | Max | 25% Quantile | 75% Quantile |
|---|---|---|---|---|---|---|
| FICO | 707 | 718 | 300 | 900 | 660 | 767 |
| Original LTV | 74 | 79 | 0 | 200 | 63 | 90 |
| Original interest rate | 6 | 5.95 | 0 | 30 | 4.9 | 6.9 |
| Original balance | 186,202 | 148,500 | 1 | 6,450,000 | 94,000 | 234,000 |

Table 4: Summary statistics for some mortgage features in full dataset (prime and subprime).

| **Feature** | **Values** |
|---|---|
| Current status | Current, 30 days delinquent, 60 days delinquent, 90+ days delinquent, Foreclosed, REO, paid off |
| Number of days delinquent | Continuous |
| Current interest rate | Continuous |
| Current interest rate − national mortgage rate | Continuous |
| Time since origination | Continuous |
| Current balance | Continuous |
| Scheduled principal payment | Continuous |
| Scheduled principal + interest payment | Continuous |
| Number of months the mortgage's interest been less than the national mortgage rate and the mortgage did not prepay | Continuous |
| Number of occurrences of current in past 12 months | 0-12 |
| Number of occurrences of 30 days delinquent in past 12 months | 0-12 |
| Number of occurrences of 60 days delinquent in past 12 months | 0-12 |
| Number of occurrences of 90+ days delinquent in past 12 months | 0-12 |
| Number of occurrences of Foreclosed in past 12 months | 0-12 |

Table 5: Loan-level performance features (from CoreLogic).

| Product type | Percent of Total | Percent of Subprime | Percent of Prime |
|---|---|---|---|
| Fixed Rate | 80.6 % | 48 % | 86.3 % |
| ARM | 11.7 % | 29 % | 8.7 % |
| GPM (graduated payment) | .01 % | 0 % | .01 % |
| Balloon Unknown | .9 % | 1 % | .9 % |
| Balloon 5 | .03 % | 0 % | .03 % |
| Balloon 7 | .03 % | .004 % | .04 % |
| Balloon 10 | .004 % | .006 % | .004 % |
| Balloon 15/30 | .2 % | 1.07 % | .005 % |
| ARM Balloon | .2 % | 1.3 % | .01 % |
| Balloon Other | .7 % | 3.3 % | .26 % |
| Two Step Unknown | .02 % | 0 % | .02 % |
| Two Step 10/20 | .003 % | 0 % | .003 % |
| GPARM | .002 % | 0 % | .002 % |
| Hybrid 2/1 | 1 % | 6.7 % | 0 % |
| Hybrid 3/1 | .63 % | 2.2 % | .35 % |
| Hybrid 5/1 | 1.9 % | .22 % | 2.2 % |
| Hybrid 7/1 | .5 % | .005 % | .64 % |
| Hybrid 10/1 | .24 % | .02 % | .28 % |
| Hybrid Other | .02 % | .02 % | .02 % |
| Other | .7 % | 4.3 % | .01 % |
| Invalid data | .18 % | .6 % | .11 % |

Table 6: Types of mortgages for full dataset, subprime subset, and prime subset.

| **Feature** | **Values** |
|---|---|
| Monthly zip code median house price per square feet (Zillow) | Continuous |
| Monthly zip code average house price (Zillow) | Continuous |
| Zillow zip code house price change since origination | Continuous |
| Monthly state unemployment rate (BLS) | Continuous |
| Yearly county unemployment rate (BLS) | Continuous |
| Original interest rate - National mortgage rate (Freddie Mac) | Continuous |
| Original interest rate - National mortgage rate at origination (Freddie Mac) | Continuous |
| Number of months where interest rate is < nat'l mortgage rate (Freddie Mac) | Continuous |
| Median income in same zip code (Powerlytics) | Continuous |
| Total number of prime mortgages in same zip code (CoreLogic) | Continuous |
| Lagged subprime default rate in same zip code (CoreLogic) | Continuous |
| Lagged prime default rate in same zip code (CoreLogic) | Continuous |
| Lagged prime paid off rate in same zip code (CoreLogic) | Continuous |

Table 7: Local and national economic risk factors. Data sources in parentheses. "Default rate" is taken to be the states foreclosure or REO.

| Current/Next | Current | 30 days | 60 days | 90+ days | Foreclosure | REO | Paid Off |
|---|---|---|---|---|---|---|---|
| Current | 93 | 4.7 | 0 | 0 | .01 | .002 | 2 |
| 30 days | 30 | 45 | 23 | 0 | .2 | .004 | 2 |
| 60 days | 11 | 16 | 35 | 32 | 5 | .01 | 1.5 |
| 90+ days | 4 | 1 | 2 | 82 | 9 | .3 | 2.2 |
| Foreclosure | 2 | .4 | .3 | 6.5 | 85 | 4 | 1.4 |
| REO | 0 | 0 | 0 | 0 | 0 | 100 | 0 |
| Paid off | 0 | 0 | 0 | 0 | 0 | 0 | 100 |

Table 8: Monthly transition matrix for subprime data. Probabilities are given in percentages.

| Current/Next | Current | 30 days | 60 days | 90+ days | Foreclosure | REO | Paid Off |
|---|---|---|---|---|---|---|---|
| Current | 97.1 | 1.3 | 0 | 0 | .001 | .0002 | 1.57 |
| 30 days | 34.6 | 44.4 | 19 | 0 | .004 | .003 | 1.82 |
| 60 days | 12 | 16.8 | 34.5 | 34 | 1.6 | .009 | 1.1 |
| 90+ days | 4.1 | 1.4 | 2.6 | 80.2 | 10 | .3 | 1.3 |
| Foreclosure | 1.9 | .3 | .1 | 6.8 | 87 | 2.5 | 1.3 |
| REO | 0 | 0 | 0 | 0 | 0 | 100 | 0 |
| Paid off | 0 | 0 | 0 | 0 | 0 | 0 | 100 |

Table 9: Monthly transition matrix for prime data. Probabilities are given in percentages.

| Current/Next | Current | 30 days | 60 days | 90+ days | Foreclosure | REO | Paid Off |
|---|---|---|---|---|---|---|---|
| Current | 96.7 | 1.6 | 0 | 0 | .002 | .0004 | 1.61 |
| 30 days | 34.2 | 44.5 | 19.3 | 0 | .02 | .003 | 1.84 |
| 60 days | 12 | 16.7 | 34.5 | 33.8 | 1.9 | .009 | 1.1 |
| 90+ days | 4.1 | 1.4 | 2.5 | 80.4 | 9.9 | .3 | 1.3 |
| Foreclosure | 1.9 | .3 | .1 | 6.8 | 86.8 | 2.6 | 1.3 |
| REO | 0 | 0 | 0 | 0 | 0 | 100 | 0 |
| Paid off | 0 | 0 | 0 | 0 | 0 | 0 | 100 |

Table 10: Monthly transition matrix for full dataset. Probabilities are given in percentages.

| Variable | Test Loss |
|---|---|
| State unemployment rate | 1.160 |
| Current outstanding balance | .303 |
| Original interest rate | .233 |
| FICO score | .204 |
| Number of times 30dd in last 12 months | .179 |
| Number of times current in last 12 months | .175 |
| Original loan balance | .175 |
| Total days delinquent $\geq$ 160 | .171 |
| Vintage year < 1995 | .171 |
| Prime mortgage flag | .171 |
| Lien type = first lien | .171 |
| Original interest rate - national mortgage rate | .170 |
| LTV ratio | .169 |
| Burnout (number of months where it was optimal to prepay but did not) | .168 |
| Time since origination | .168 |
| Current interest rate - national mortgage rate | .168 |
| Number of times 60dd in last 12 months | .168 |
| Number of times foreclosure in last 12 months | .168 |
| Number of days delinquent | .168 |
| Product type = Fixed Rate Loan | .168 |
| Not a convertible loan | .168 |
| Number of times 90+ days delinquent in last 12 months | .168 |
| Lagged prime prepayment rate in same zip code | .168 |
| Zillow zip code housing price change since origination | .168 |
| No pool insurance | .168 |
| Channel = corresponded lender | .168 |
| Documentation = full documentation | .168 |
| Number of units $\leq$ 5 | .168 |
| Loan type = conventional loan | .168 |
| ⋮ | ⋮ |

Table 11: This table reports a leave-one-out analysis for the explanatory power of the variables (performed using 5-layer neural network). For each variable, a leave-one-out test is performed: the variable is removed from the model and the model's negative average log-likelihood is evaluated on the test dataset in the absence of the covariate (this is the test loss). Specifically, the leave-one-out variable is set to 0 for all data samples in the test dataset and the model's log-likelihood is calculated using the reduced variable vector. Then, the variable is replaced in the model, and a leave-one-out test is performed on a new variable. Removing the variable of course reduces the accuracy of the model, and the test loss becomes larger. If a particular variable has strong explanatory power, the test loss will significantly increase. The test loss when no variables are dropped (complete model) is .167.

| Variable | Gradient |
|---|---|
| Current Outstanding Balance | 0.1878 |
| Original Loan Balance | 0.0856 |
| Original Interest Rate | 0.0503 |
| Current Interest Rate - National Mortgage Rate | 0.0478 |
| Original Interest Rate - National Mortgage Rate | 0.0463 |
| Zillow Zip Code Housing Price Change Since Origination | 0.0386 |
| Number of Times 30 Days Delinquent in Last 12 Months | 0.0384 |
| Scheduled Interest and Principle Due | 0.0364 |
| Number of Times 60 Days Delinquent in Last 12 Months | 0.0362 |
| Zillow zip code median house price change since origination | 0.0346 |
| Time Since Origination | 0.0306 |
| ARM First Rate Reset Period | 0.0295 |
| FICO Score | 0.0293 |
| Lagged Prime Prepayment Rate in Same Zip Code | 0.0292 |
| Number of Times 90+ Days Delinquent in Last 12 Months | 0.0237 |
| Current Interest Rate - Original Interest Rate | 0.0228 |
| State Unemployment Rate | 0.0214 |
| Number of Days Delinquent | 0.0195 |
| ARM periodic rate cap | 0.0191 |
| Lagged Prime Default Rate in Same Zip Code | 0.0190 |
| Total Number of Prime Mortgages in Same Zip Code | 0.0190 |
| Number of Times Current in Last 12 Months | 0.0145 |
| Original Appraised Value | 0.0132 |
| Original Interest Rate - National Mortgage Rate at Origination | 0.0129 |
| LTV Ratio | 0.0116 |
| Lagged Default Rate for Subprime Mortgages in Same Zip Code | 0.0115 |
| ⋮ | ⋮ |

Table 12: Variable sensitivity analysis. We report the average absolute gradient for transition current $\rightarrow$ paid off. Performed using 5-layer neural network.

| Variable | Gradient |
|---|---|
| Number of Times 30 Days Delinquent in Last 12 Months | 0.0650 |
| FICO Score | 0.0445 |
| Number of Times 60 Days Delinquent in Last 12 Months | 0.0334 |
| Current Outstanding Balance | 0.0320 |
| Original Loan Balance | 0.0285 |
| Original Interest Rate | 0.0235 |
| Zillow Zip Code Housing Price Change Since Origination | 0.0187 |
| Original Interest Rate - National Mortgage Rate | 0.0170 |
| Number of Times 90+ Days Delinquent in Last 12 Months | 0.0145 |
| Lagged Prime Default Rate in Same Zip Code | 0.0116 |
| Number of Times Foreclosed in Last 12 Months | 0.0109 |
| Zillow zip code median house price change since origination | 0.0108 |
| Number of Days Delinquent | 0.0095 |
| Number of Times Current in Last 12 Months | 0.0088 |
| Time Since Origination | 0.0087 |
| Current Interest Rate - Original Interest Rate | 0.0087 |
| Lagged Prime Prepayment Rate in Same Zip Code | 0.0074 |
| ARM Rate Reset Frequency | 0.0070 |
| Total Number of Prime Mortgages in Same Zip Code | 0.0068 |
| Current Interest Rate - National Mortgage Rate | 0.0065 |
| State Unemployment Rate | 0.0060 |
| Scheduled Interest and Principle Due | 0.0050 |
| LTV Ratio | 0.0050 |
| Lagged Default Rate for Subprime Mortgages in Same Zip Code | 0.0050 |
| Original Term of the Loan | 0.0041 |
| ⋮ | ⋮ |

Table 13: Variable sensitivity analysis. We report the average absolute gradient for transition current $\rightarrow$ 30 days delinquent. Performed using 5-layer neural network.

| Most Important Pairs of Variables | Gradient |
|---|---|
| Original interest rate, State unemployment rate | $1.33 \times 10^{-3}$ |
| Original interest Rate, Number of times current in past 12 months | $1.21 \times 10^{-3}$ |
| Original interest rate, Original term of the loan | $0.87 \times 10^{-3}$ |
| FICO score, Original interest rate | $0.86 \times 10^{-3}$ |
| Number of times current in past 12 months, Original term of the loan | $0.69 \times 10^{-3}$ |
| State unemployment rate, Original term of the loan | $0.67 \times 10^{-3}$ |
| ⋮ | ⋮ |

Table 14: Analysis of pairwise variable interactions. We report the average absolute gradient for transition current → paid off. Performed using 5-layer neural network.

| Most Important Triplets of Variables | Gradient |
|---|---|
| Original interest rate, FICO score, State unemployment rate | $7.52 \times 10^{-4}$ |
| Original interest Rate, State unemployment rate, Original interest rate - National mortgage rate | $5.31 \times 10^{-4}$ |
| Original loan balance, Original interest rate, State unemployment rate | $4.78 \times 10^{-4}$ |
| Original loan balance, Original interest rate, FICO score | $4.53 \times 10^{-4}$ |
| Current outstanding balance, Original interest rate, State unemployment rate | $3.72 \times 10^{-4}$ |
| Original loan balance, FICO score, State unemployment rate | $3.71 \times 10^{-4}$ |
| ⋮ | ⋮ |

Table 15: Analysis of interactions between three variables. We report the average absolute gradient for transition current → paid off. Performed using 5-layer neural network.

| Most Important Pairs of Variables | Gradient |
|---|---|
| FICO score, Original term of the loan | $1.68 \times 10^{-3}$ |
| Original interest rate, Original term of the loan | $1.28 \times 10^{-3}$ |
| Number of times current in last 12 months, Original term of the loan | $1.27 \times 10^{-3}$ |
| State unemployment rate, Original term of the loan | $0.91 \times 10^{-3}$ |
| Original term of the loan, Scheduled Principle Due (missing value) | $0.66 \times 10^{-3}$ |
| Original term of the loan, Number of IO months (missing value | $0.66 \times 10^{-3}$ |
| $\vdots$ | $\vdots$ |

Table 16: Analysis of pairwise variable interactions. We report the average absolute gradient for transition current $\to$ 30 days delinquent. Performed using 5-layer neural network.

| Most Important Triplets of Variables | Gradient |
|---|---|
| FICO score, Original interest rate, Number of times current in last 12 months | $2.10 \times 10^{-4}$ |
| FICO score, Original interest rate,<br>Original interest rate - National mortgage rate | $1.82 \times 10^{-4}$ |
| FICO score, Current outstanding balance, Original interest rate | $1.60 \times 10^{-4}$ |
| FICO score, Original loan balance, Original interest rate | $1.57 \times 10^{-4}$ |
| FICO score, Current outstanding balance, Original loan balance | $1.30 \times 10^{-4}$ |
| Current Outstanding balance, Original loan balance, Original interest rate | $1.27 \times 10^{-4}$ |
| $\vdots$ | $\vdots$ |

Table 17: Analysis of interactions between three variables. We report the average absolute gradient for transition current $\to$ 30 days delinquent. Performed using 5-layer neural network.

| Model | In-sample Loss | Out-of-sample Loss | Out-of-sample Loss | LR Score |
|---|---|---|---|---|
| | w/o Dropout | w/o Dropout | with Dropout | |
| 0 hidden layer | .1840 | .1805 | .1805 | N/A |
| 1 hidden layer | .1680 | .1700 | .1685 | $1.006 \times 10^8$ |
| 3 hidden layer | .1644 | .1679 | .1671 | $2.264 \times 10^7$ |
| 5 hidden layer | .1639 | .1684 | .1670 | $3.145 \times 10^6$ |
| 7 hidden layer | .1638 | .1688 | .1673 | $6.290 \times 10^5$ |
| Ensemble | .1640 | .1659 | .1654 | N/A |

Table 18: In-sample and out-of-sample loss (negative average log-likelihood) for neural networks of different depth. The ensemble is composed of eight 5-layer networks. The LR Score is the likelihood ratio test statistic, given by twice the difference between the in-sample log-likelihood of the alternative model and the in-sample log-likelihood of the null model. We test a more complex model (alternative) against the simpler one (null); for example, $1.006 \times 10^8$ is the score for a test of the 1 hidden layer network against the 0 hidden layer network. All tests reported are significant at the 99% level.

| Model | C→P | 30dd→P | 60dd→P | 90+dd→P | F→P |
|---|---|---|---|---|---|
| 0 hidden layer | .65 | .77 | .68 | .59 | .57 |
| 1 hidden layer | .72 | .79 | .71 | .76 | .68 |
| 3 hidden layer | .74 | .81 | .73 | .79 | .72 |
| 5 hidden layer | .74 | .81 | .73 | .79 | .73 |
| Ensemble | .76 | .83 | .74 | .79 | .74 |

Table 19: Out-of-sample AUC for various transitions to paid off (i.e., prepayment) for neural networks of different depth. The ensemble is composed of eight 5-layer networks. “P” stands for paid off, “dd” stands for days delinquent, and “F” stands for foreclosure. The AUC for transition $u \to P$ is the AUC for the two-way classification of whether the mortgage is in state $P$ or not in state $P$ at a 1-month horizon conditional on the mortgage currently being in state $u$.

| State/Portfolio | 0-Layer Neural Network | 5-Layer Neural Network |
|---|---|---|
| REO | 0.00 | 0.00 |
| Paid off | 0.83 | 0.36 |
| Current | 98.71 | 99.25 |
| 30 dd | 0.46 | 0.39 |
| 60 dd | 0.00 | 0.00 |
| 90+ dd | 0.00 | 0.00 |
| Foreclosure | 0.01 | 0.00 |

Table 20: Percent of portfolio which is in each state at a 1 month time horizon. Portfolios have size $N = 20,000$ and are chosen from an available pool of $100,000$ mortgages.

| State/Portfolio | 0-Layer Neural Network | 5-Layer Neural Network |
|---|---|---|
| REO | 0.03 | 0.02 |
| Paid off | 8.14 | 4.06 |
| Current | 89.09 | 93.28 |
| 30 dd | 1.54 | 1.60 |
| 60 dd | 0.36 | 0.36 |
| 90+ dd | 0.54 | 0.49 |
| Foreclosure | 0.30 | 0.19 |

Table 21: Percent of portfolio which is in each state at a 1 year time horizon. Portfolios have size $N = 20,000$ and are chosen from an available pool of $100,000$ mortgages.

| | 0-Layer Neural Network | 5-Layer Neural Network |
|---|---|---|
| Avg. Actual Prepayments | 1723.8 | 1723.8 |
| Avg. Predicted Prepayments | 2853.8 | 1456.9 |
| Avg. Absolute Gap | 1186.0 | 278.5 |
| Avg. Standardized Gap | 2.4 | 1.9 |

Table 22: Comparison of out-of-sample pool-level distribution. The table reports averages for 50 random test portfolios. "Avg. Predicted Prepayments" is the average across 50 test portfolios of the mean of the forecast distribution. The "Avg. Absolute Gap" is the absolute difference between the predicted and the actual number of prepayments. The "Avg. Standardized Gap" is the difference between the predicted and the actual number of prepayments measured in multiples of the forecast standard deviation.

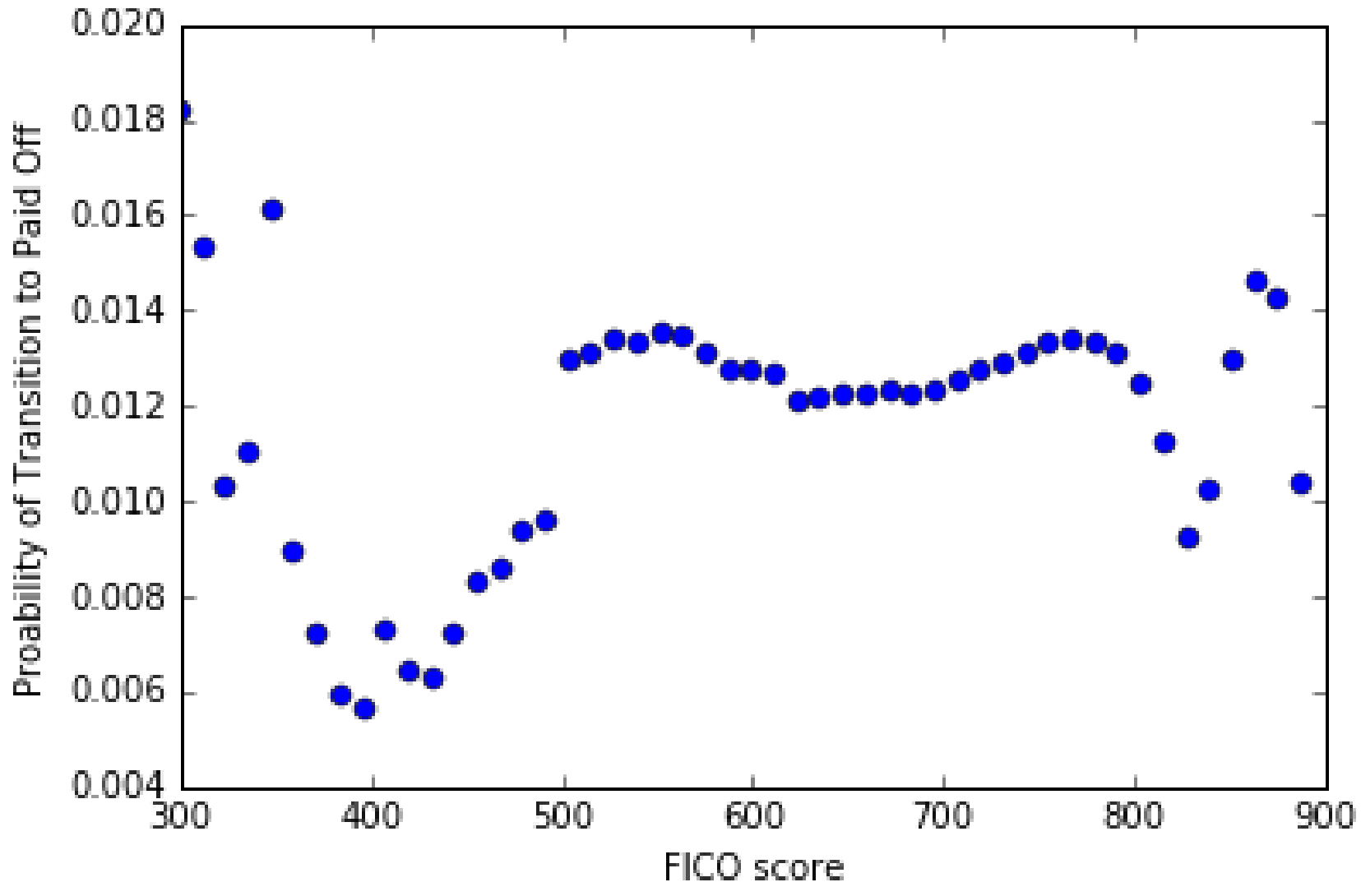

Figure 4: Empirical monthly prepayment rate versus FICO score. The figure shows that the prepayment rate has a significant nonlinear relationship with the FICO score of the borrower. The propensity to prepay is less for borrowers with lower FICO scores but it plateaus once the score crosses a threshold of about 500 points. This reinforces the need for a model family that is capable of learning nonlinear functions of the data.

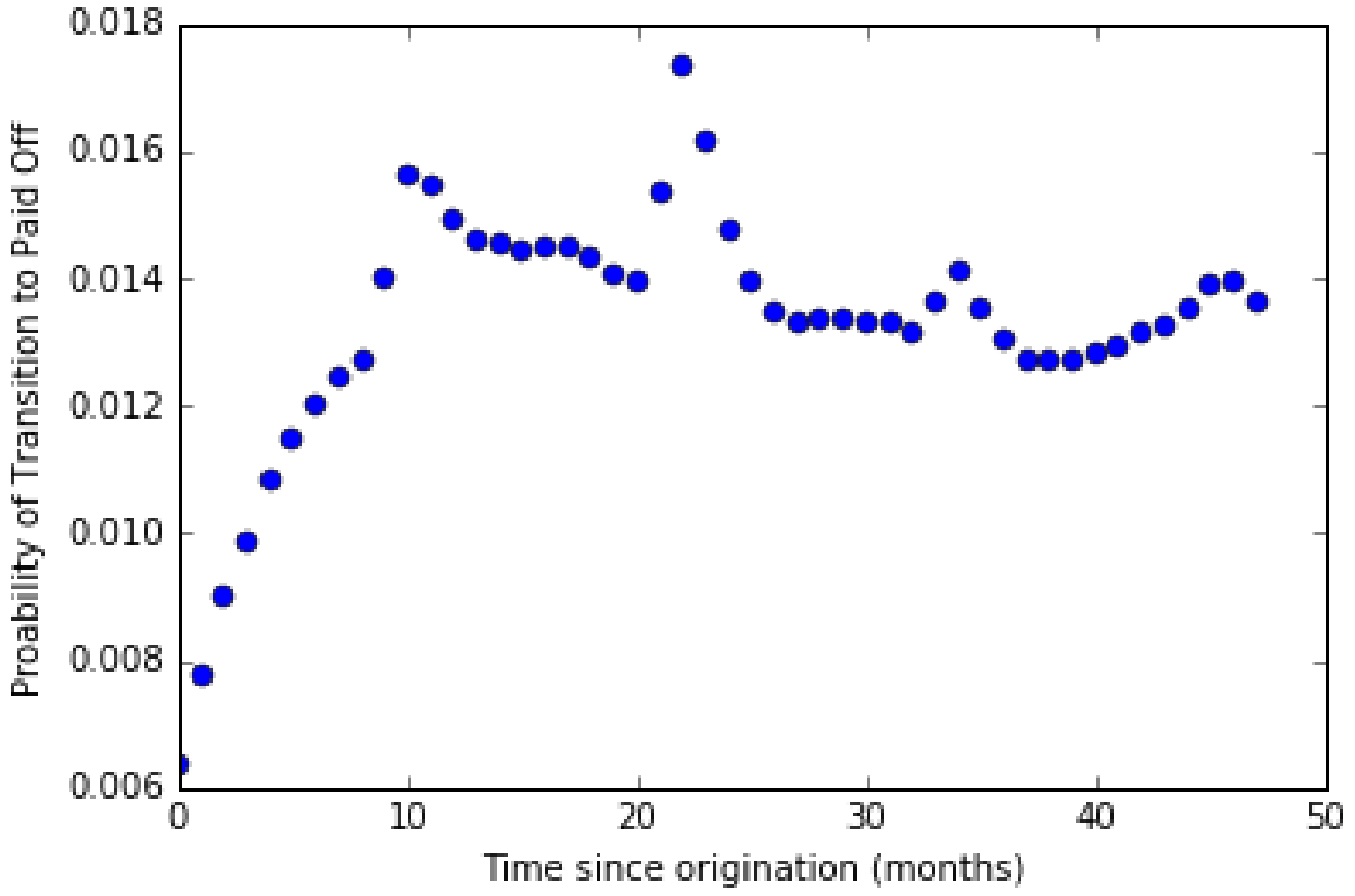

Figure 5: Empirical monthly prepayment rate versus time since origination (loan age). The figure shows that the prepayment rate has a significant nonlinear relationship with the age of the mortgage. Several spikes in the rate occur at 1, 2, and 3 years. These might be due to the expiration of prepayment penalties or ARM and hybrid mortgages having rate resets. Many of the subprime mortgages started with low teaser rates and would later jump to higher rates; borrowers would refinance to avoid these rate jumps.

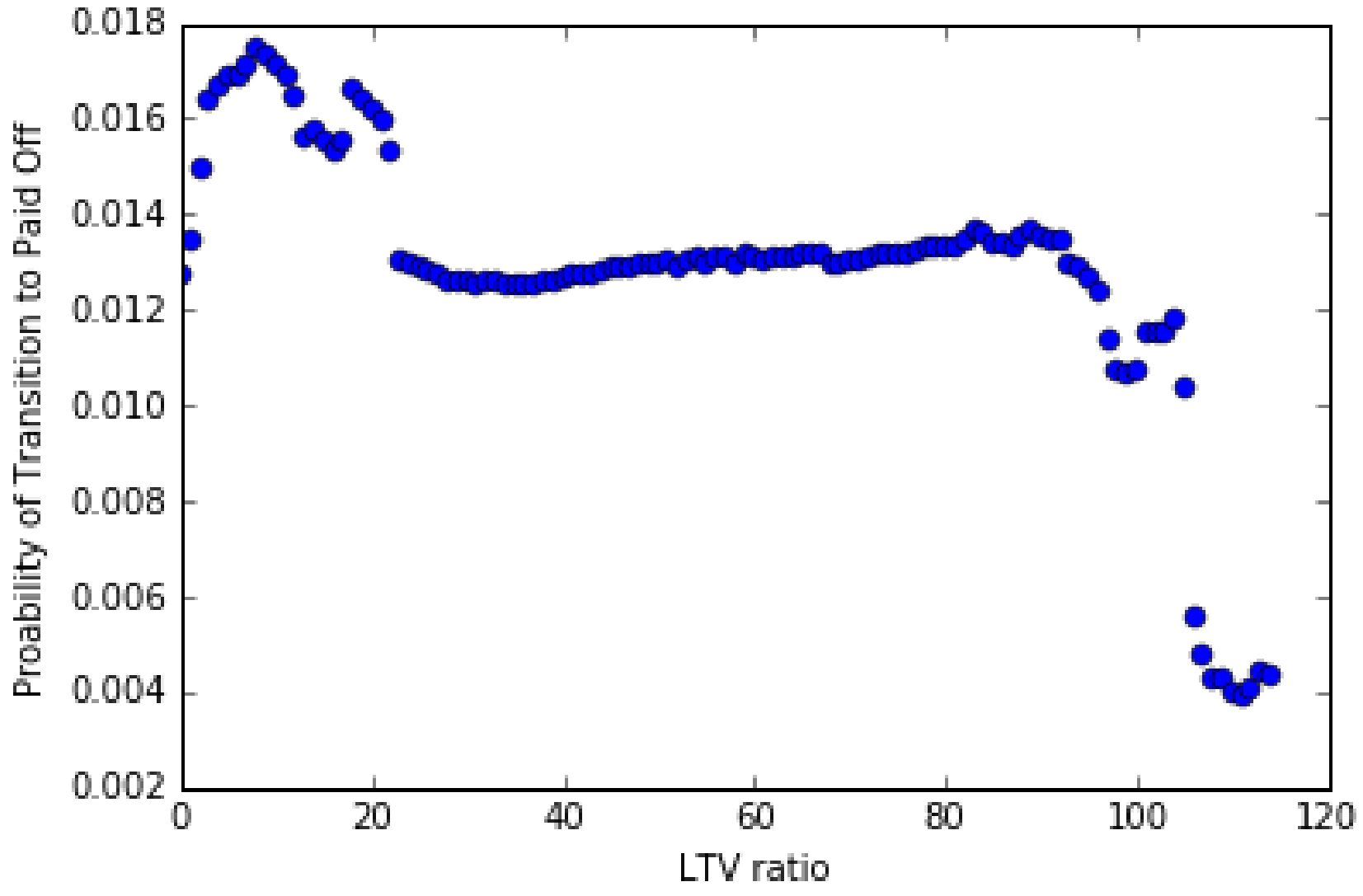

Figure 6: Empirical monthly prepayment rate versus loan-to-value (LTV) ratio at origination. The figure shows that the prepayment rate has a significant nonlinear relationship with the LTV ratio. One should expect this curve to have a downward slope since a loan with high LTV will have lesser opportunities to refinance due to a large loan amount relative to the value of the asset. This trend is observed in the data, albeit with significant nonlinearity as seen in the figure.

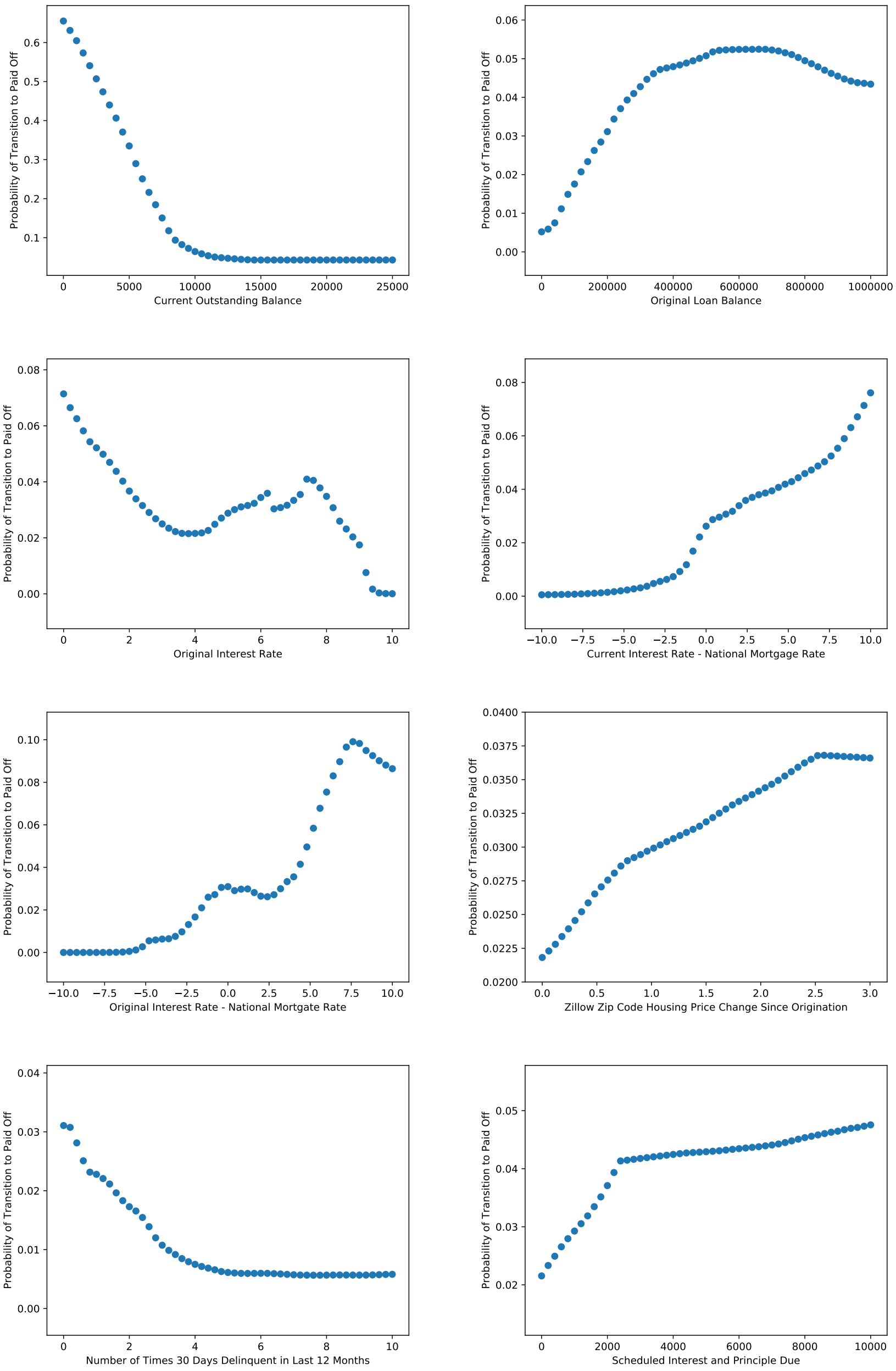

Figure 7: Nonlinear Relationships between Prepayment and Covariates.

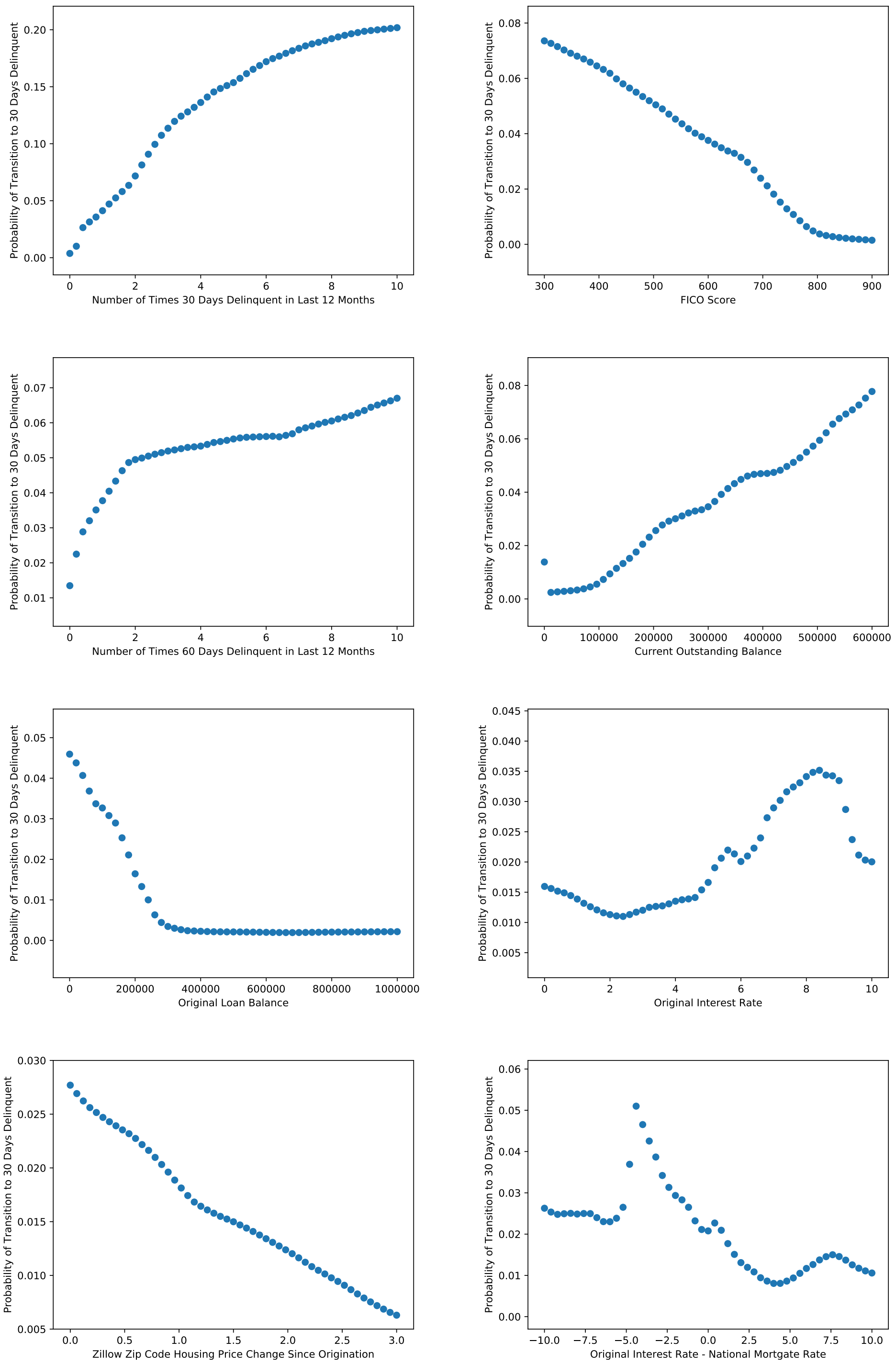

Figure 8: Nonlinear Relationships between Delinquency and Covariates.

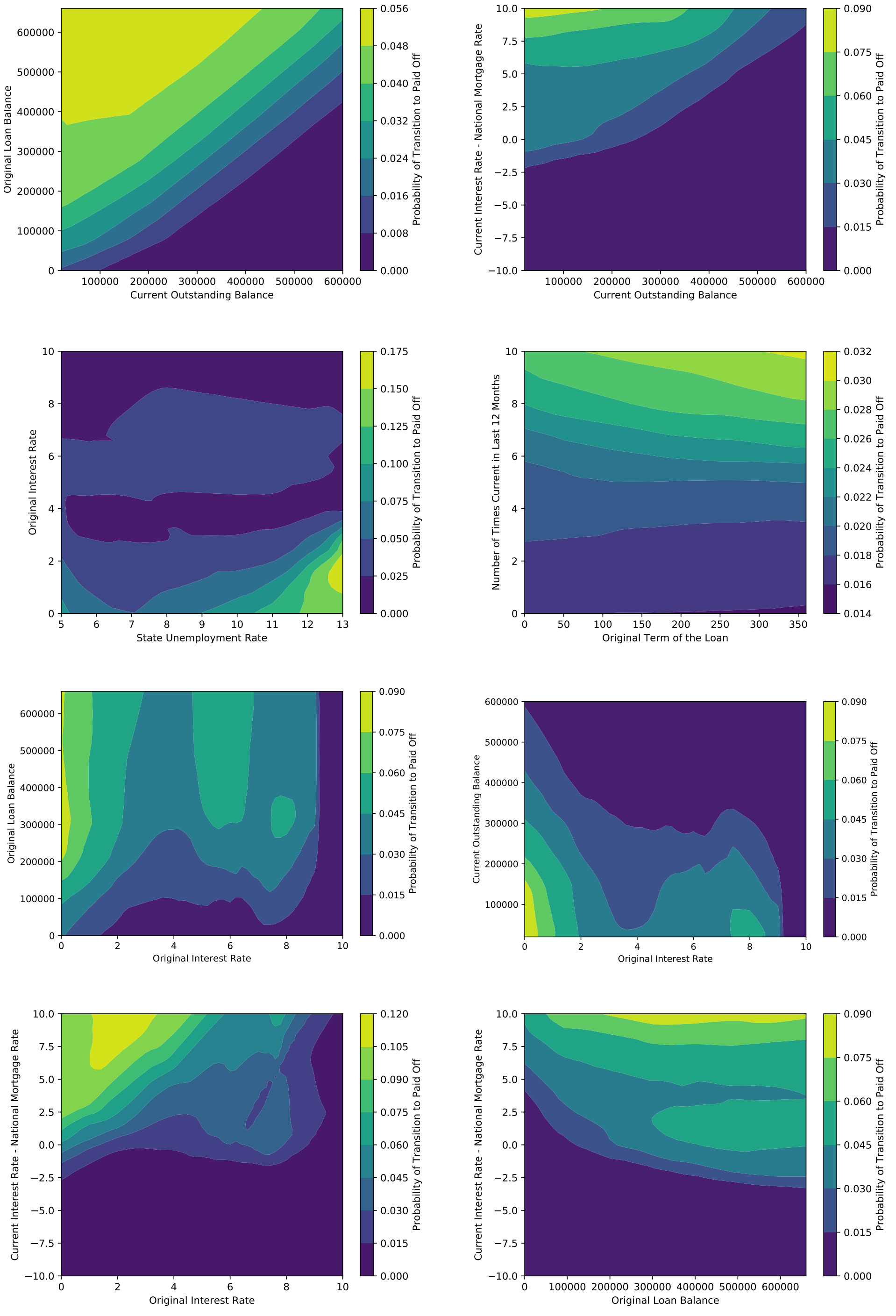

Figure 9: Relationship between prepayment and pairs of covariates.

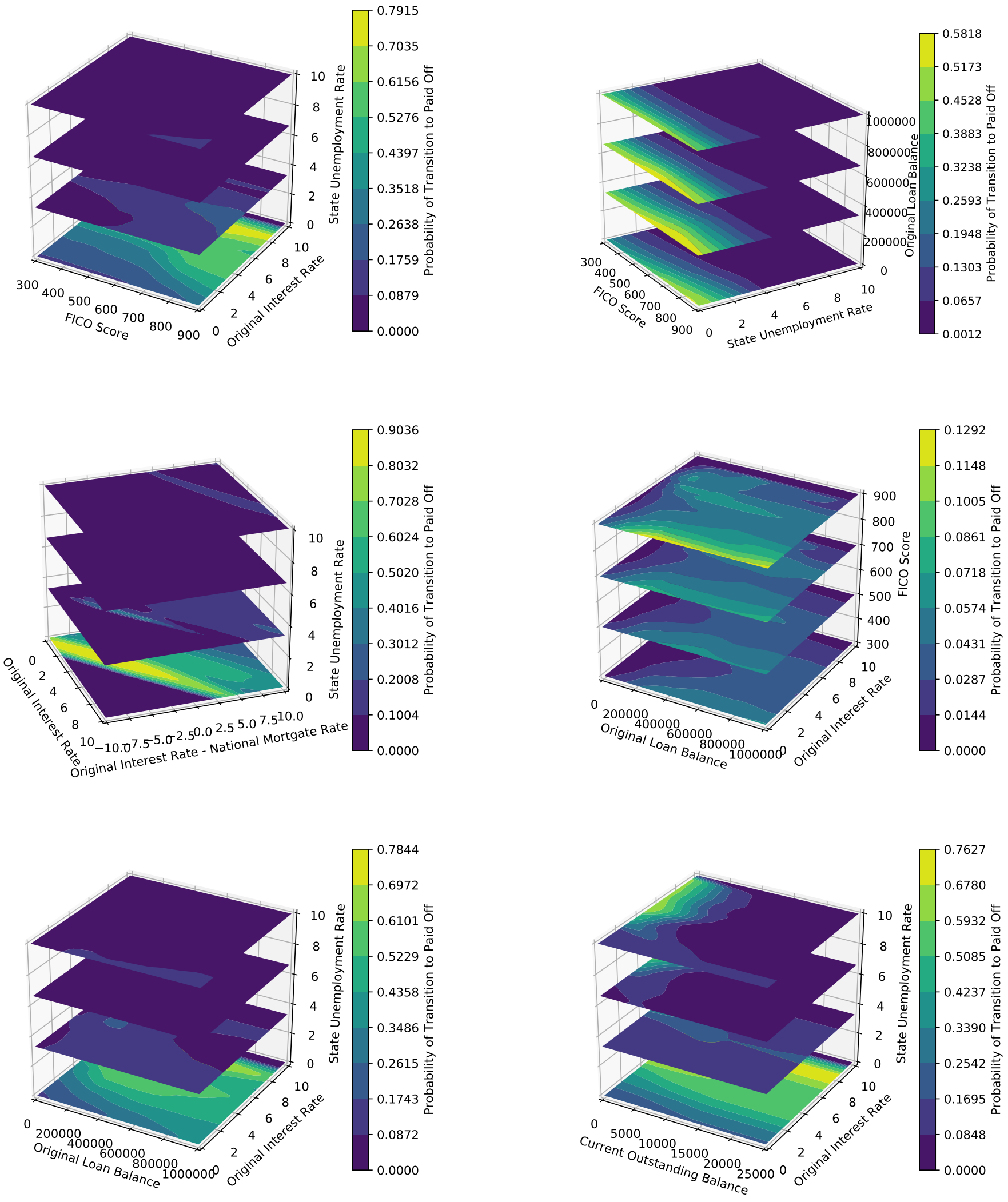

Figure 10: Relationship between Prepayment and Triplets of Covariates

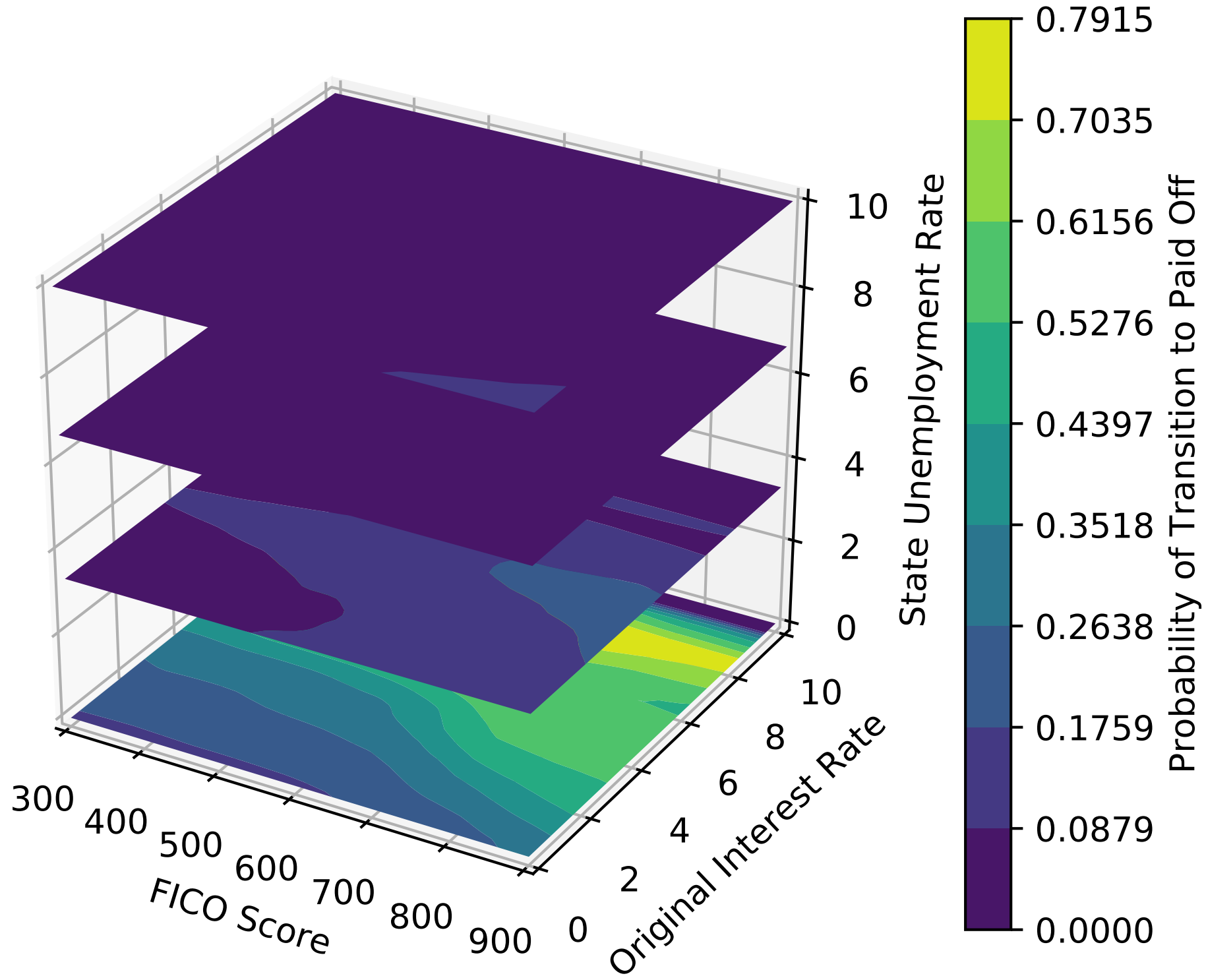

Figure 11: Relationship between original interest rate, original interest rate - national mortgage rate, and state unemployment rate.

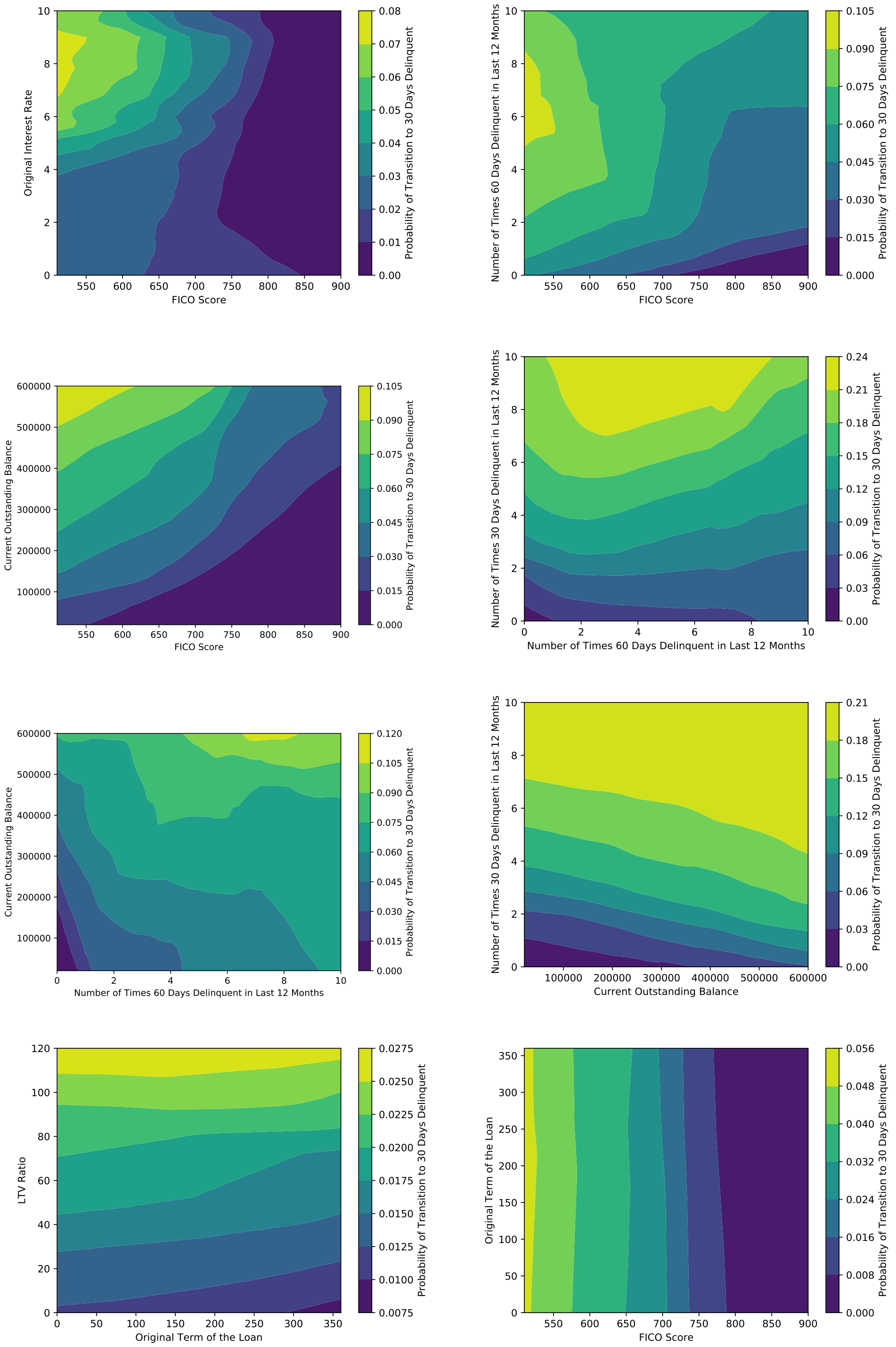

Figure 12: Relationship between delinquency and pairs of covariates.

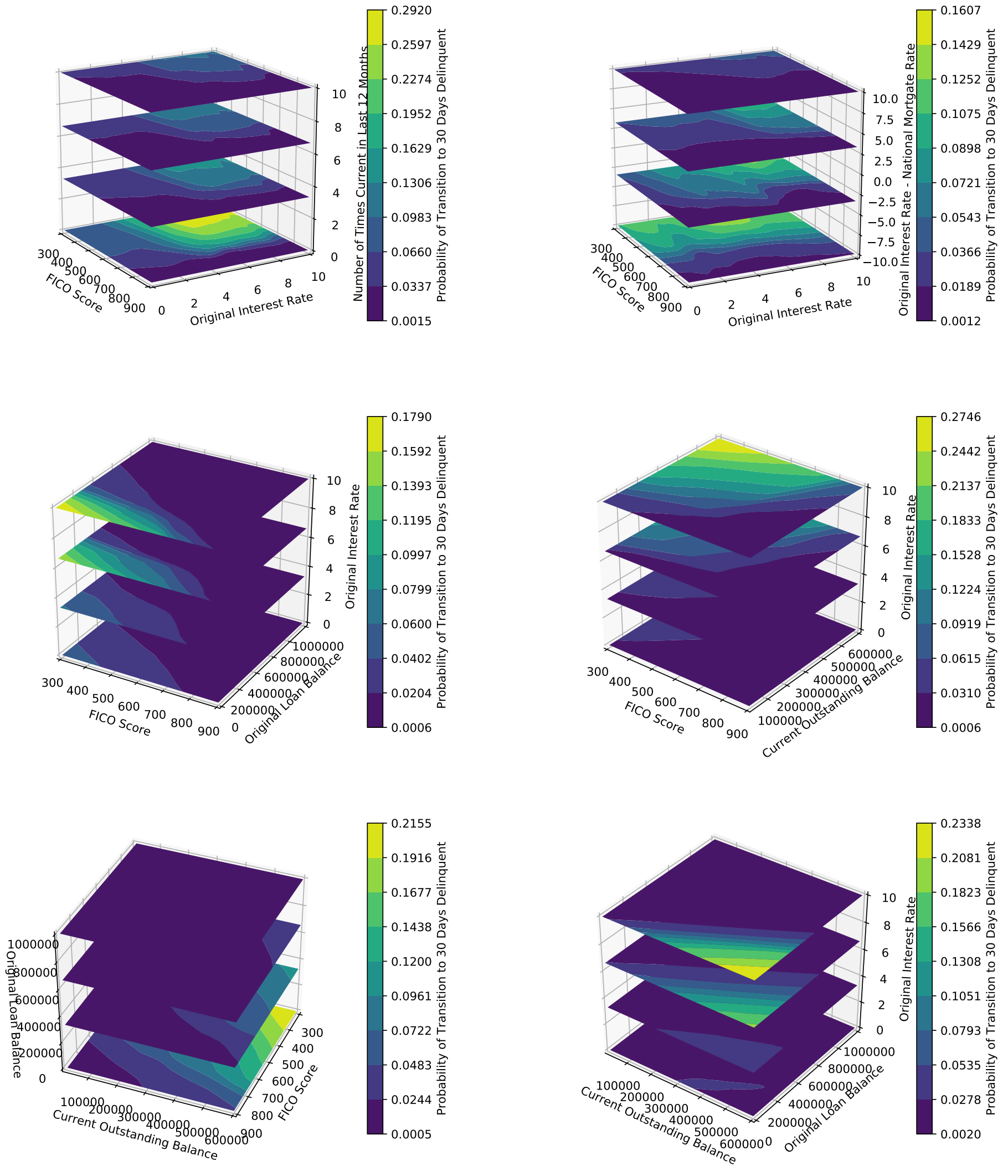

Figure 13: Relationship between Delinquency and Triplets of Covariates

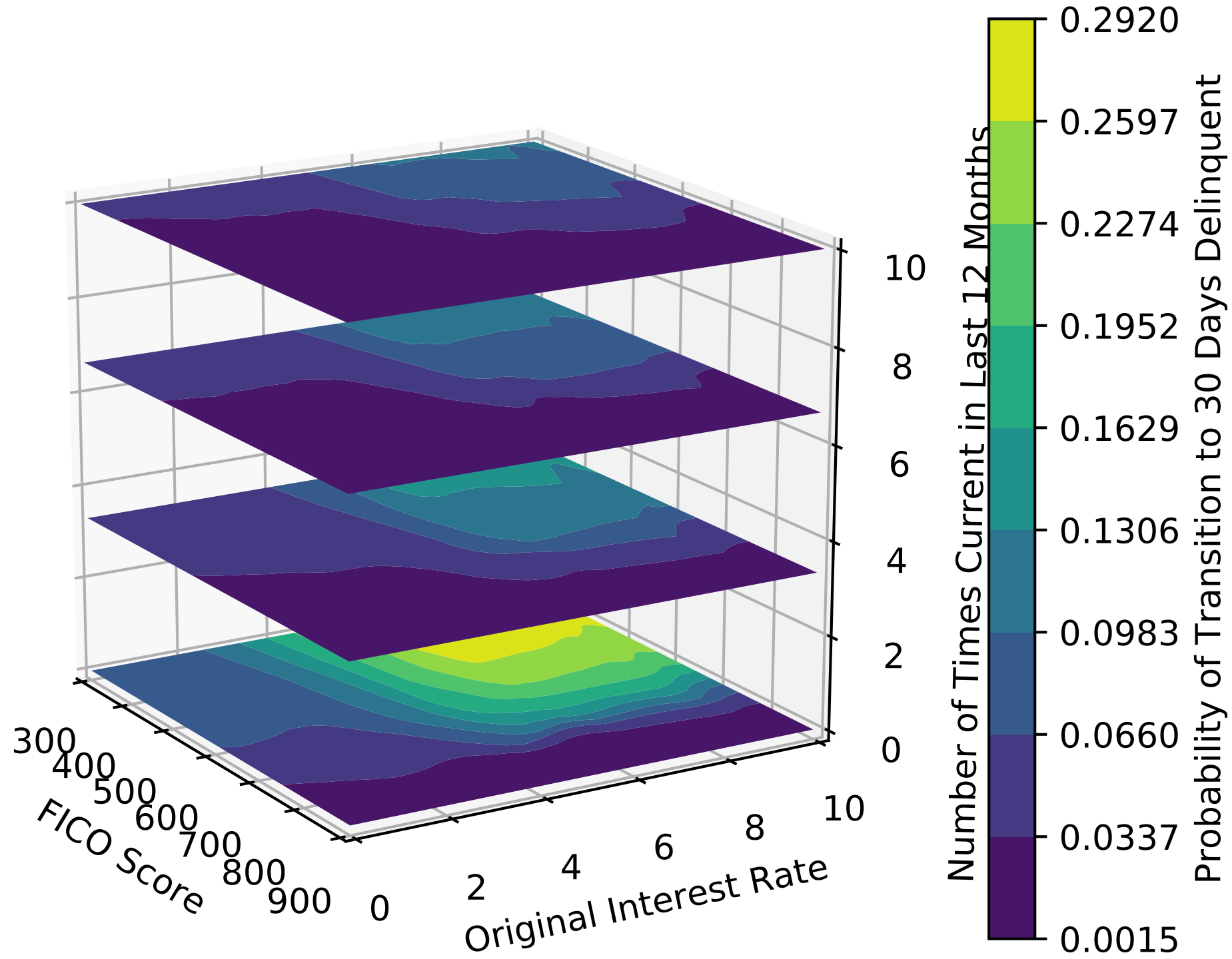

Figure 14: Relationship between FICO score, Original interest rate, and Number of times current in last 12 months.

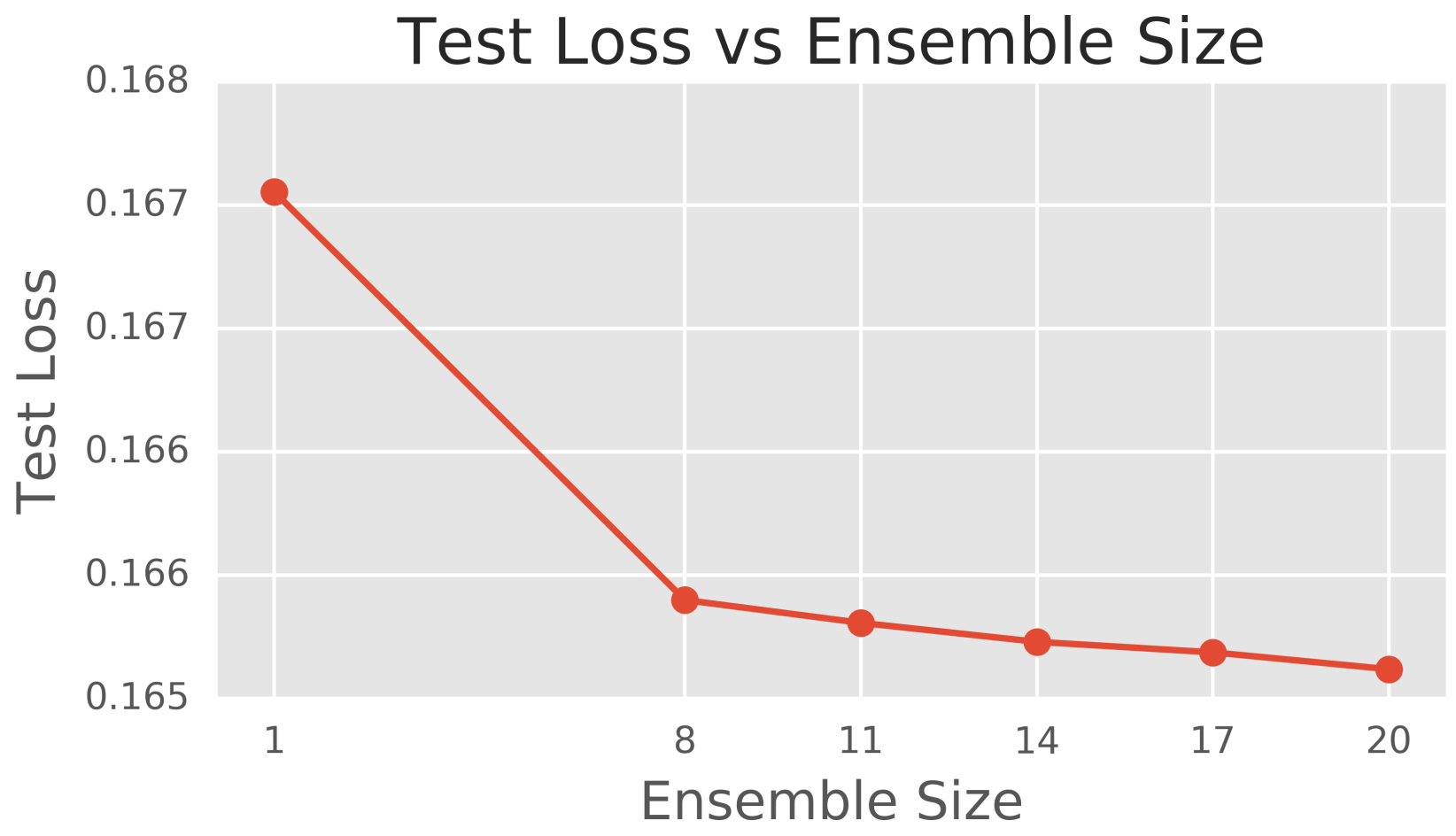

Figure 15: Out-of-sample loss (negative average log-likelihood) versus number of neural networks in the ensemble. The figure shows the improvement in the out-of-sample loss, an indicator of the performance of the ensemble, as the number of independently trained models in the ensemble are increased. Note that each model in an ensemble is a 5-layer neural network that is trained with bootstrapped data and random initialization chosen independently of that for other models. The predictions from all models within an ensemble are averaged to produce a low-variance estimate of transition probabilities, so the computational effort increases linearly with the ensemble size. The figure shows that the gains beyond an ensemble size of 8 are marginal and may not justify using bigger ensembles due to the computational burden.

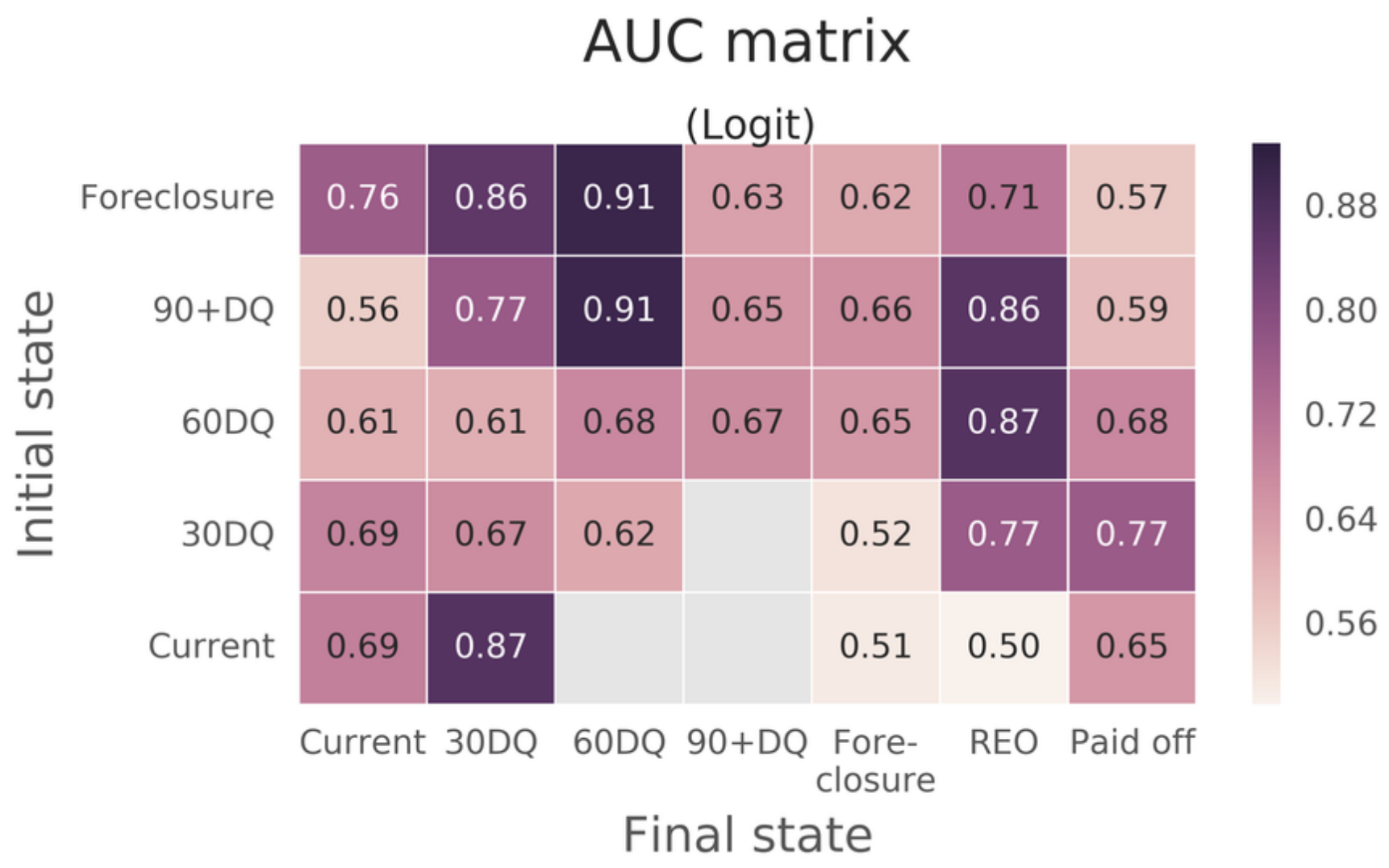

Figure 16: Out-of-sample AUCs for the 0-layer neural network (i.e., linear logistic regression) model. The AUC matrix above offers the most granular view into model performance.For mortgages in state $u$ in the current month, the AUC for event $u \to v$ is the AUC for the two-way classification of whether the mortgage will be in state $v$ or not next month. A higher value, depicted by a darker color, indicates better performance.

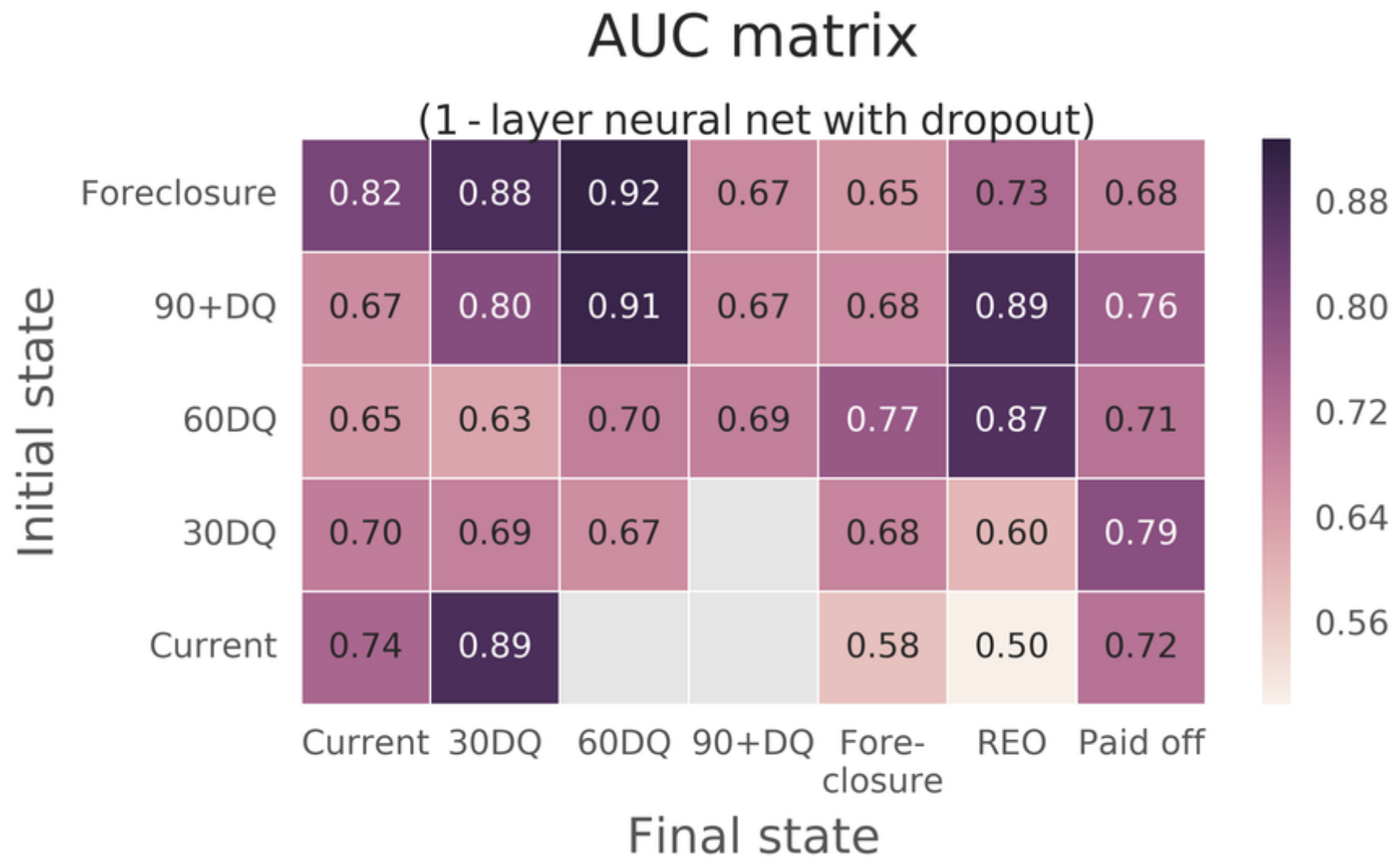

Figure 17: Out-of-sample AUCs for the 1-layer neural network. The AUC matrix above offers the most granular view into model performance. For mortgages in state $u$ in the current month, the AUC for event $u \to v$ is the AUC for the two-way classification of whether the mortgage will be in state $v$ or not next month. A higher value, depicted by a darker color, indicates better performance. We see marked improvement in the AUC values in going from the 0-layer model to the 1-layer network, especially for transitions to foreclosure and paid off as well as for the transitions from the delinquent states to current.

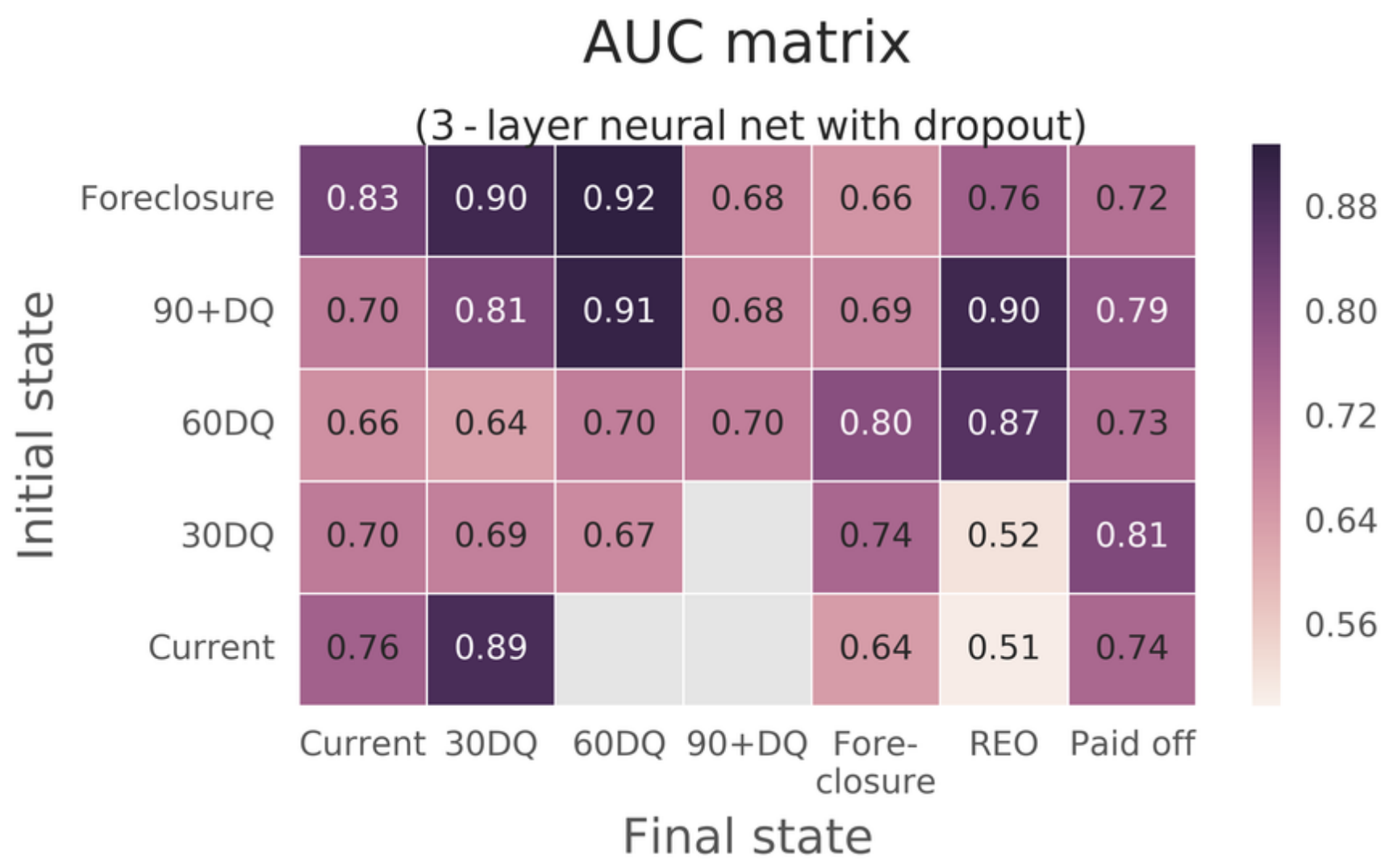

Figure 18: Out-of-sample AUCs for the 3-layer neural network. The AUC matrix above offers the most granular view into model performance. For mortgages in state $u$ in the current month, the AUC for event $u \to v$ is the AUC for the two-way classification of whether the mortgage will be in state $v$ or not next month. A higher value, depicted by a darker color, indicates better performance. We see further improvement in the AUC values in going from the shallow networks to the deeper networks.

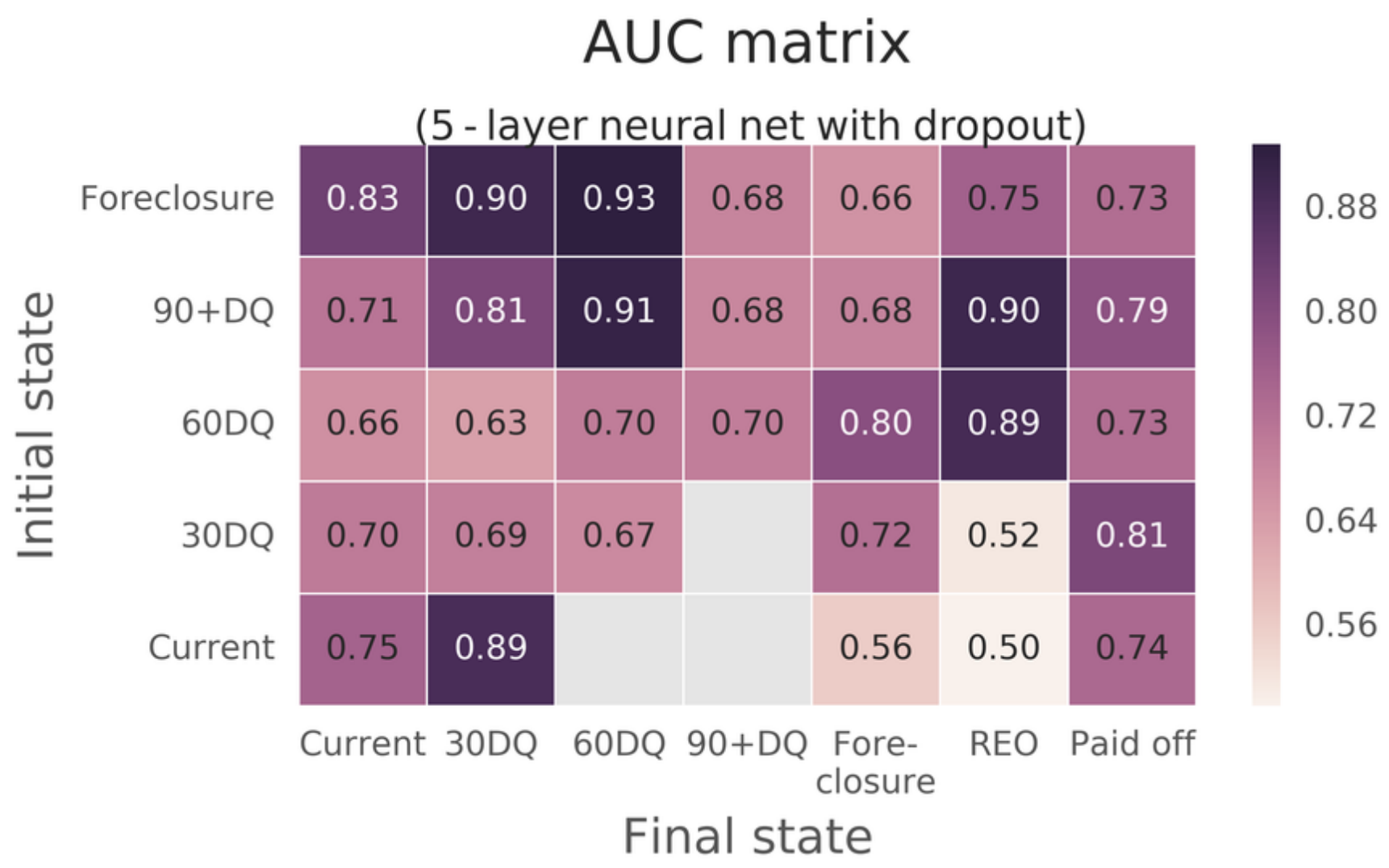

Figure 19: Out-of-sample AUCs for the 5-layer neural network. The AUC matrix above offers the most granular view into model performance. For mortgages in state $u$ in the current month, the AUC for event $u \to v$ is the AUC for the two-way classification of whether the mortgage will be in state $v$ or not next month. A higher value, depicted by a darker color, indicates better performance. We see further improvement in the AUC values in going from the 3-layer network to the 5-layer network.

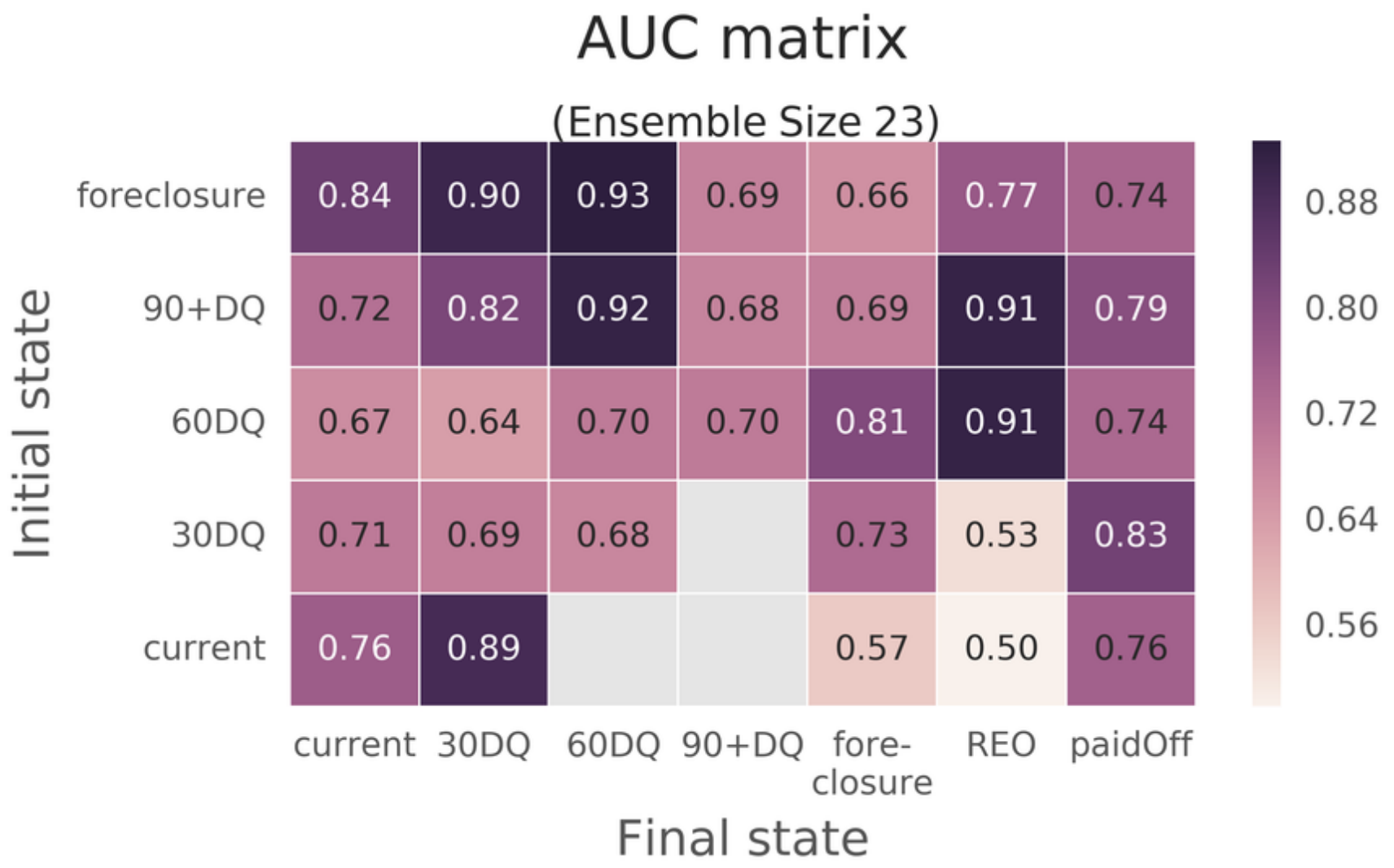

Figure 20: Out-of-sample AUCs for an ensemble of independently trained 5-layer neural networks. The AUC matrix above offers the most granular view into model performance. For mortgages in state $u$ in the current month, the AUC for event $u \to v$ is the AUC for the two-way classification of whether the mortgage will be in state $v$ or not next month. A higher value, depicted by a darker color, indicates better performance. In going from the 5-layer neural network to their ensemble, *every* transition in the matrix sees an improvement in prediction.

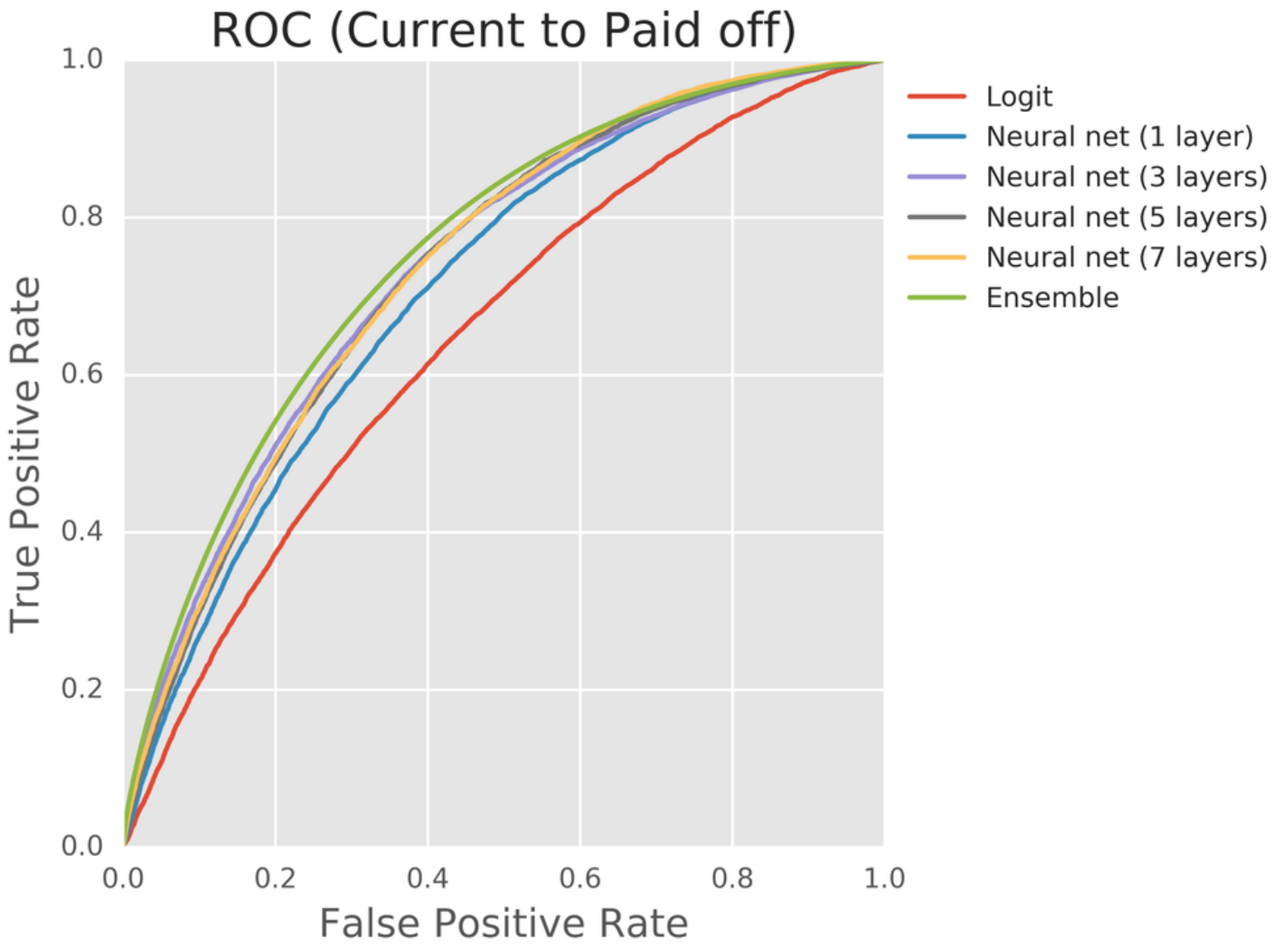

Figure 21: Out-of-sample ROC curves for various models for the transition current → paid off. The ROC curve corresponding to the ensemble dominates the curves for the individual networks, which in turn dominate the curve for the 0-layer (logistic regression) model. This implies that for those mortgages that are presently in the current state, predicting whether the state next month would be paid off or not is best predicted by the ensemble, followed by the networks with at least one layer, and then by the 0-layer model. Further, the gap between the curves for the 0-layer model and those for the deep neural networks indicates the significant gain in predictive power due to the modeling of more complex nonlinear relationships obtained by adding multiple hidden layers to the model.

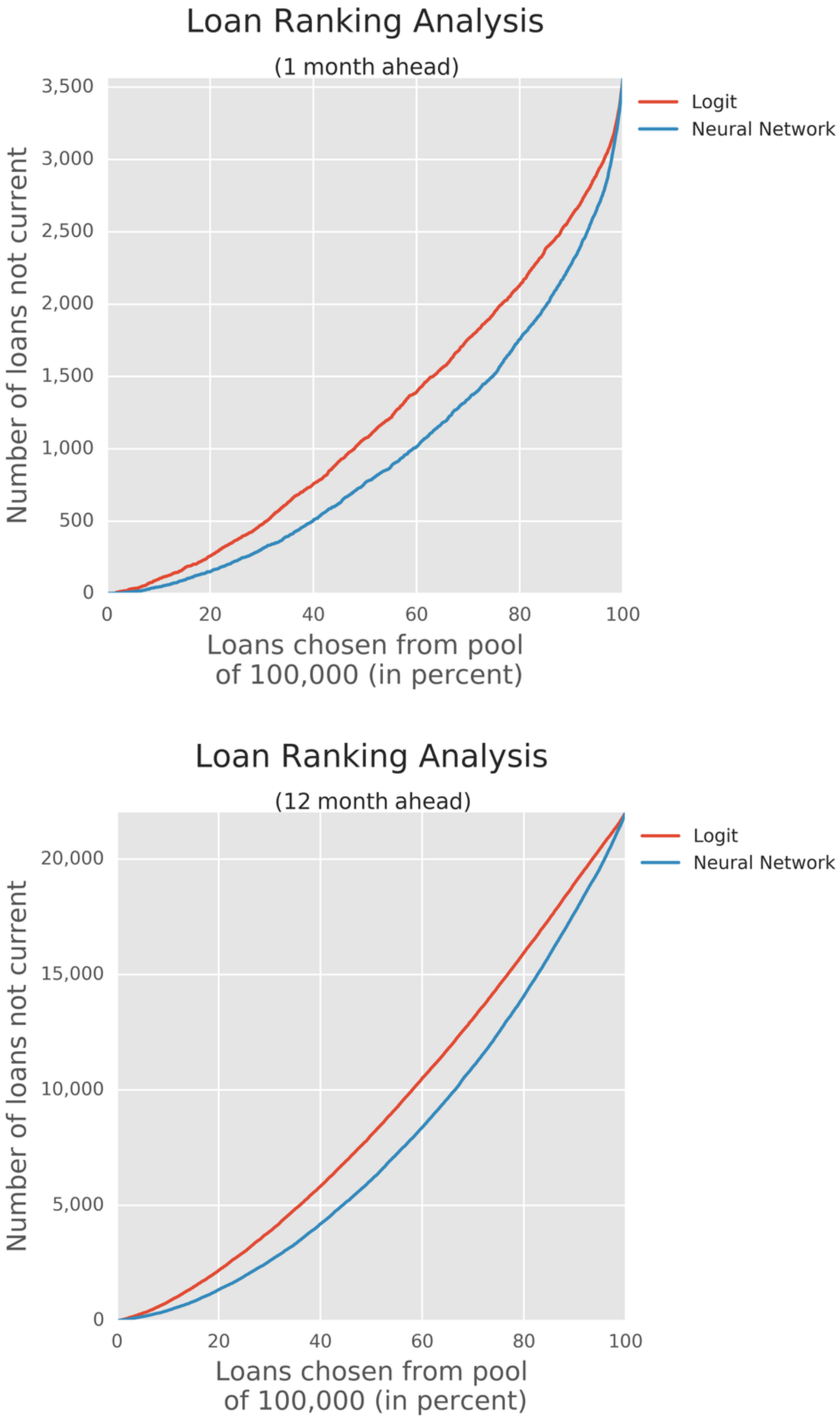

Figure 22: Comparison of out-of-sample performance for 5-layer network portfolio and 0-layer network portfolio for 1-month (top) and 1-year (bottom) ahead horizon. From a random pool of $100,000$ loans, an investor selects $N$ loans so as to maximize the number of loans (out of these $N$) that remain current after one (twelve) months. This requires ranking the loans on their probability of remaining current in the next month (year) and then selecting the top $N$ loans. This ranking is obtained for two models, and the number of loans selected, $N$, is varied from 0 to $100,000$. The figure shows for each portfolio size $N$ (expressed as percent of the pool size) on the x-axis the corresponding number of loans that are not current in the subsequent month (year) on the y-axis. The portfolio constructed using the 5-layer neural network yields superior performance for all portfolio sizes. Note that the curves intersect at the end points by design, since the portfolios selected for $N = 0$ (no loans) and $N = 100,000$ (entire pool of loans) are identical.

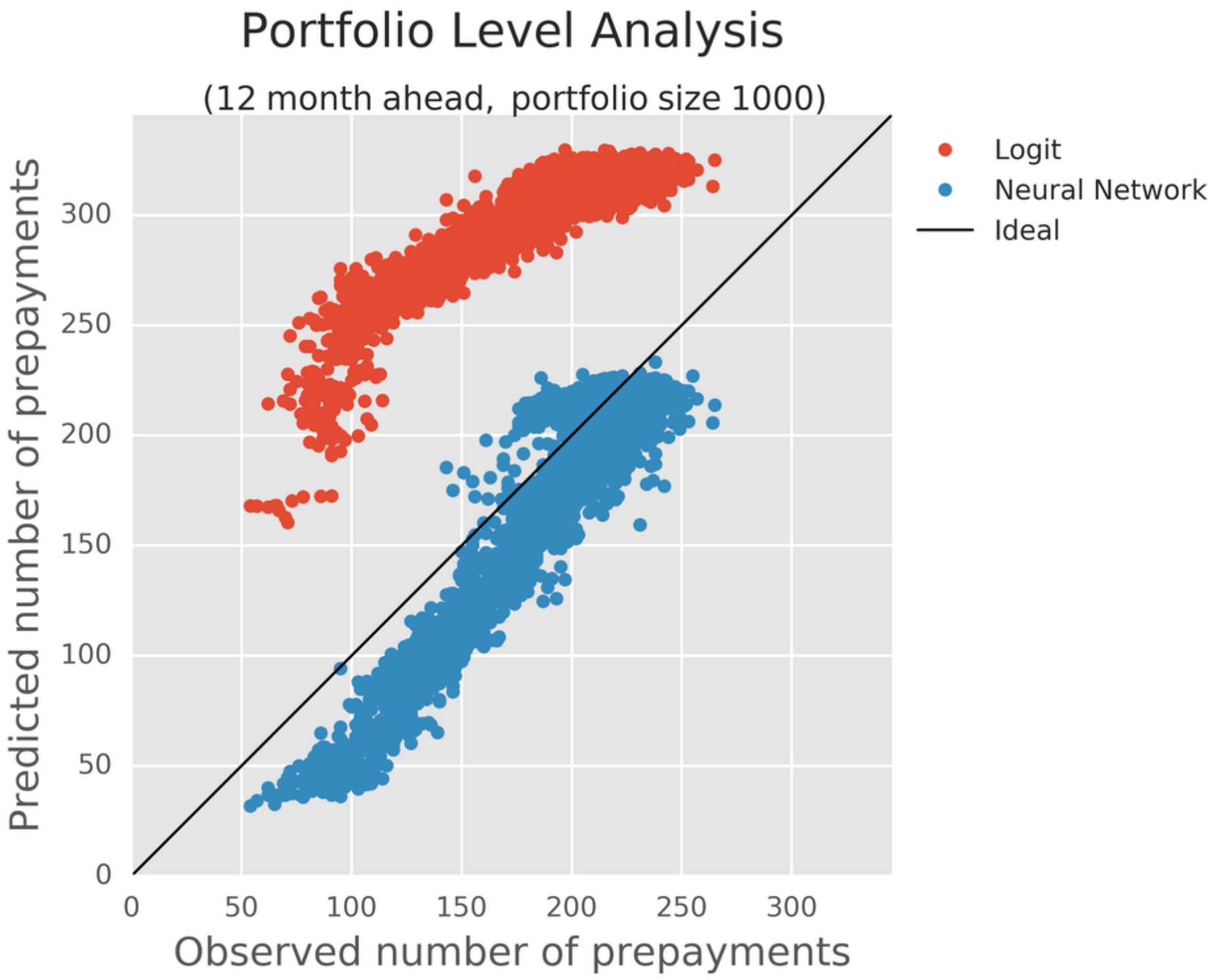

Figure 23: Comparison of out-of-sample pool-level predictions of the 5-layer network and the 0-layer model. A pool of 2 million mortgages is grouped into $2,000$ portfolios by ordering loans according to the borrowers' FICO score and then sequentially packaging every 1,000 loans into individual portfolios. For each such portfolio, the figure shows the observed number of prepayments in the next 12 months on the x-axis and the predicted number of prepayments in the next 12 months from the two models, the 5-layer neural network and the logistic regression model, on the y-axis. The $x = y$ line (in black) shows the ideal but hypothetical scenario under which the predicted and the observed number of prepayments coincide. It is seen that the predictions from the 5-layer neural network are much closer to this ideal line than those from the 0-layer model.

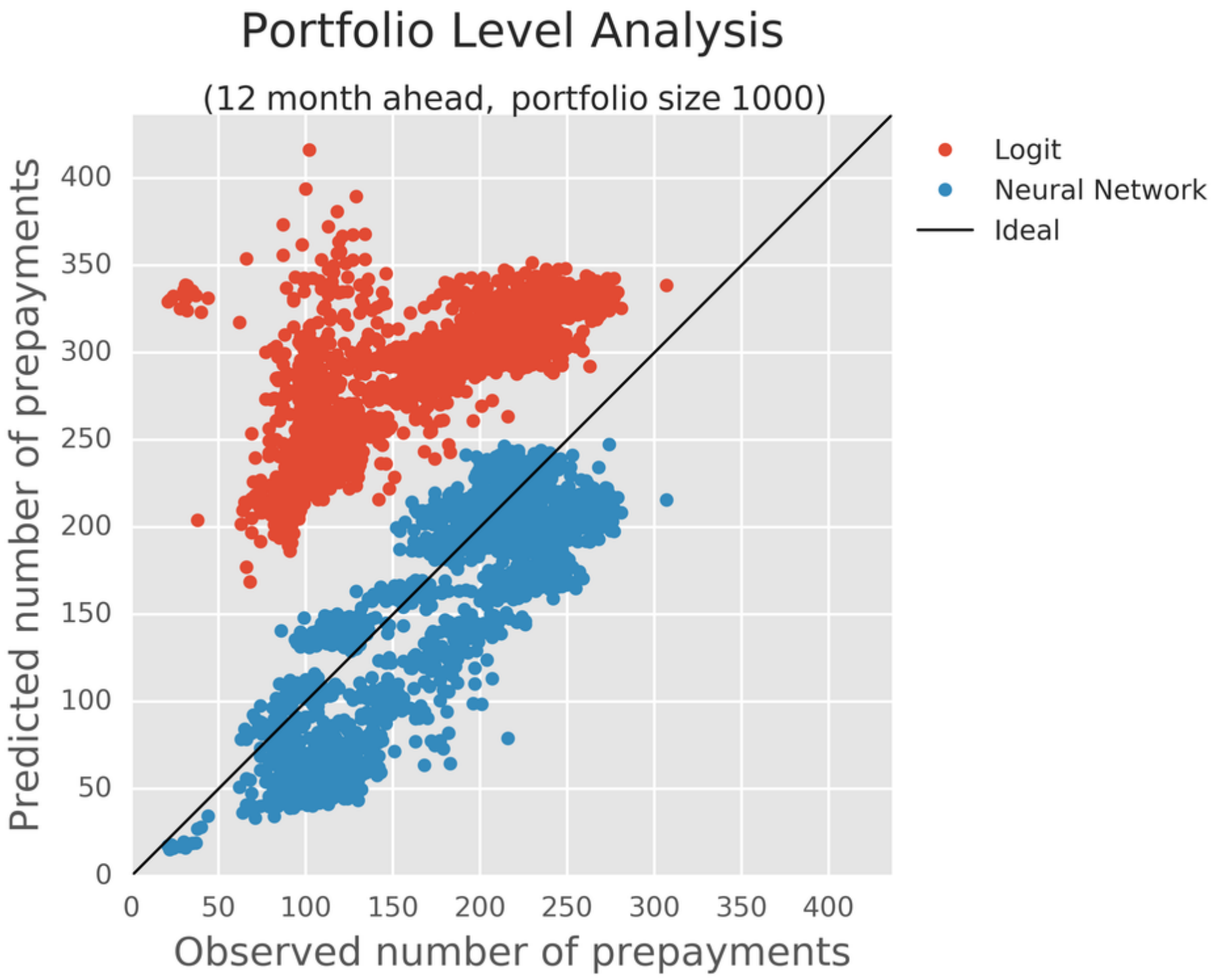

Figure 24: Comparison of out-of-sample pool-level predictions of the 5-layer neural network and the logistic regression model. A pool of 2 million mortgages is grouped into 2,000 portfolios by ordering loans according to their initial interest rate and then sequentially packaging every 1,000 loans into individual portfolios. For each such portfolio, the figure shows the observed number of prepayments in the next 12 months on the x-axis and the predicted number of prepayments in the next 12 months from the two models, the 5-layer neural network and the logistic regression model, on the y-axis. The $x = y$ line (in black) shows the ideal but hypothetical scenario under which the predicted and the observed number of prepayments coincide. It is seen that the predictions from the 5-layer neural network are much closer to this ideal line than those from the logistic regression model.

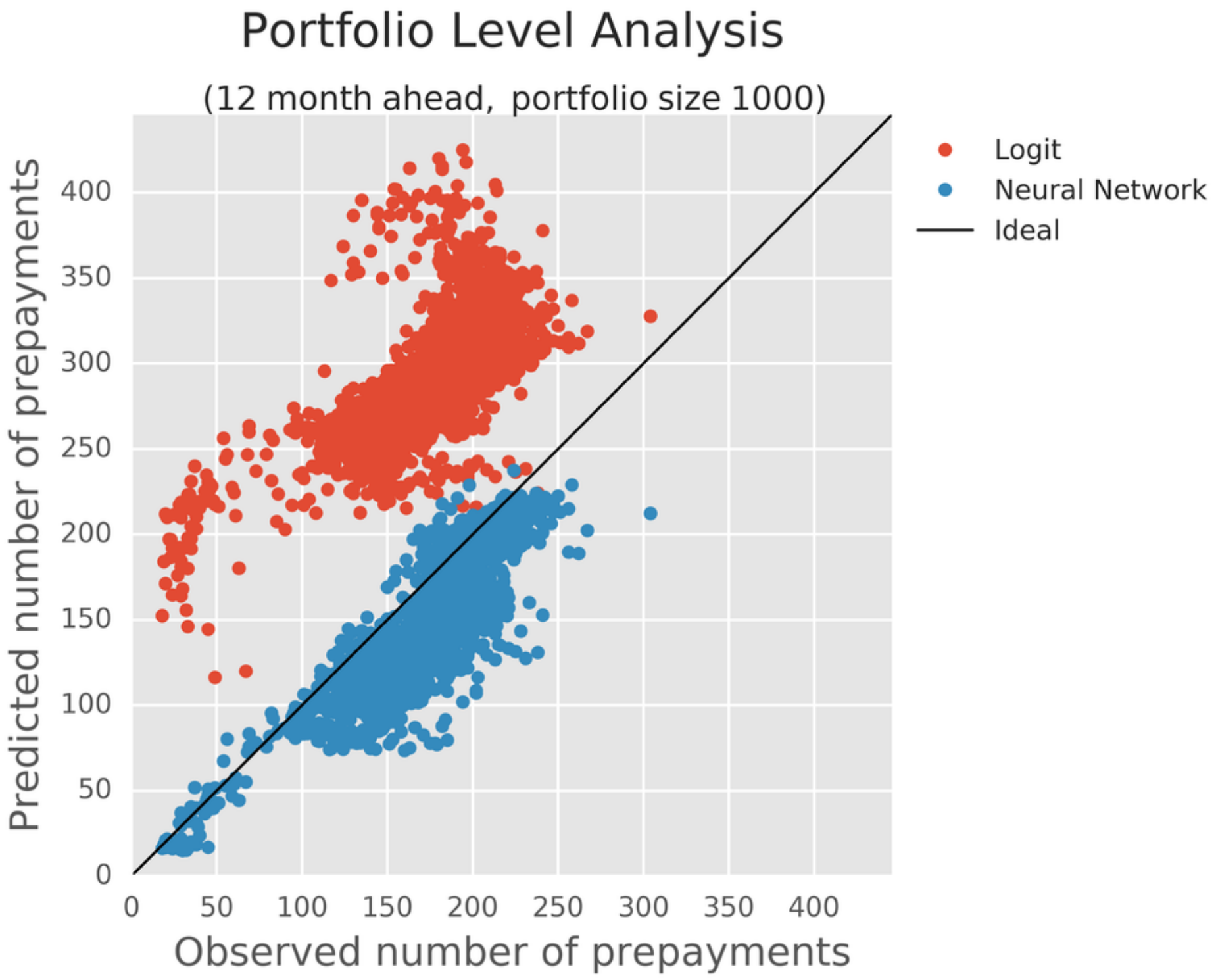

Figure 25: Comparison of out-of-sample pool-level predictions of the 5-layer neural network and the logistic regression model. A pool of 2 million mortgages is grouped into $2,000$ portfolios by ordering loans according to their loan-to-value (LTV) ratio and then sequentially packaging every 1,000 loans into individual portfolios. For each such portfolio, the figure shows the observed number of prepayments in the next 12 months on the x-axis and the predicted number of prepayments in the next 12 months from the two models, the 5-layer neural network and the logistic regression model, on the y-axis. The $x = y$ line (in black) shows the ideal but hypothetical scenario under which the predicted and the observed number of prepayments coincide. It is seen that the predictions from the 5-layer neural network are much closer to this ideal line than those from the logistic regression model.

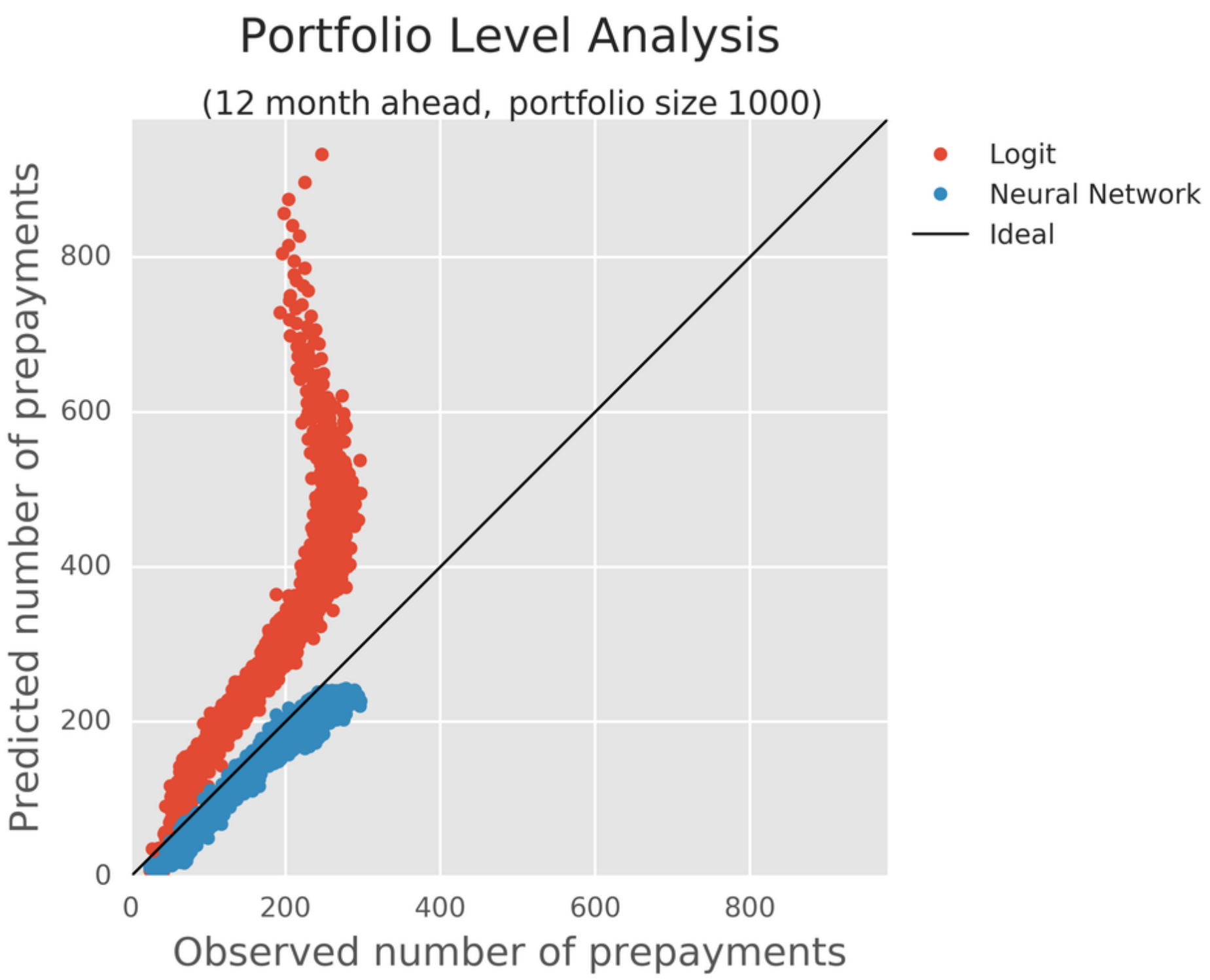

Figure 26: Comparison of out-of-sample pool-level predictions of the 5-layer neural network and the logistic regression model. A pool of 2 million mortgages is grouped into 2,000 portfolios by ordering loans according to their probability of being current after 12 months and then sequentially packaging every 1,000 loans into individual portfolios. For each such portfolio, the figure shows the observed number of prepayments in the next 12 months on the x-axis and the predicted number of prepayments in the next 12 months from the two models, the 5-layer neural network and the logistic regression model, on the y-axis. The $x = y$ line (in black) shows the ideal but hypothetical scenario under which the predicted and the observed number of prepayments coincide. It is seen that the predictions from the 5-layer neural network are much closer to this ideal line than those from the logistic regression model. It is important to note here that the loans were ordered on their probability of being current after 12 months, where this probability is estimated using the logistic regression model. If the estimated probabilities were accurate, the portfolios so obtained would have large variations in quality with the observed number of prepayments covering the entire x-axis (as in previous plots) as well as the logistic regression model would see an increasing curve; however, neither of these trends is observed, implying that the predicted probabilities are inaccurate.

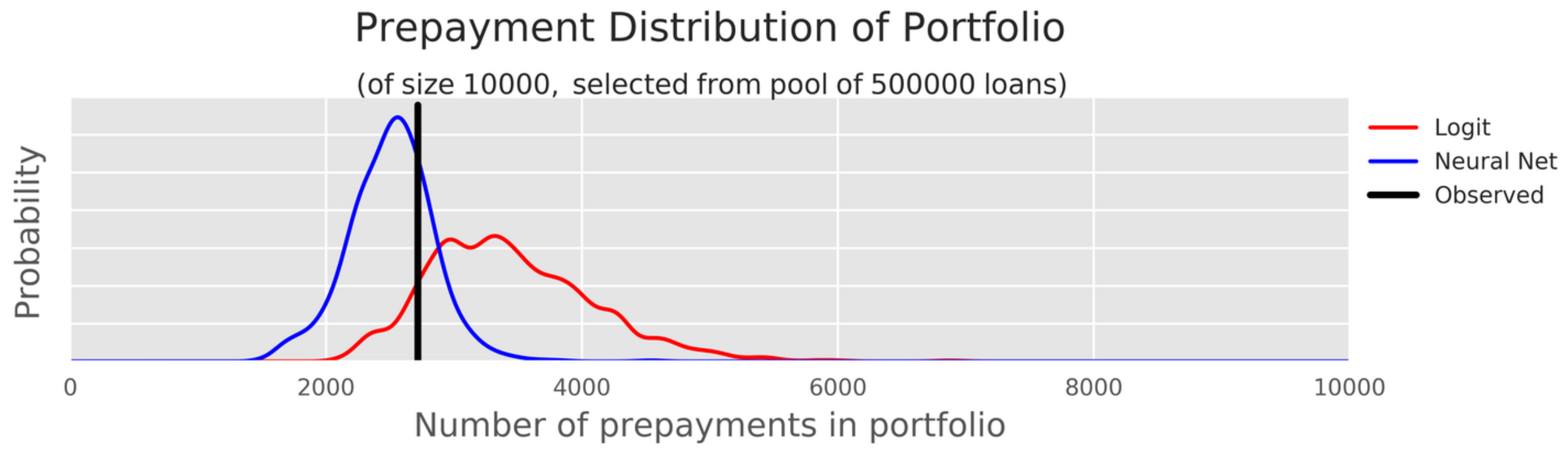

Figure 27: Comparison of out-of-sample pool-level distribution from the 5-layer neural network and the logistic regression model. The distribution of the number of prepayments at a 12-month horizon is obtained by simulating several trajectories for the time-varying covariates and then computing the transition probabilities for each loan for every trajectory; this approach is described in Section 3. For a wide range of portfolios, we observe that the gap between the mean of the distribution and the actual number of prepayments is lesser for the neural network model than for the logistic regression model.

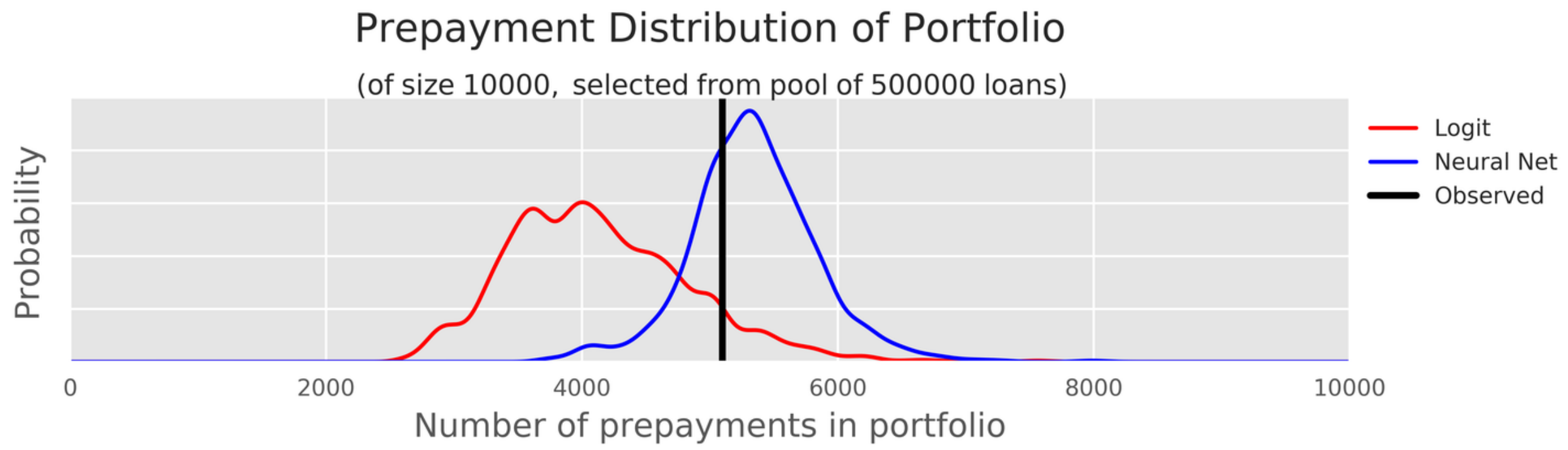

Figure 28: Comparison of out-of-sample pool-level distribution from the 5-layer neural network and the logistic regression model. The distribution of the number of prepayments at a 12-month horizon is obtained by simulating several trajectories for the time-varying covariates and then computing the transition probabilities for each loan for every trajectory; this approach is described in Section 3. For a wide range of portfolios, we observe that the gap between the mean of the distribution and the actual number of prepayments is lesser for the neural network model than for the logistic regression model.